\documentclass[12pt]{report}

\makeatletter
\renewenvironment{abstract}{%
    \if@twocolumn
    \section*{\abstractname}%
    \else
    \small
    \begin{center}%
        {\bfseries \abstractname\vspace{-.5em}\vspace{\z@}}%
    \end{center}%
    \quotation
    \fi}
{\if@twocolumn\else\endquotation\fi}
\makeatother

\usepackage{geometry}
\usepackage[T1]{fontenc}
\usepackage[title]{appendix}
\usepackage{amsfonts}
\usepackage{amsmath}
\usepackage{balance}
\usepackage{booktabs}
\usepackage{chngcntr}
\usepackage{color}
\usepackage{float}
\usepackage{graphicx}
\usepackage{listings}
\usepackage{makecell}
\usepackage{mdwlist}
\usepackage{multicol}
\usepackage{multirow}
\usepackage{paralist}
\usepackage{rotating}
\usepackage{subcaption}
\usepackage{tabularx}
\usepackage{tabulary}
\usepackage{times}
\usepackage{todonotes}
\usepackage{url}
\usepackage{xfrac}
\usepackage{xspace}
\usepackage{lmodern}
\usepackage[british]{babel}
\usepackage[useregional]{datetime2}
\usepackage{setspace}
\usepackage{hyperref}
\usepackage[nottoc,notlot,notlof]{tocbibind}
\usepackage{soul}

\presetkeys{todonotes}{inline}{}

\hypersetup{
    colorlinks,
    citecolor=black,
    filecolor=black,
    linkcolor=black,
    urlcolor=black
}

\DTMlangsetup[en-GB]{showdayofmonth=false}

\newcommand{\excision}{\textsc{Excision}\xspace}

\definecolor{lightgray}{rgb}{.6,.6,.6}
\definecolor{darkgray}{rgb}{.4,.4,.4}

\lstdefinelanguage{JSON}{
  keywords={break, case, catch, continue, debugger, default, delete, do, else, false, finally, for, function, if, in, instanceof, new, null, return, switch, this, throw, true, try, typeof, var, void, while, with},
  morecomment=[l]{//},
  morecomment=[s]{/*}{*/},
  morestring=[b]',
  morestring=[b]",
  ndkeywords={class, export, boolean, throw, implements, import, this},
  keywordstyle=\color{black}\bfseries,
  ndkeywordstyle=\color{darkgray}\bfseries,
  identifierstyle=\color{black},
  commentstyle=\color{lightgray}\ttfamily,
  stringstyle=\color{black}\ttfamily,
  sensitive=true
}

\lstdefinelanguage{JavaScript}{
  keywords={break, case, catch, continue, debugger, default, delete, do, else, false, finally, for, function, if, in, instanceof, new, null, return, switch, this, throw, true, try, typeof, var, void, while, with},
  morecomment=[l]{//},
  morecomment=[s]{/*}{*/},
  morestring=[b]',
  morestring=[b]",
  ndkeywords={class, export, boolean, throw, implements, import, this},
  keywordstyle=\color{black}\bfseries,
  ndkeywordstyle=\color{darkgray}\bfseries,
  identifierstyle=\color{black},
  commentstyle=\color{lightgray}\ttfamily,
  stringstyle=\color{red}\ttfamily,
  sensitive=true
}

\newcommand{\origintracer}{\textsc{OriginTracer}\xspace}

\newcommand \redbold[1]{\textcolor{red}{\textbf{#1}}}
\newcommand \bluebold[1]{\textcolor{blue}{\textbf{#1}}}

\begin{document}

\newgeometry{top=2in,bottom=1in}

\begin{titlepage}
    \centering

    {\bf\Huge Understanding and Mitigating the Security Risks of Content Inclusion in Web Browsers \par}

    \vspace{0.5cm}

    {\large A dissertation presented in partial fulfillment of \par}
    {\large the requirements for the degree of \par}

    \vspace{0.5cm}

    {\huge Doctor of Philosophy \par}

    \vspace{0.5cm}

    {\large in the field of \par}

    \vspace{0.5cm}

    {\huge Information Assurance \par}

    \vspace{0.5cm}

    {\large by \par}

    \vspace{0.5cm}

    {\huge Sajjad Arshad \par}

    \vspace{0.5cm}

    {\large Khoury College of Computer Sciences \par}
    {\large Northeastern University \par}

    \vspace{1.5cm}

    {\large Ph.D. Committee \par}

    \vspace{0.5cm}

    \large {
    \begin{tabular}{ll}
        William Robertson & Advisor, Northeastern University \\
        Engin Kirda & Advisor, Northeastern University \\
        Guevara Noubir & Internal, Northeastern University \\
        Gianluca Stringhini & External, Boston University \\
    \end{tabular}
    }

    \vfill

    {\large \today\par}
\end{titlepage}

\doublespacing

\begin{abstract}

Thanks to the wide range of features offered by web browsers, modern websites
include various types of content such as JavaScript and Cascading Style Sheets
(CSS) in order to create interactive user interfaces. Browser vendors also
provided extensions to enhance web browsers with additional useful capabilities
that are not necessarily maintained or supported by default.

However, included content can introduce security risks to users of these
websites, unbeknownst to both website operators and users. In addition, the
browser's interpretation of the resource URLs may be very different from how the
web server resolves the URL to determine which resource should be returned to
the browser. The URL may not correspond to an actual server-side file system
structure at all, or the web server may internally rewrite parts of the URL.
This semantic disconnect between web browsers and web servers in interpreting
relative paths (\textit{path confusion}) could be exploited by \textit{Relative
Path Overwrite (RPO)}. On the other hand, even tough extensions provide useful
additional functionality for web browsers, they are also an increasingly popular
vector for attacks. Due to the high degree of privilege extensions can hold,
extensions have been abused to inject advertisements into web pages that divert
revenue from content publishers and potentially expose users to malware.

In this thesis, I propose novel research into understanding and mitigating the
security risks of content inclusion in web browsers to protect website
publishers as well as their users. First, I introduce an in-browser approach
called \textsc{Excision} to automatically detect and block malicious third-party
content inclusions as web pages are loaded into the user's browser or during the
execution of browser extensions. Then, I propose \textsc{OriginTracer}, an
in-browser approach to highlight extension-based content modification of web
pages. Finally, I present the first in-depth study of style injection
vulnerability using RPO and discuss potential countermeasures.

\end{abstract}

\chapter*{Acknowledgments}

I would like to thank my advisors, William Robertson and Engin Kirda, for their
support and valuable insights during my Ph.D. career. I am also thankful for
working alongside my brilliant colleagues from whom I learned a lot: Amin
Kharraz, Tobias Lauinger, Kaan Onarlioglu, Christo Wilson, Muhammad Ahmad
Bashir, Abdelberi Chaabane, Michael Weissbacher, and Mansour Ahmadi.

Special thanks to my brother, Kazem, for encouraging me to study computer
science in the first place, my brother, Saleh, for supporting me during the time
I was living far from home, Ali Mirheidari for pulling me into the web security
world, and Reza Mirzazade farkhani for pushing me toward playing CTFs.

Last but not least, I thank my whole family, specially my mom and my sisters,
for their support and patience during all these years.

\restoregeometry

\pagebreak

\tableofcontents

\pagebreak

\listoffigures

\pagebreak
 
\listoftables

\pagebreak

\doublespacing

\chapter{Introduction}

Linking to the web content has been one of the defining features of the World
Wide Web since its inception, and this feature remains strongly evident today.
For instance, recent research~\cite{ccs2012jsinclusion} reveals that more than
93\% of the most popular websites include JavaScript from external sources.
Developers typically include third-party content for convenience and performance
-- e.g., many JavaScript libraries are hosted on fast content delivery networks
(CDNs) and are likely to already be cached by users -- or to integrate with
advertising networks, analytics frameworks, and social media. Content inclusion
has also been used by entities other than the website publishers themselves. For
example, ad injection has been adopted by ISPs and browser extension authors as
a prominent technique for monetization~\cite{adinjection_profit}. Browser
extensions enhance browsers with additional useful capabilities that are not
necessarily maintained or supported by the browser vendor. Instead, this code is
typically written by third parties and can perform a wide range of tasks, from
simple changes in the appearance of web pages to sophisticated tasks such as
fine-grained filtering of content. To achieve these capabilities, browser
extensions possess more privilege than other third-party code that runs in the
browser. For instance, extensions can access cross-domain content, and perform
network requests that are not subject to the same origin policy.

However, the inherent feature of content-sharing on the Web is also an Achilles
heel when it comes to security. Advertising networks, as one example, have
emerged as an important vector for adversaries to distribute attacks to a wide
audience~\cite{sp2013linchpins,ccs2012madtracer,www2014shortening,ccs2013spiderweb,imc2014malvertising}.
Moreover, users are more susceptible to \emph{malvertising} in the presence of
ad injection~\cite{usenixsec2015webeval,sp2015adinjection,www2015adinjection}.
In general, linking to third-party content is essentially an assertion of trust
that the content is benign. This assertion can be violated in several ways,
however, due to the dynamic nature of the Web. Since website operators cannot
control external content, they cannot know \emph{a priori} what links will
resolve to in the future. The compromise of linked content or pure malfeasance
on the part of third parties can easily violate these trust assumptions. This is
only exacerbated by the transitive nature of trust on the Web, where requests
for content can be forwarded beyond the first, directly observable origin to
unknown parties.

Furthermore, since extensive capabilities of browser extensions allow a
comparatively greater degree of control over the browser, they provide a unique
opportunity to attack users and their data, the underlying system, and even the
Internet at large. For this reason, newer browser extension frameworks such as
Chromium's have integrated least privilege separation via isolated worlds and a
fine-grained permissions system to restrict the capabilities of third-party
extensions~\cite{tr2008chromium_security_architecture}. However, extension
security frameworks are not a panacea. In practice, their effectiveness is
degraded by over-privilege and a lack of understanding of the threats posed by
highly-privileged extensions on the part of
users~\cite{usenixwebapps2011app_permissions}. Indeed, despite the existence of
extension security frameworks, it has recently been shown that extension-based
advertisement injection has become a popular and lucrative technique for
dishonest parties to monetize user web browsing. These extensions simply inject
or replace ads in web pages when users visit a website, thus creating or
diverting an existing revenue stream to the third party. Users often are not
aware of these incidents and, even if this behavior is noticed, it can be
difficult to identify the responsible party.

Web browsers also load internal resources using either absolute URLs or relative
ones. Before a web browser can issue a request for such a resource to the
server, it must expand the relative path into an absolute URL. Web browsers
basically treat URLs as file system-like paths. However, the browser's
interpretation of the URL may be very different from how the web server resolves
the URL to determine which resource should be returned to the browser. The URL
may not correspond to an actual server-side file system structure at all, or the
web server may internally rewrite parts of the URL. This semantic disconnect
between web browsers and web servers in interpreting relative paths
(\textit{path confusion}) could be exploited by \textit{Relative Path Overwrite
(RPO)}. When an injection vulnerability is present in a page, an attacker could
manipulate the URL such that the web page references itself as the stylesheet,
which turns a simple text injection vulnerability into a style sink~\cite{rpo}.
The general threat model of RPO resembles that of Cross-Site Scripting (XSS).
Typically, the attacker's goal is to steal sensitive information from a
third-party site or make unauthorized transactions on the site, such as gaining
access to confidential financial information or transferring money out of a
victim's account.

\section{Thesis Contributions}

Due to the increasing reliance of users on web browsers for day to day
activities, I believe it is important to characterize the extent of security
risks of content inclusion on the Web. In this thesis, I investigate the
feasibility and effectiveness of novel approaches to measure and reduce the
security risks for website publishers as well as their users. I show that our
novel techniques are complementary to the existing defenses. To support my
claim, I propose the following:

First, I present a novel in-browser approach called \excision that automatically
detects and blocks malicious third-party content before it can attack the user's
browser. The approach leverages a high-fidelity in-browser vantage point that
allows it to construct a precise inclusion sequence for every third-party
resource. We also describe a prototype of \excision for the Chromium browser
that can effectively prevent inclusions of malicious content. Furthermore, we
evaluate the effectiveness and performance of our prototype, and show that it is
able to automatically detect and block malicious third-party content inclusions
in the wild -- including malicious resources not previously identified by
popular malware blacklists -- without a significant impact on browser
performance. Finally, we evaluate the usability of our prototype and show that
most users did not notice any significant quality impact on their browsing
experience.

Then, I introduce a novel in-browser approach to provenance tracking for web
content at the granularity of DOM elements, and present semantics for provenance
propagation due to script and extension execution. The approach leverages a
high-fidelity in-browser vantage point that allows it to construct a precise
provenance label set for each DOM element introduced into a web page. We also
implement a prototype called \origintracer that uses content provenance to
identify and highlight third-party content injection -- e.g., unwanted
advertisements -- by extensions to notify users of their presence and the
originating principal. Furthermore, we evaluate the effectiveness, performance,
and usability of our prototype, and show that it is able to significantly assist
users in identifying ad injection by extensions in the wild without degrading
browser performance or the user experience.

Finally, I present the first automated and large-scale study of the prevalence
and significance of RPO vulnerabilities in the wild. To date, little is known
about how widespread RPO vulnerabilities are on the Web. Especially since the
attack is more recent and less well-known than traditional XSS, we believe it is
important to characterize the extent of the threat and quantify its enabling
factors. Our measurement methodology tests how often these preconditions hold in
the wild in order to quantify the vulnerability and exploitability of pages with
respect to RPO attacks. We enumerate a range of factors that prevent a
vulnerable page from being exploited, and discuss how these could be used to
mitigate these vulnerabilities.

\section{Thesis Structure}

The remainder of this thesis is organized as follows.
Chapter~\ref{chap:related_work} presents the related work. The design and
implementation of \excision for detecting malicious third-party content
inclusions are introduced in Chapter~\ref{chap:inclusion}.
Chapter~\ref{chap:adinjection} presents the architecture and evaluation of
\origintracer to identify ad injection in browser extensions. We propose our
methodology for large-scale analysis of style injection by relative path
overwrite in Chapter~\ref{chap:rpo}. Finally, Chapter~\ref{chap:conclusion}
concludes the thesis.

\chapter{Related Work}
\label{chap:related_work}

In this chapter, we place our proposed approaches in the context of related
work.

\section{Content Isolation and Containment}

Several recent research
projects~\cite{sp2008op,usenixosdi2010illinois,usenixsec2009gazelle} attempted
to improve the security of browsers by isolating browser components in order to
minimize data sharing among software components. The main issue with these
approaches is that they do not perform any isolation between JavaScript loaded
from different domains and web applications, letting untrusted scripts access
the main web application's code and data. Efforts such as
AdJail~\cite{usenixsec2010adjail} attempt to protect privacy by isolating ads
into an iframe-based sandbox. However, this approach restricts contextual
targeting advertisement in which ad scripts need to have access to host page
content.

Another approach is to search and restrict third-party code included in web appl
ications~\cite{ndss2010capabilityleaks,usenixsec2009gatekeeper,csf2009langisojs}
. For example, ADsafe~\cite{adsafe} removes dangerous JavaScript features (e.g.,
\texttt{eval}), enforcing a whitelist of allowed JavaScript functionality
considered safe. It is also possible to protect against malicious JavaScript ads
by enforcing policies at
runtime~\cite{asiaccs2009lightjs,usenixosdi2006browsershield}. For example,
Meyerovich et al.~\cite{sp2010conscript} introduce a client-side framework that
allows web applications to enforce fine-grained security policies for DOM
elements. AdSentry~\cite{acsac2011adsentry} provides a shadow JavaScript engine
that runs untrusted ad scripts in a sandboxed environment.

\section{Blacklisting Malicious Domains}

There are multiple approaches to automatically detecting malicious web domains.
Madtracer~\cite{ccs2012madtracer} has been proposed to automatically capture
malvertising cases. But, this system is not as precise as our approach in
identifying the causal relationships among different domains.
EXPOSURE~\cite{ndss2011exposure} employs passive DNS analysis techniques to
detect malicious domains. SpiderWeb~\cite{ccs2013spiderweb} is also a system
that is able to detect malicious web pages by crowd-sourcing redirection chains.
Segugio~\cite{dsn2015segugio} tracks new malware-control domain names in very
large ISP networks. WebWitness~\cite{usenixsec2015webwitness} automatically
traces back malware download paths to understand attack trends. While these
techniques can be used to automatically detect malicious websites and update
blacklists, they are not online systems and may not be effectively used to
detect malicious third-party inclusions since users expect a certain level of
performance while browsing the Web.

Another effective detection approach is to produce blacklists of malicious sites
by scanning the Internet that can be efficiently checked by the browser (e.g.,
Google Safe Browsing~\cite{gsb}). Blacklist construction requires extensive
infrastructure to continuously scan the Internet and bypass cloaking and general
malware evasion attempts in order to reliably identify malware distribution
sites, phishing pages, and other Web malice. These blacklists sometimes lag the
introduction of malicious sites on the Internet, or fail to find these malicious
sites. However, they are nevertheless effective, and we view the approach we
propose as a complementary technique to established blacklist generation and
enforcement techniques.

\section{Browser Extension Security}

Browser extension security has become a hot topic. The Chromium extension
framework substantially improved the ability of users to limit the amount of
privilege conferred upon potentially vulnerable
extensions~\cite{tr2008chromium_security_architecture}, and follow-on work has
studied the success of this
approach~\cite{usenixwebapps2011app_permissions,ndss2012chrome_extensions}.
Other works have broadly studied malicious extensions that attempt to exfiltrate
sensitive user data~\cite{raid2007spyshield,springer2008browserspy}. For
instance, Arjun et al.~showed that many extensions in the Chrome Web Store are
over-privileged for the actual services they purport to
provide~\cite{sp2011ibex}.

A line of work has focused on the problem of ad injection via browser
extensions. Thomas et al.~\cite{sp2015adinjection} proposed a detection
methodology in which, they used a priori knowledge of a legitimate DOM structure
to report the deviations from that structure as potential ad injections.
However, this approach is not purely client-side and requires cooperation from
content publishers. Expector~\cite{www2015adinjection} inspects a browser
extension and determines if it injects advertisements into websites.
Hulk~\cite{usenixsec2014hulk} is a dynamic analysis system that automatically
detects Chrome extensions that perform certain types of malicious behaviors,
including ad injection. WebEval~\cite{usenixsec2015webeval} is an automatic
system that considers an extension's behaviors, code, and author reputation to
identify malicious extensions distributed through the Chrome Web Store. Web
Tripwires~\cite{nsdi2008tripwires} were also proposed to detect in-flight page
changes performed in order to inject advertisements.

In contrast, our work does not attempt to automatically classify extensions that
engage in content modification as malicious or not, but rather focuses on
enabling users to make informed decisions as to whether extensions engage in
desirable behavior or not.

\section{Provenance Tracking}

A significant amount of work has examined the use of provenance in various
contexts. For instance, one line of work has studied the collection of
provenance information for generic applications up to entire
systems~\cite{middleware2012spade,ccs2009sprov,acsac2012hifi}. However, to our
knowledge, no system considers the provenance of fine-grained web content
comprising the DOM. Provenance tracking is also related to information flow
control (IFC), for which a considerable body of work exists at the operating
system level~\cite{sosp2005asbestos,sosp2007flume,nsdi2008dstar}, the language
level~\cite{popl1999jflow,usenixsec2007sif}, as well as the
web~\cite{osdi2012hails,sacmat2010e2eaccctrl}. In contrast to our work, IFC is
focused more on enforcing principled security guarantees for new applications
rather than tracking and indicating data provenance for existing ones. Numerous
systems have examined the use of dynamic taint analysis, a related concept to
provenance. Some prior
work~\cite{ndss2015infoflows,usenixatc2007spyware_analysis} focuses on tracking
information flow within the browser, Sabre~\cite{acsac2009sabre} detects whether
extensions access sensitive information within the browser, and DSI
enforcement~\cite{ndss2009dsi} defends against XSS attacks by preserving the
integrity of document structure in the browser. While there is certainly an
overlap between dynamic taint analysis and provenance, taint analysis is most
often focused on simple reachability between sources and sinks, while provenance
is concerned with precisely tracking principals that influenced data.

Finally, there is a line of work that examines provenance on the web. Some prior
work~\cite{propr2007provenance_web,ldow2009provenance_web,websci2010provenance_web}
concerns coarse-grained ontologies for describing the origins of data on the
web, and does not consider provenance at a fine-grained scale within the
browser. ESCUDO~\cite{icdcs2010escudo} only considers the principals that can be
controlled by web applications, and it does not handle plug-ins and browser
extensions. LeakTracker~\cite{acns2012leaktracker} performs principal-based
tracking on web pages to study privacy violations related to JavaScript
libraries, but it only tracks injection of scripts into the page, and does not
provide any provenance information for other types of DOM elements.

\section{Relative Path Overwrite}

The first account of RPO is attributed to a blog post by Gareth
Heyes~\cite{rpo}, introducing self-referencing a PHP script with server-side URL
rewriting. Furthermore, the post notes that certain versions of Internet
Explorer allow JavaScript execution from within a CSS context in the
\textit{Compatibility View} mode~\cite{cvlist}, escalating style injection to
XSS~\cite{csscripting}. Another blog post by Dalili~\cite{rpo_iis} extends the
technique to IIS and ASP.Net applications, and shows how URL-encoded slashes are
decoded by the server but not the browser, allowing not only self-reference but
also the inclusion of different resources. Kettle~\cite{prssi} coins the term
Path Relative StyleSheet Import (PRSSI) for a specific subset of RPO attacks,
introduces a PRSSI vulnerability scanner for Burp Suite~\cite{burpsuite}, and
proposes countermeasures. Terada~\cite{rpo_techniques} provides more
exploitation techniques for various browsers or certain web applications,
and~\cite{rpo_gadgets} discusses an example chaining several vulnerabilities to
result in a combination of RPO and a double style injection attack. Gil shows
how attackers can deceive web cache servers by using
RPO~\cite{blackhatusa2017webcache,webcache_deception}. Some of the attacks
discussed in the various blog posts are custom-tailored to specific sites or
applications, whereas others are more generic and apply to certain web server
configurations or frameworks.

We are not aware of any scholarly work about RPO, or any research about how
prevalent RPO vulnerabilities are on the Web. To the best of our knowledge, Burp
Suite~\cite{burpsuite} is the first and only tool that can detect PRSSI
vulnerabilities based on RPO in web applications. However, in contrast to our
work, it does not determine if the vulnerability can be exploited. Furthermore,
we are the first to provide a comprehensive survey of how widespread RPO style
vulnerabilities and exploitabilities are in the wild.

\section{Client-side Attacks}

Script-based attacks has been studied extensively, such as systematic analysis
of XSS sanitization frameworks~\cite{esorics2011xss}, detecting XSS
vulnerabilities in Rich Internet Applications~\cite{asiaccs2012flashover},
large-scale detection of DOM-based
XSS~\cite{ccs2013domxss,blackhatasia2014domxss}, and bypassing XSS mitigations
by Script Gadgets~\cite{blackhatusa2017script_gadgets,ccs2017script_gadgets}. An
array of XSS prevention mechanisms have been proposed, such as XSS
Filter~\cite{xss_filter}, XSS-Guard~\cite{dimva2008xss_guard},
SOMA~\cite{ccs2008soma}, BluePrint~\cite{sp2009blueprint}, Document Structure
Integrity~\cite{ndss2009doc_integrity}, XSS Auditor~\cite{www2010xss_auditor},
NoScript~\cite{noscript}, Context-Sensitive Auto-Sanitization
(CSAS)~\cite{ccs2011csas}, DOM-based XSS filtering using runtime taint
tracking~\cite{usenixsec2014client_side_xss}, preventing script injection
through software design~\cite{commacm2014tangled_web}, Strict
CSP~\cite{ccs2016cspisdead}, and DOMPurify~\cite{esorics2017dompurify}. However,
the vulnerability measurements and proposed countermeasures of these works on
script injection do not apply to RPO-based style injection.

\chapter{Detection of Malicious Third-Party Content Inclusions}
\label{chap:inclusion}

\section{Introduction}
\label{inclusion:sec:intro}

While the Same Origin Policy (SOP) enforces a modicum of origin-based separation
between code and data from different principals, developers have clamored for
more flexible sharing models provided by, e.g., Content Security Policy
(CSP)~\cite{csp_spec}, Cross-Origin Resource Sharing (CORS)~\cite{cors_spec},
and postMessage-based cross-frame communication. These newer standards permit
greater flexibility in performing cross-origin inclusions, and each come with
associated mechanisms for restricting communication to trusted origins. However,
recent work has shown that these standards are difficult to apply securely in
practice~\cite{ndss2013postman,raid2014csp}, and do not necessarily address the
challenges of trusting remote inclusions on the dynamic Web. In addition to the
inapplicability of some approaches such as CSP, third parties can leverage their
power to bypass these security mechanisms. For example, ISPs and browser
extensions are able to tamper with HTTP traffic to modify or remove CSP rules in
HTTP responses~\cite{usenixsec2015webeval,sp2015adinjection}.

In this chapter, we propose an in-browser approach called \excision to
automatically detect and block malicious third-party content inclusions as web
pages are loaded into the user's browser or during the execution of browser
extensions. Our approach does not rely on examination of the content of the
resources; rather, it relies on analyzing the sequence of inclusions that leads
to the resolution and loading of a terminal remote resource. Unlike prior
work~\cite{ccs2012madtracer}, \excision resolves \emph{inclusion sequences}
through instrumentation of the browser itself, an approach that provides a
high-fidelity view of the third-party inclusion process as well as the ability
to interdict content loading in real-time. This precise view also renders
ineffective common obfuscation techniques used by attackers to evade detection.
Obfuscation causes the detection rate of these approaches to degrade
significantly since obfuscated third-party inclusions cannot be traced using
existing techniques~\cite{ccs2012madtracer}. Furthermore, the in-browser
property of our system allows users to browse websites with a higher confidence
since malicious third-party content is prevented from being included while the
web page is loading.

We implemented \excision as a set of modifications to the Chromium browser, and
evaluated its effectiveness by analyzing the Alexa Top 200K over a period of 11
months. Our evaluation demonstrates that \excision achieves a 93.39\% detection
rate, a false positive rate of 0.59\%, and low performance overhead. We also
performed a usability test of our research prototype, which shows that \excision
does not detract from the user's browsing experience while automatically
protecting the user from the vast majority of malicious content on the Web. The
detection results suggest that \excision could be used as a complementary system
to other techniques such as CSP.

The rest of this chapter is organized as follows.
Section~\ref{inclusion:sec:background} outlines the necessary background.
Section~\ref{inclusion:sec:design} presents the architecture of \excision, while
Section~\ref{inclusion:sec:implementation} discusses the implementation of our
system. We present an evaluation of the effectiveness, usability, and
performance of our prototype in Section~\ref{inclusion:sec:analysis}. Finally, a
discussion about our system is presented in
Section~\ref{inclusion:sec:discussion}, and Section~\ref{inclusion:sec:summary}
summarizes the chapter.

\section{Background}
\label{inclusion:sec:background}

In the following, we first discuss the threats posed by third-party content and
then motivate our work.

\subsection{Threats}

While the inclusion of third-party content provides convenience for web
developers and allows for integration into advertising distribution, analytics,
and social media networks, it can potentially introduce a set of serious
security threats for users. For instance, advertising networks and social media
have been and continue to be abused as a vector for injection of malware.
Website operators, or publishers, have little control over this content aside
from blind trust or security through isolation. Attacks distributed through
these vectors -- in the absence of isolation -- execute with the same privileges
as all other JavaScript within the security context of the enclosing DOM. In
general, malicious code could launch drive-by downloads~\cite{www2010jsand},
redirect visitors to phishing sites, generate fraudulent clicks on
advertisements~\cite{ccs2012madtracer}, or steal user
information~\cite{ccs2010cssattack}.

Moreover, ad injection has become a new source of income for ISPs and browser
extension authors~\cite{adinjection_profit}. ISPs inject advertisements into web
pages by tampering with their users' HTTP traffic~\cite{isp_adinjection}, and
browser extension authors have recently started to inject or replace ads in web
pages to monetize their work. Ad injection negatively impacts both website
publishers and users by diverting revenue from publishers and exposing users to
malvertising~\cite{sp2015adinjection,www2015adinjection}. In addition to ad
injection, malicious browser extensions can also pose significant risks to users
due to the special privileges they have~\cite{usenixsec2014hulk}.

\subsection{Motivation}

Publishers can try to isolate untrusted third-party content using iframes
(perhaps enhanced with HTML5 sandboxing features), language-based sandboxing, or
policy
enforcement~\cite{adsafe,ndss2010capabilityleaks,usenixsec2009gatekeeper,usenixsec2010adjail,csf2009langisojs}.
However, these approaches are not commonly used in practice; some degrade the
quality of ads (from the advertiser's perspective), while others are non-trivial
to deploy. Publishers could attempt to use Content Security Policy
(CSP)~\cite{csp_spec} to define and enforce access control lists for remote
inclusions in the browser. However, due to the dynamic nature of the web, this
approach (and similar access control policy-based techniques) has problems.
Recent studies~\cite{ndss2013postman,raid2014csp} indicate that CSP is difficult
to apply in practice. A major reason for this is the unpredictability of the
origins of inclusions for third-party resources, which complicates the
construction of a correct, yet tight, policy.

For example, when websites integrate third-party advertisements, multiple
origins can be contacted in order to deliver an ad to the user's browser. This
is often due to the practice of re-selling ad space (a process known as ad
syndication) or through real-time ad auctions. Either of these approaches can
result in ads being delivered through a series of JavaScript code
inclusions~\cite{imc2011adexchange}. As a consequence, a long inclusion sequence
of distinct origins will be observed that -- critically -- does not remain
constant on successive loads of the enclosing web page. Additionally, the
growing number of browser extensions makes it a non-trivial task for website
operators to enumerate the set of benign origins from which browser extensions
might include a resource. Therefore, defining an explicit whitelist of CSP rules
is a challenging task.

\begin{figure}[t]
    \centering
    \includegraphics[width=0.8\textwidth]{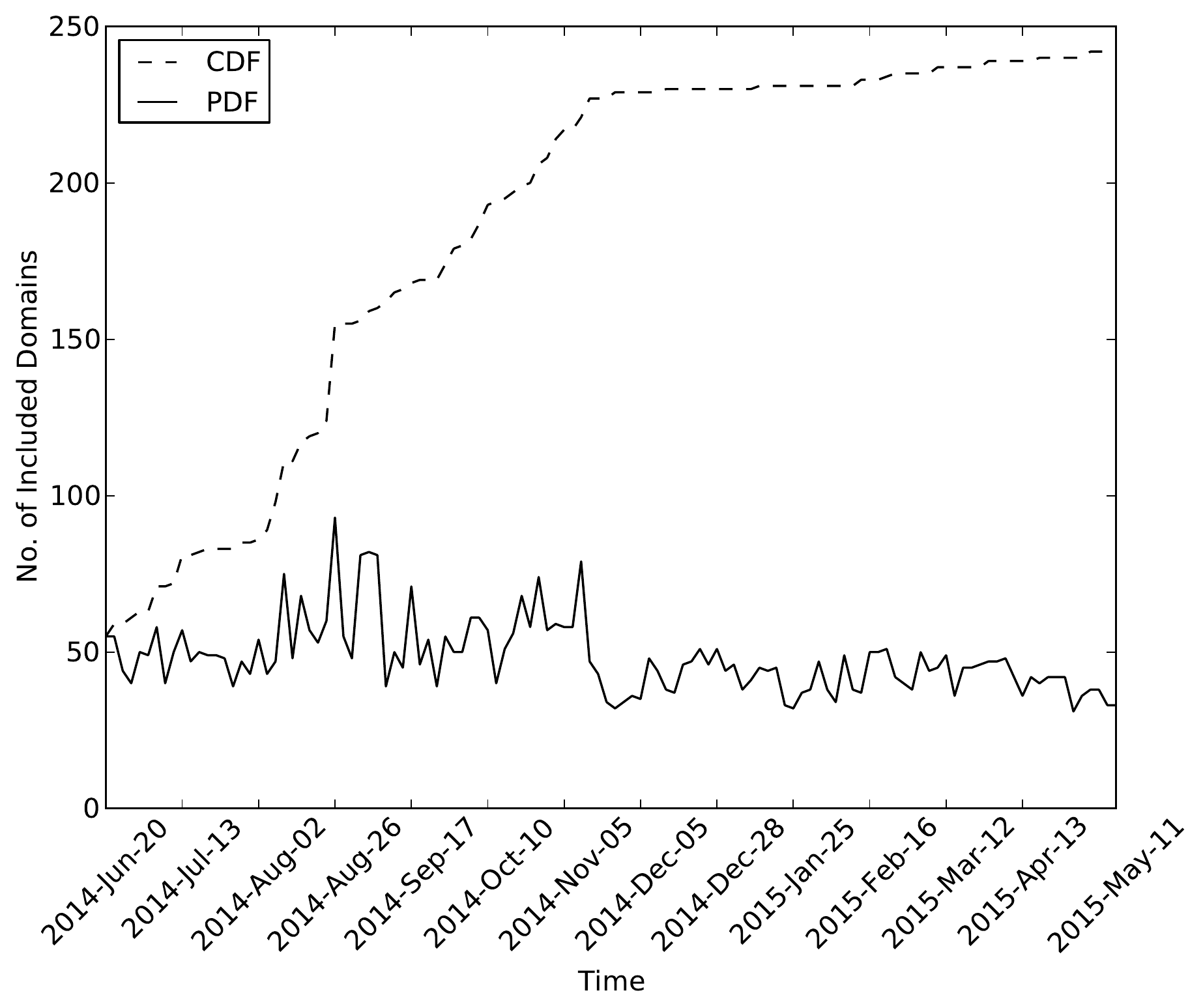}
    \caption{Unique number of included domains in \texttt{theverge.com} over 11 months. Measurements were collected as part of the data set described in Section~\ref{inclusion:sec:analysis}; the sampling frequency was approximately once every three days}
    \label{inclusion:fig:included_domains}
\end{figure}

To illustrate, Figure~\ref{inclusion:fig:included_domains} shows the unique number of
domains as well as the cumulative number of unique domains included by
\texttt{theverge.com} over a period of 11 months. The unique number of domains
increases roughly linearly over this period; clearly, constructing an effective
access control policy that tightly captures the set of allowable inclusions
while avoiding false positives that would lead to either lost revenue or broken
functionality is difficult.

Even if website publishers can keep pace with origin diversity over time with a
comprehensive list of CSP rules, ISPs and browser extensions are able to tamper
with in-transit HTTP traffic and modify CSP rules sent by the websites. In
addition, in browsers such as Chrome, the web page's CSP does not apply to
extension scripts executed in the page's context~\cite{csp_content_scripts};
hence, extensions are able to include arbitrary third-party resources into the
web page.

Given the challenges described above, we believe that existing techniques such
as CSP can be evaded and, hence, there is a need for an automatic approach to
protect users from malicious third-party content. We do not necessarily advocate
such an approach in isolation, however. Instead, we envision this approach as a
complementary defense that can be layered with other techniques in order to
improve the safety of the Web.

\section{Design}
\label{inclusion:sec:design}

\begin{figure*}[t]
    \centering
    \includegraphics[width=0.8\textwidth]{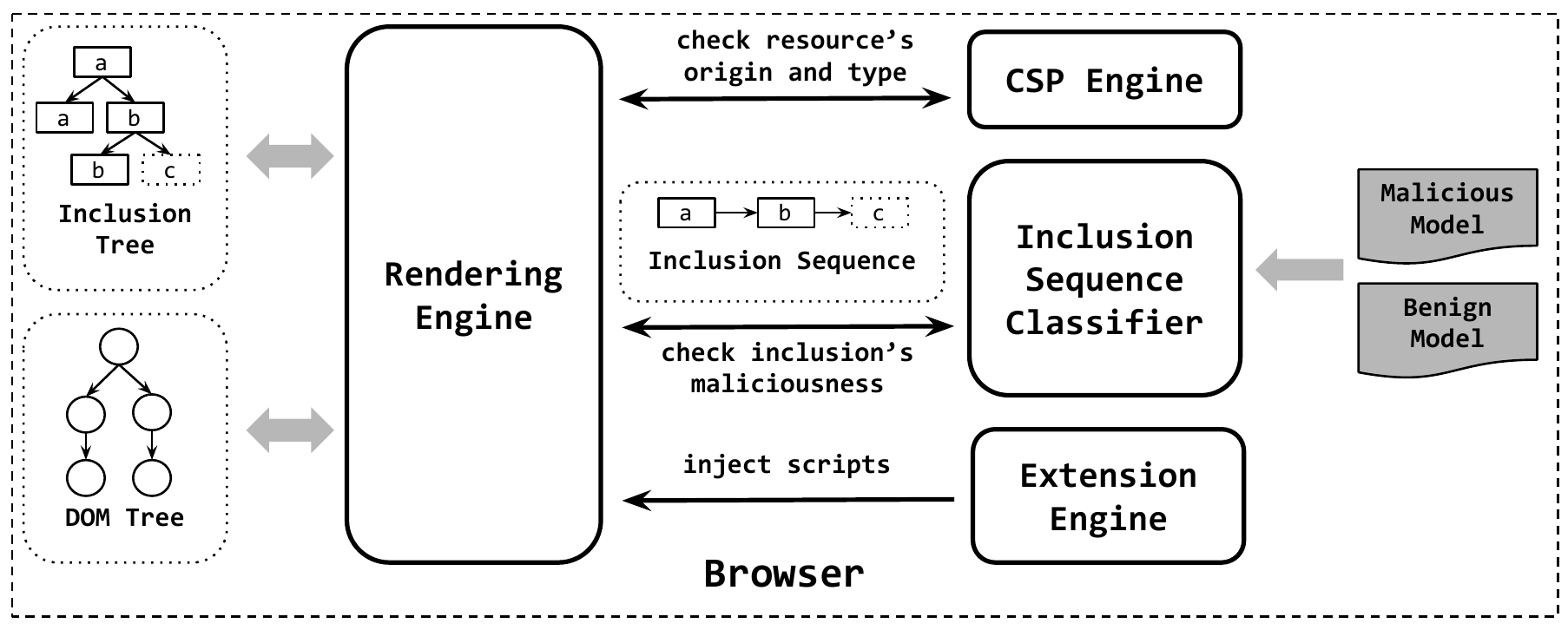}
    \caption{An overview of \excision}
    \label{inclusion:fig:architecture}
\end{figure*}

In this section, we describe \excision, our approach for detecting and blocking
the inclusion of malicious third-party content in real-time. An overview of our
system is shown in Figure~\ref{inclusion:fig:architecture}. \excision operates
by extracting resource \emph{inclusion trees} from within the browser. The
inclusion tree precisely records the inclusion relationships between different
resources in a web page. When the user requests a web page, the browser
retrieves the corresponding HTML document and passes it to the rendering engine.
The rendering engine incrementally constructs an inclusion tree for the DOM and
begins extracting external resources such as scripts and frames as it reaches
new HTML tags. For inclusion of a new resource, the rendering engine consults
the CSP engine and the \emph{inclusion sequence classifier} in order to decide
whether to include the resource. If the resource's origin and type are
whitelisted in the CSP rules, the rendering engine includes the resource without
consulting the inclusion sequence classifier and continues parsing the rest of
the HTML document. Otherwise, it extracts the \emph{inclusion sequence} (path
through the page's inclusion tree) for the resource and forwards this to the
inclusion sequence classifier. Using pre-learned models, the classifier returns
a decision about the malice of the resource to the rendering engine. Finally,
the rendering engine discards the resource if it was identified as malicious.
The same process occurs for resources that are included dynamically during the
execution of extension content scripts after they are injected into the page.

\subsection{Inclusion Trees and Sequences}
\label{inclusion:sec:inclusion_tree}

A website can include resources in an HTML document from any origin so long as
the inclusion respects the same origin policy, its standard exceptions, or any
additional policies due to the use of CSP, CORS, or other access control
framework. A first approximation to understanding the inclusions of third-party
content for a given web page is to process its DOM tree~\cite{domtree} while the
page loads. However, direct use of a web page's DOM tree is unsatisfactory
because the DOM does not in fact reliably record the inclusion relationships
between resources referenced by a page. This follows from the ability for
JavaScript to manipulate the DOM at run-time using the DOM API.

\begin{figure}[t]
   \centering
   \includegraphics[width=0.8\textwidth]{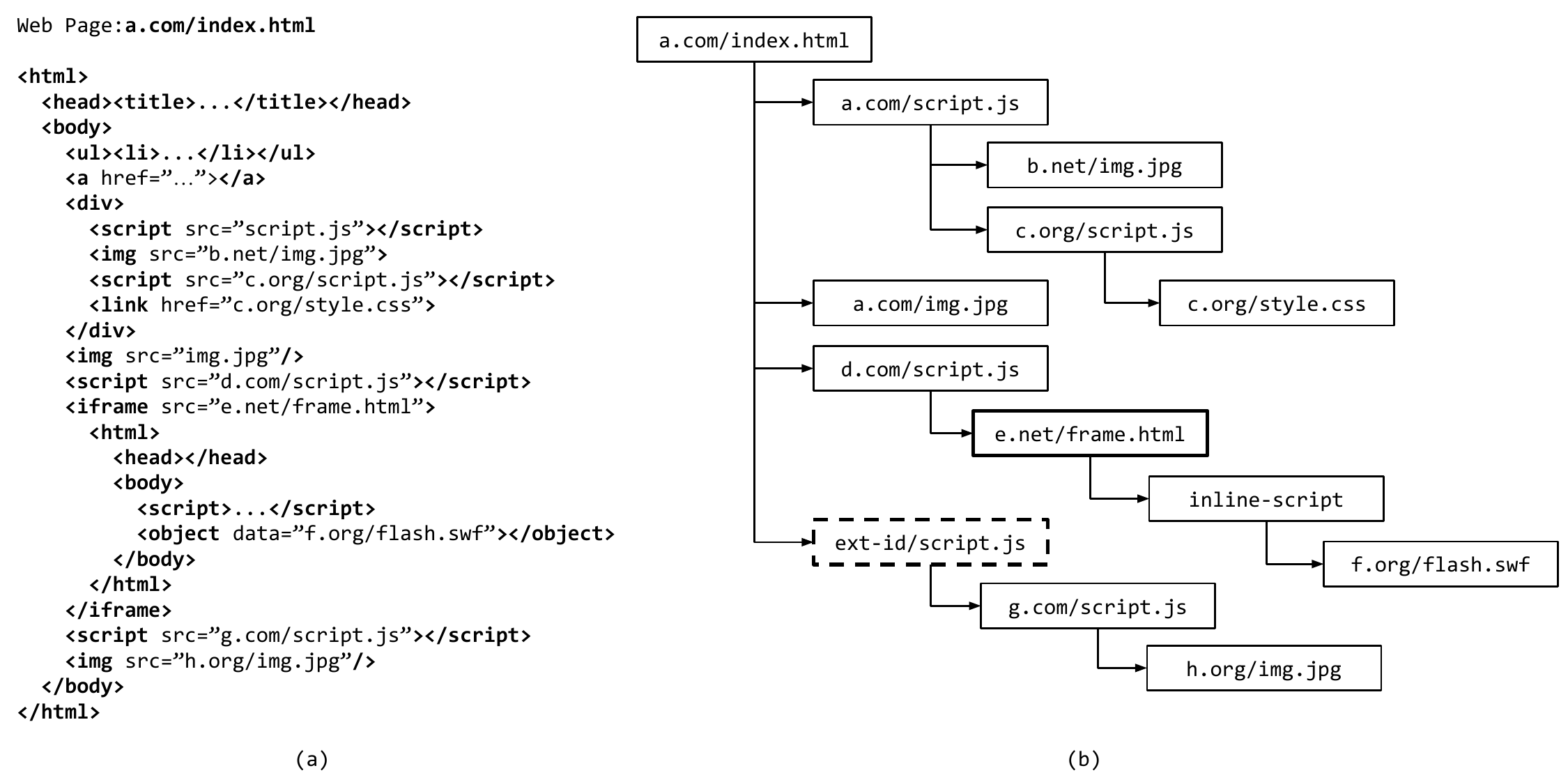}
   \caption{(a) DOM Tree, and (b) Inclusion Tree}
   \label{inclusion:fig:dom_inclusion_tree}
\end{figure}

Instead, in this work we define an \emph{inclusion tree} abstraction extracted
directly from the browser's resource loading code. Unlike a DOM tree, the
inclusion tree represents how different resources are included in a web page
that is invariant with respect to run-time DOM updates. It also discards
irrelevant portions of the DOM tree that do not reference remote content. For
each resource in the inclusion tree, there is an \emph{inclusion sequence} that
begins with the root resource (i.e., the URL of the web page) and terminates
with the corresponding resource. Furthermore, browser extensions can also
manipulate the web page by injecting and executing JavaScript code in the page's
context. Hence, the injected JavaScript is considered a direct child of the root
node in the inclusion tree. An example of a DOM tree and its corresponding
inclusion tree is shown in Figure~\ref{inclusion:fig:dom_inclusion_tree}. As shown in
Figure~\ref{inclusion:fig:dom_inclusion_tree}b, \texttt{f.org/flash.swf} has been
dynamically added by an \texttt{inline script} to the DOM tree, and its
corresponding inclusion sequence has a length of 4 since we remove the inline
resources from inclusion sequence. Moreover, \texttt{ext-id/script.js} is
injected by an extension as the direct child of the root resource. This script
then included \texttt{g.com/script.js}, which in turn included
\texttt{h.org/img.jpg}.

\begin{figure}
    \centering
    \includegraphics[width=0.8\textwidth]{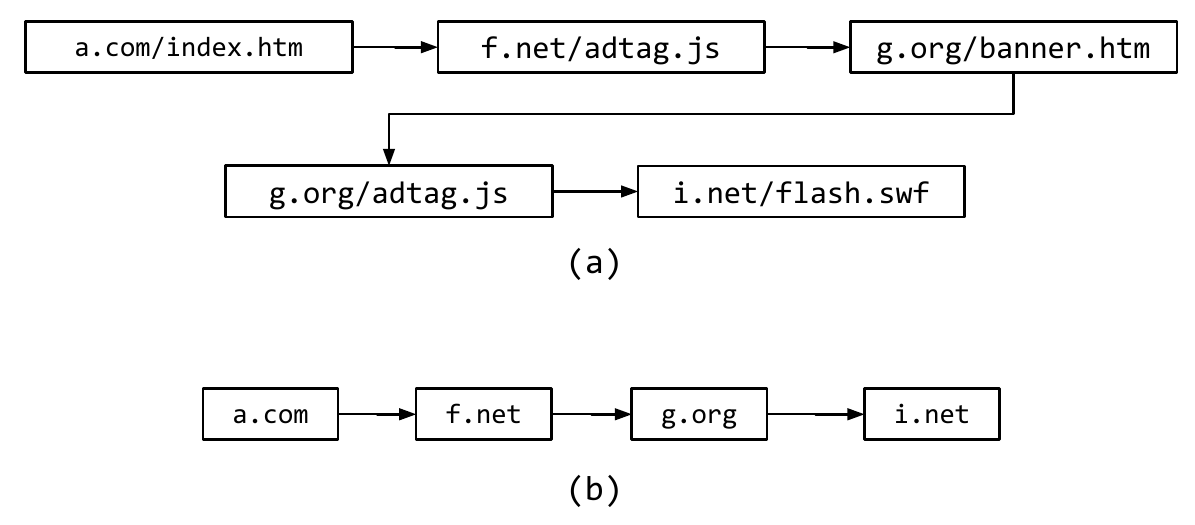}
    \caption{(a) URL Inclusion Sequence, and (b) Domain Inclusion Sequence}
    \label{inclusion:fig:inclusion_sequence}
\end{figure}

When we consider the full URL for constructing an inclusion sequence, the
resulting sequence is called a \emph{URL Inclusion Sequence}.
Figure~\ref{inclusion:fig:inclusion_sequence}a shows the URL inclusion sequence
of the resource \texttt{i.net/flash.swf}. However, some malware campaigns change
their URL patterns frequently to avoid detection. This can be done by changing
the URL path and the parameter values~\cite{ccs2012madtracer}. To overcome this
problem and capture the high-level relationships between different websites, we
only consider a domain part of the URL to build the \emph{Domain Inclusion
Sequence}. Figure~\ref{inclusion:fig:inclusion_sequence}b shows the domain inclusion
sequence corresponding to the aforementioned URL inclusion sequence. As
depicted, if consecutive URLs in a sequence have the same domains, we merge them
into one node. From now on, by inclusion sequence, we refer to a domain
inclusion sequence unless we mention URL inclusion sequence explicitly.

\subsection{Inclusion Sequence Classification}
\label{inclusion:sec:classification}

Given an inclusion sequence, \excision must classify it as benign or malicious
based on features extracted from the sequence. The task of the \emph{inclusion
sequence classifier} is to assign a class label from the set
\texttt{\{benign,malicious\}} to a given sequence based on previously learned
models from a labeled data set. In our definition, a malicious sequence is one
that starts from the root URL of a web page and terminates in a URL that
delivers malicious content. For classification, we used hidden Markov models
(HMM)~\cite{ieee1989hmm}. Models are comprised of states, each of which holds
transitions to other states based on a probability distribution. Each state can
probabilistically emit a symbol from an alphabet. There are other sequence
classification techniques such as Na\"{i}ve Bayes~\cite{ecml1998naivebayes}, but
we used an HMM for our classifier because we also want to model the
inter-dependencies between the resources that compose an inclusion sequence.

In the training phase, the system learns two HMMs from a training set of labeled
sequences, one for the benign class and one for the malicious class. We
estimated the HMM parameters by employing the Baum-Welch algorithm which finds
the maximum likelihood estimate of these parameters based on the set of observed
sequences. In our system, we empirically selected 20 for the number of states
that are fully connected to each other. In the subsequent detection phase, we
compute the likelihood of a new sequence given the trained models using the
forward-backward algorithm and assign the sequence to the class with the highest
likelihood. Training hidden Markov models is computationally expensive. However,
computing the likelihood of a sequence is instead very efficient, which makes it
a suitable method for real-time classification~\cite{ieee1989hmm}.

\subsection{Classification Features}
\label{inclusion:sec:features}

Let $r_0 \rightarrow r_1 \rightarrow \dots \rightarrow r_n$ be an inclusion
sequence as described above. Feature extraction begins by converting the
inclusion sequence into sequences of feature vectors. After analyzing the
inclusion trees of several thousand benign and malicious websites for a period
of 11 months, we identified 12 feature types from three categories. For each
feature type, we compute two different features: individual and relative
features. An individual feature value is only dependent on the current resource,
but a relative feature value is dependent on the current resource and its
preceding (or parent) resources. Consequently, we have 24 features for each
resource in an inclusion sequence. Individual features can have categorical or
continuous values. All continuous feature values are normalized on
$\left[0,1\right]$ and their values are discretized. In the case of continuous
individual features, the relative feature values are computed by comparing the
individual value of the resource to its parent's individual value. The result of
the comparison is \texttt{less}, \texttt{equal}, or \texttt{more}. We use the
value \texttt{none} for the root resource. To capture the high-level
relationships between different inclusions, we only consider the domain part of
the URL for feature calculation.

\subsubsection{DNS-based Features}

The first feature category that we consider is based on DNS properties of the
resource domain.

\begin{table}[t]
    \centering
    \footnotesize
    \label{inclusion:tab:tld}
    \caption{TLD values}
    \begin{subtable}{.43\textwidth}
        \centering
        \caption{Individual}
        \label{inclusion:tab:tld:individual}
        \begin{tabular}{ll}
        \toprule
        \textbf{Value} & \textbf{Example} \\
        \midrule
        none & IPs, Extensions \\
        gen & *.com, *.org \\
        gen-subdomain & *.us.com \\
        cc & *.us, *.de, *.cn \\
        cc-subdomain & *.co.uk, *.com.cn \\
        cc-int & *.xn{-{}-}p1ai (ru) \\
        other & *.biz, *.info \\
        \bottomrule
        \end{tabular}
    \end{subtable}
    \begin{subtable}{.54\textwidth}
        \centering
        \caption{Relative}
        \label{inclusion:tab:tld:relative}
        \begin{tabular}{ll}
        \toprule
        \textbf{Value} & \textbf{Example} \\
        \midrule
        none & root resource \\
        \{got,lost\}-tld & Ext. $\rightarrow$ *.de, *.us $\rightarrow$ IP \\
        gen-to-\{cc,other\} & *.org $\rightarrow$ \{*.de, *.info\} \\
        cc-to-\{gen,other\} & *.uk $\rightarrow$ \{*.com, *.biz\} \\
        other-to-\{gen,cc\} & *.info $\rightarrow$ \{*.net, *.uk\} \\
        same-\{gen,cc,other\} & *.com $\rightarrow$ *.com \\
        diff-\{gen,cc,other\} & *.info $\rightarrow$ *.biz \\
        \bottomrule
        \end{tabular}
    \end{subtable} 
\end{table}

\paragraph{Top-Level Domain.}

For this feature, we measure the types of TLDs from which a resource is included
and how it changes along the inclusion sequence. For every resource in an
inclusion sequence, we assign one of the values in
Table~\ref{inclusion:tab:tld:individual} as an individual feature. For the relative
feature, we consider the changes that occur between the top-level domain of the
preceding resource and the resource itself. Table~\ref{inclusion:tab:tld:relative} shows
15 different values of the relative TLD feature.

\paragraph{Type.}

This feature identifies the types of resource domains and their changes along the
inclusion sequence. Possible values of individual and relative features are
shown in Table~\ref{inclusion:tab:type:individual} and Table~\ref{inclusion:tab:type:relative}
respectively.

\paragraph{Level.}

A domain name consists of a set of labels separated by dots. We say a domain
name with $n$ labels is in level $n-1$. For example, \texttt{www.google.com} is
in level 2. For IP addresses and extension scripts, we consider their level to
be 1. For a given domain, we compute the individual feature by dividing the level
by a maximum value of 126.

\paragraph{Alexa Ranking.}

We also consider the ranking of a resource's domain in the Alexa Top 1M
websites. To compute the normalized ranking as an individual feature, we divide
the ranking of the domain by one million. For IP addresses, extensions, and
domains that are not in the top 1M, we use the value \texttt{none}.

\begin{table}[t]
    \centering
    \footnotesize
    \label{inclusion:tab:type}
    \caption{Type values}
    \begin{subtable}{.43\textwidth}
        \centering
        \caption{Individual}
        \label{inclusion:tab:type:individual}
        \begin{tabular}{ll}
            \toprule
            \textbf{Value} & \textbf{Example} \\
            \midrule
            ipv6 & 2607:f0d0::::4 \\
            ipv4-private & 192.168.0.1 \\
            ipv4-public & 4.2.2.4 \\
            extension & Ext. Scripts \\
            dns-sld & google.com \\
            dns-sld-sub & www.google.com \\
            dns-non-sld & abc.dyndns.org \\
            dns-non-sld-sub & a.b.dyndns.org \\
            \bottomrule
        \end{tabular}
    \end{subtable}
    \begin{subtable}{.54\textwidth}
        \centering
        \caption{Relative}
        \label{inclusion:tab:type:relative}
        \begin{tabular}{ll}
            \toprule
            \textbf{Value} & \textbf{Example} \\
            \midrule
            none & root resource \\
            same-site & w.google.com $\rightarrow$ ad.google.com \\
            same-sld & 1.dyndns.org $\rightarrow$ 2.dyndns.org \\
            same-company & ad.google.com $\rightarrow$ www.google.de \\
            same-eff-tld & bbc.co.uk $\rightarrow$ london.co.uk \\
            same-tld & bbc.co.uk $\rightarrow$ london.uk \\
            different & google.com $\rightarrow$ facebook.net \\
            \bottomrule
            \\
        \end{tabular}
    \end{subtable} 
\end{table}

\subsubsection{String-based Features}

We observed that malicious domain names often make liberal use of digits and
hyphens in combination with alphabetical characters. So, in this feature
category, we characterize the string properties of resource domains. For IP
addresses and extension scripts, we assign the value 1 for individual features.

\paragraph{Non-Alphabetic Characters.}

For this feature, we compute the individual feature value by dividing the number
of non-alphabetical characters over the length of domain.

\paragraph{Unique Characters.}

We also measure the number of unique characters that are used in a domain. The
individual feature is the number of unique characters in the domain divided by
the maximum number of unique characters in the domain name, which is 38 (26
alphabetics, 10 digits, hyphen, and dot).

\paragraph{Character Frequency.}

For this feature, we simply measure how often a single character is seen in a
domain. To compute an individual feature value, we calculate the frequency of
each character in the domain and then divide the average of these frequencies by
the length of the domain to normalize the value.

\paragraph{Length.}

In this feature, we measure the length of the domain divided by the maximum
length of a domain, which is 253.

\paragraph{Entropy.}

In practice, benign domains are typically intended to be memorable to users.
This is often not a concern for attackers, as evidenced by the use of domain
generation algorithms~\cite{ndss2011exposure}. Consequently, we employ Shannon
entropy to measure the randomness of domains in the inclusion sequence. We
calculate normalized entropy as the absolute Shannon entropy divided by the
maximum entropy for the domain name.

\subsubsection{Role-based Features}

We observed that identifying the role of resources in the inclusion sequences
can be helpful in detecting malicious resources. For example, recent
work~\cite{www2014shortening} reveals that attackers misuse ad networks as well
as URL shortening services for malicious intent. So far, we consider three roles
for a resource:
\begin{inparaenum}[\itshape i)\upshape]
     \item ad-network,
     \item content delivery network (CDN), and
     \item URL shortening service.
\end{inparaenum}
In total, we have three features in this category, as each domain can
simultaneously perform multiple roles. Both individual and relative features in
this category have binary values. For the individual feature, the value is
\texttt{Yes} if the domain has the role, and \texttt{No} otherwise. For the
relative feature, we assign a value \texttt{Yes} if at least one of the
preceding domains have the corresponding role, and \texttt{No} otherwise. For
extension scripts, we assign the value \texttt{No} for all of the features. To
assign the roles, we compiled a list of common domains related to these roles that
contains 5,767 ad-networks, 48 CDNs, and 461 URL shortening services.

\section{Implementation}
\label{inclusion:sec:implementation}

In this section, we discuss our prototype implementation of \excision for
detecting and blocking malicious third-party content inclusions. We implemented
\excision as a set of modifications to the Chromium browser. In order to
implement our system, we needed to modify Blink and the Chromium extension
engine to enable \excision to detect and block inclusions of malicious content
in an online and automatic fashion while the web page is loading. The entire set
of modifications consists of less than 1,000 lines of C++ and several lines of
JavaScript\footnote{https://github.com/sajjadium/Excision}. While our
implementation could be adopted as-is by any browser vendors that use
WebKit-derived engines, the design presented here is highly likely to be
portable to other browsers.

\subsection{Enhancements to the Blink}

Blink is primarily responsible for parsing HTML documents, managing script
execution, and fetching resources from the network. Consequently, it is ideally
suited for constructing the inclusion tree for a web page, as well as blocking
the inclusion of malicious content.

\subsubsection{Tracking Resource Inclusion}

Static resource inclusions that are hard-coded by publishers inside the page's
HTML are added to the inclusion tree as the direct children of the root node.
For dynamic inclusions (e.g.,~via the \texttt{document.createElement()} and
\texttt{document.write()} DOM API functions), the system must find the script
resource responsible for the resource inclusion. To monitor dynamic resource
inclusions, the system tracks the start and termination of script execution. Any
resources that are included in this interval will be considered as the children
of that script resource in the inclusion tree.

\subsubsection{Handling Events and Timers}

Events and timers are widely used by web developers to respond to user
interactions (e.g.,~clicking on an element) or schedule execution of code after
some time has elapsed. To capture the creation and firing of events and timers,
the system tracks the registration of callback functions for the corresponding
APIs.

\subsection{Enhancements to the Extension Engine}

The Chromium extension engine handles the loading, management, and execution of
extensions. To access the page's DOM, the extension injects and executes
\emph{content scripts} in the page's context which are regular JavaScript
programs.

\subsubsection{Tracking Content Scripts Injection and Execution}

Content scripts are usually injected into web pages either via the extension's
manifest file using the \texttt{content\_scripts} field or at runtime via the
\texttt{executeScript} API. Either way, content scripts are considered direct
children of the root node in the inclusion tree. Therefore, in order to track
the inclusion of resources as a result of content script execution, the
extension engine was modified to track the injection and execution of content
scripts.

\subsubsection{Handling Callback Functions}

Like any other JavaScript program, content scripts rely heavily on callback
functions. For instance, \texttt{onMessage} and \texttt{sendMessage} are used by
content scripts to exchange messages with their background pages. To track the
execution of callback functions, two JavaScript files were modified in the
extension engine which are responsible for invocation and management of callback
functions.

\section{Analysis}
\label{inclusion:sec:analysis}

In this section, we evaluate the security benefits, performance, and usability
of the \excision prototype. We describe the data sets we used to train and
evaluate the system, and then present the results of the experiments.

\subsection{Data Collection}

\begin{table}[t]
    \centering
    \footnotesize
    \caption{Summary of crawling statistics}
    \begin{tabular}{lrr}
    \toprule
    \textbf{Item} & \textbf{Website Crawl} & \textbf{Extension Crawl} \\
    \midrule
    Websites Crawled & 234,529 & 20 \\
    Unavailable Websites & 7,412 & 0 \\
    \midrule
    Unique Inclusion Trees & 47,789,268 & 35,004 \\
    Unique Inclusion Sequences & 27,261,945 & 61,489 \\
    \midrule
    Unique URLs & 546,649,590 & 72,064 \\
    Unique Domains & 1,368,021 & 1,144 \\
    Unique Sites & 459,615 & 749 \\
    Unique SLDs & 419,119 & 723 \\
    Unique Companies & 384,820 & 719 \\
    Unique Effective TLDs & 1,115 & 21 \\
    Unique TLDs & 404 & 21 \\
    Unique IPs & 9,755 & 3 \\
    \bottomrule
    \end{tabular}
    \label{inclusion:tab:crawling_statistics}
\end{table}

To collect inclusion sequences, we performed two separate crawls for websites
and extensions. The summary of crawling statistics are presented in
Table~\ref{inclusion:tab:crawling_statistics}.

\subsubsection{Website Crawl}

We built a crawler based on an instrumented version of
PhantomJS~\cite{phantomjs}, a scriptable open source browser based on WebKit,
and crawled the home pages of the Alexa Top 200K. We performed our data
collection from June 20th, 2014 to May 11th, 2015. The crawl was parallelized by
deploying 50 crawler instances on five virtual machines, each of which crawled a
fixed subset of the Alexa Top 200K websites. To ensure that visited websites did
not store any data on the clients, the crawler ran a fresh instance of PhantomJS
for each visit. Once all crawlers finished crawling the list of websites, the
process was restarted from the beginning. To thwart cloaking
techniques~\cite{sec2011deseo} utilized by attackers, the crawlers presented a
user agent for IE 6.0 on Windows and employed Tor to send HTTP requests from
different source IP addresses. We also address JavaScript-based browser
fingerprinting by modifying the internal implementation of the
\texttt{navigator} object to return a fake value for the \texttt{appCodeName},
\texttt{appName}, \texttt{appVersion}, \texttt{platform}, \texttt{product},
\texttt{userAgent}, and \texttt{vendor} attributes.

\subsubsection{Extension Crawl}

To collect inclusion sequences related to extensions, we used 292 Chrome
extensions reported in prior work~\cite{www2015adinjection} that injected ads
into web pages. Since ad-injecting extensions mostly target shopping websites
(e.g., Amazon), we chose the Alexa Top 20 shopping websites for crawling to
trigger ad injection by those 292 extensions. We built a crawler by
instrumenting Chromium~43 and collected data for a period of one week from June
16th to June 22nd, 2015. The system loaded every extension and then visited the
home pages of the Alexa Top 20 shopping websites using Selenium
WebDriver~\cite{selenium}. This process was repeated after crawling the entire
set of extensions. In addition, our crawler triggered all the events and timers
registered by content scripts.

\subsection{Building Labeled Datasets}
\label{inclusion:sec:building-labelled-dataset}

\begin{table}[t]
    \centering
    \footnotesize
    \caption{Data sets used in the evaluation}
    \begin{tabular}{lcccc}
    \toprule
    \multirow{2}{*}{\textbf{Dataset}} & \multicolumn{2}{c}{\textbf{No. of Inclusion Sequences}} & \multicolumn{2}{c}{\textbf{No. of Terminal Domains}} \\
    \cmidrule[0.5pt](ll){2-3}
    \cmidrule[0.5pt](ll){4-5}
    & \textbf{Website Crawl} & \textbf{Ext. Crawl} & \textbf{Website Crawl} & \textbf{Ext. Crawl} \\
    \midrule
    Benign & 3,706,451 & 7,372 & 35,044 & 250 \\
    Malicious & 25,153 & 19 & 1,226 & 2 \\
    \bottomrule
    \end{tabular}
    \label{inclusion:tab:dataset_statistics}
\end{table}

To classify a given inclusion sequence as benign or malicious, we trained two
hidden Markov models for benign and malicious inclusion sequences from our data
set. We labeled collected inclusion sequences as either benign or malicious
using VirusTotal~\cite{virustotal}. VirusTotal's URL scanning service aggregates
reports of malicious URLs from most prominent URL scanners such as Google Safe
Browsing~\cite{gsb} and the Malware Domain List. The malicious data set contains
all inclusion sequences where the last included resource's domain is reported
malicious by at least three out of the 62 URL scanners in VirusTotal. On the
other hand, the benign data set only contains inclusion sequences that do not
contain any domain in the entire sequence that is reported as malicious by any
URL scanner in VirusTotal. To build benign data set, we considered reputable
domains such as well-known search engines and advertising networks as benign
regardless of whether they are reported as malicious by any URL scanner in
VirusTotal. Table~\ref{inclusion:tab:dataset_statistics} summarizes the data
sets\footnote{https://github.com/sajjadium/Excision}. The unique number of
inclusion sequences and terminal domains are shown separately for the website
and extension data sets. The terminal domains column is the number of unique
domains that terminate inclusion sequences.

\subsection{Detection Results}
\label{inclusion:sec:detection_results}

To evaluate the accuracy of our classifier, we used 10-fold cross-validation, in
which we first partitioned each data set into 10 equal-sized folds, trained the
models on nine folds, and then validated the resulting models with the remaining
fold. The process was repeated for each fold and, at the end, we calculated the
average false positive rate and false negative rate. When splitting the data set
into training and testing sets, we made sure that inclusion sequences with
different lengths were present in both. We also ensured that both sets contained
extension-related inclusion sequences.

\begin{figure}[t]
    \centering
    \includegraphics[width=0.8\textwidth]{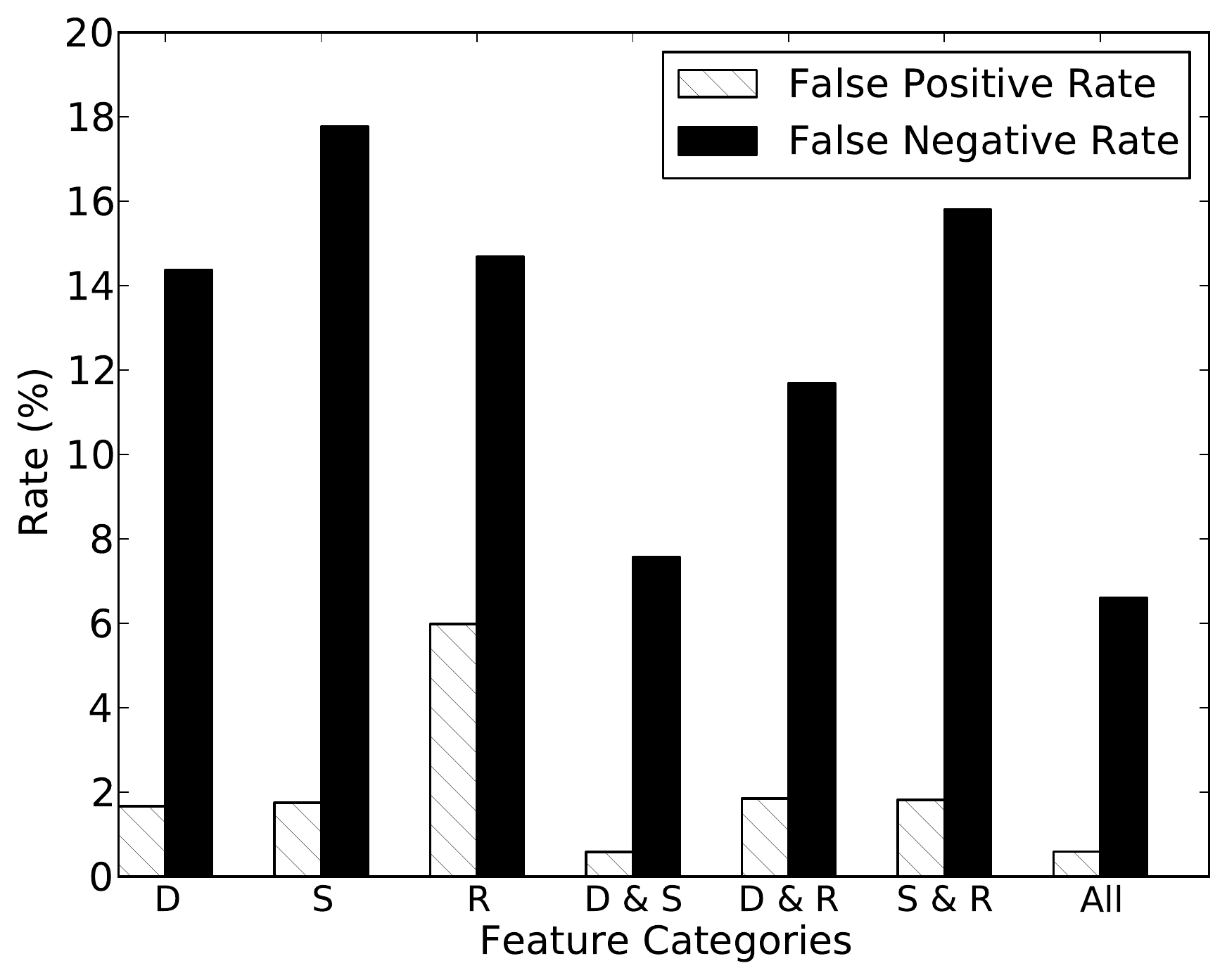}
    \caption{Effectiveness of features for classification (D = DNS, S = String, R = Role)}
    \label{inclusion:fig:detection_rate}
\end{figure}

The results show that our classifier achieved a false positive rate of 0.59\%
and false negative rate of 6.61\% (detection rate of 93.39\%). Most of the false
positives are due to inclusion sequences that do not appear too often in the
training sets. Hence, users are unlikely to experience many false positives in a
real browsing environment (as will be shown in our usability analysis in
Section~\ref{inclusion:sec:usability}).

To quantify the contribution of different feature categories to the
classification, we trained classifiers using different combinations of feature
categories and compared the results.
Figure~\ref{inclusion:fig:detection_rate} shows the false positive rate and
false negative rate of every combination with a 10-fold cross-validation
training scheme. According to Figure~\ref{inclusion:fig:detection_rate},
the best false positive and false negative rates were obtained using the
combination of all feature categories.

\subsection{Comparison with URL Scanners}
\label{inclusion:sec:comparison}

To evaluate the ability of our system in detecting unreported suspicious
domains, we ran our classifier on inclusion sequences collected from June 1st
until July 14th, 2015. We compared our detection results with reports from URL
scanners in VirusTotal and detected 89 new suspicious domains. We believe that
these domains are in fact dedicated malicious domains that play the role of
redirectors and manage malicious traffic flows as described in prior
work~\cite{sp2013linchpins}. These domains did not deliver malicious resources
themselves, but they consistently included resources from other domains that
were flagged as malicious by URL scanners. Out of 89 suspicious domains, nearly
44\% were recently registered in 2015, and more than 23\% no longer resolve to
an IP address.

\begin{figure}[t]
    \centering
    \includegraphics[width=0.8\textwidth]{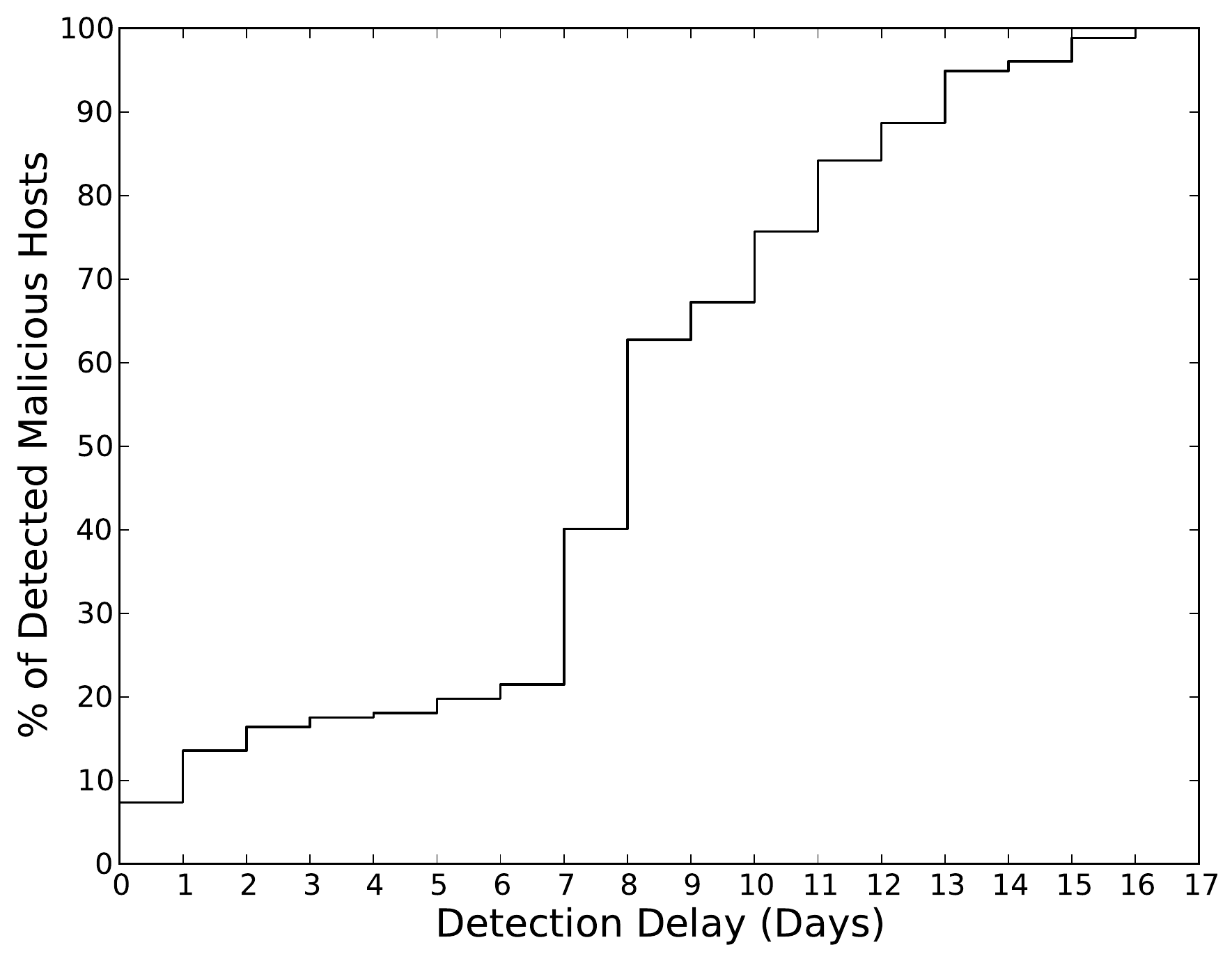}
    \caption{Early detection results}
    \label{inclusion:fig:early_detection}
\end{figure}

Furthermore, we detected 177 domains that were later reported by URL scanners
after some delay. Figure~\ref{inclusion:fig:early_detection} shows the
early detection results of our system. A significant number of these domains
were not reported until some time had passed after \excision initially
identified them. For instance, nearly 78\% of the malicious domains were not
reported by any URL scanner during the first week.

\subsection{Performance}
\label{inclusion:sec:performance}

To assess the performance of \excision, we used Selenium to automatically visit
the Alexa Top 1K with both original and modified Chromium browsers. In order to
measure our prototype performance with a realistic set of extensions, we
installed five of the most popular extensions in the Chrome Web Store: Adblock
Plus, Google Translate, Google Dictionary, Evernote Web Clipper, and
Tampermonkey.

For each browser, we visited the home pages of the entire list of websites and
recorded the total elapsed time. Due to the dynamic nature of ads and their
influence on page load time, we repeated the experiment 10 times and measured
the average elapsed time. On average, the elapsed times were 3,065 and 3,438
seconds for the original and modified browsers, respectively. Therefore,
\excision incurred a 12.2\% overhead on browsing time on average, which
corresponds to a noticeable overhead that is nevertheless acceptable for many
users (see Section~\ref{inclusion:sec:usability}). To measure the overhead
incurred by \excision on browser startup time, we launched the modified browser
10 times and measured the average browser launch time. \excision caused a 3.2
seconds delay on browser startup time, which is ameliorated by the fact that
this is a one-time performance hit.

\subsection{Usability}
\label{inclusion:sec:usability}

We conducted an experiment to evaluate the impact of \excision on the user's
browsing experience. We conducted the study on 10 students that self-reported as
expert Internet users. We provided each participant with a list of 50 websites
that were selected randomly from the Alexa Top 500 and then asked them to visit
at least three levels down in each website. Participants were asked to report
the number of visited pages and the list of domains reported as malicious by our
system. In addition, participants were asked to record the number of errors they
encountered while they browsed the websites. Errors were considered to occur
when the browser crashed, the appearance of a web page was corrupted, or page
load times were abnormally long. Furthermore, in order to ensure that benign
extensions were not prevented from executing as expected in the presence of our
system, the browser was configured to load the five popular extensions listed in
Section~\ref{inclusion:sec:performance} and participants were asked to report
any problem while using the extensions.

The results of the study show that out of 5,129 web pages visited by the
participants, only 83 errors were encountered and the majority of web pages
loaded correctly. Most of these errors happened due to relatively high load
times. In addition, none of the participants reported any broken extensions.
Furthermore, 31 malicious inclusions were reported by our tool that were
automatically processed (without manual examination, for privacy reasons) using
VirusTotal. Based on the results, we believe that our proof-of-concept prototype
is compatible with frequently used websites and extensions, and can be improved
through further engineering to work completely free of errors.

\paragraph{Ethics.}

In designing the usability experiment, we made a conscious effort to avoid
collecting personal or sensitive information. In particular, we restricted the
kinds of information we asked users to report to incidence counts for each of
the categories of information, except for malicious URLs that were reported by
our tool. Malicious URLs were automatically submitted to VirusTotal to obtain a
malice classification before being discarded, and were not viewed by us or
manually inspected. In addition, the participants were asked to avoid browsing
websites requiring a login or involving sensitive subject matter.

\section{Discussion}
\label{inclusion:sec:discussion}

Our study shows that detecting malicious third-party inclusions is possible by
analyzing resource inclusion sequences. According to the evaluation results,
\excision can detect a large number of malicious inclusions with a low false
positive rate of 0.59\%. However, due to the in-browser and real-time nature of
our system, we cannot easily incorporate other useful features such as domain
registration information or a global view of Web inclusions into our detection
system. For domain registration information, we would need to regularly fetch
domain \emph{whois} records; as these databases are rate-limited, this is not
currently feasible. In this work, we crafted a feature set that is suited for an
online, in-browser system to detects malicious inclusion sequences as web pages
load. But, attackers might try to exploit features we adopt to avoid detection
by \excision. For example, they might choose more meaningful names for their
domains or improve their domains' Alexa rankings with SEO
techniques~\cite{sec2011deseo}. However, these attempts are not very effective
since \excision rely on the business relationship between the hosts inside the
inclusion sequences for finding malicious resource in addition to the individual
hosts characteristics. Attackers need to change the sequence of inclusions to
evade our system which is not a trivial task and it increases the difficulty of
the attack significantly.

Moreover, we envision that both web users and website administrators can benefit
from using \excision. \excision protects users from attacks by preventing
browsers from including a malicious resource into web pages. Furthermore,
\excision allows website administrators to have more control over the content
that is delivered to their visitors when they sell space to ad networks.
Administrators do not need to write comprehensive CSP rules to control dynamic
content that is managed by third-party content providers. In addition to website
administrators and web users, the models learned by \excision can be used by ad
networks, URL scanners, and large organizations as well. They could passively
crawl various websites to identify compromised websites and malicious origins,
and this information could be used to augment blacklists and reputation-based
services (e.g.,~Google Safebrowsing) and also update corporate firewall policies
to prevent other clients from loading resources from those malicious origins.

\section{Chapter Summary}
\label{inclusion:sec:summary}

In this chapter, we presented \excision, an in-browser system to automatically
detect and block malicious third-party content inclusions before they can attack
the user's browser. Our system is implemented as a set of modifications to the
Chromium browser and does not perform any blacklisting to detect malicious
third-party inclusions. Our evaluation over an 11 month crawl of the Alexa Top
200K demonstrates that the prototype implementation of \excision achieved a
93.39\% detection rate with a false positive rate of 0.59\%. We also evaluated
the performance and usability of \excision when browsing popular websites, and
showed that the approach is capable of improving the security of users on the
Web by detecting 31 malicious inclusions during a user study without
significantly degrading the user experience.

\chapter{Identifying Ad Injection in Browser Extensions}
\label{chap:adinjection}

\section{Introduction}
\label{adinjection:sec:intro}

While ad injection cannot necessarily be categorized as an outright malicious
activity on its own, it is highly likely that many users in fact \emph{do not
want or expect} browser extensions to inject advertisements or other content
into Web pages. Moreover, it can have a significant impact on the security and
privacy of both users as well as website publishers. For example, recent studies
have shown that ad-injecting extensions not only serve ads from ad networks
other than the ones with which the website publishers intended, but they also
attempt to trick users into installing malware by inserting rogue elements into
the web page~\cite{sp2015adinjection,www2015adinjection}.

To address this problem, several automatic approaches have been proposed to
detect malicious behaviors (e.g., ad injection) in browser
extensions~\cite{www2015adinjection,usenixsec2014hulk,usenixsec2015webeval}. In
addition, centralized distribution points such as Chrome Web Store and Mozilla
Add-ons are using semi-automated techniques for review of extension behavior to
detect misbehaving extensions. However, there is no guarantee that analyzing the
extensions for a limited period of time leads to revealing the ad injection
behaviors. Finally, a client-side detection methodology has been proposed
in~\cite{sp2015adinjection} that reports any deviation from a legitimate DOM
structure as potential ad injections. However, this approach requires a priori
knowledge of a legitimate DOM structure as well as cooperation from content
publishers.

Although ad injection can therefore potentially pose significant risks, this
issue is not as clear-cut as it might first seem. Some users might legitimately
want the third-party content injected by the extensions they install, even
including injected advertisements. This creates a fundamental dilemma for
automated techniques that aim to identify clearly malicious or unwanted content
injection, since such techniques cannot intuit user intent and desires in a
fully automatic way.

To resolve this dilemma, we present \origintracer, an in-browser approach to
highlight extension-based content modification of web pages. \origintracer
monitors the execution of browser extensions to detect content modifications
such as the injection of advertisements. Content modifications are visually
highlighted in the context of the web page in order to
\begin{inparaenum}[\itshape i)\upshape]
     \item notify users of the presence of modified content, and
     \item inform users of the \emph{source} of the modifications.
\end{inparaenum}

With this information, users can then make an informed decision as to whether
they actually want these content modifications from specific extensions, or
whether they would rather uninstall the extensions that violate their
expectations.

\origintracer assists users in detecting content injection by distinguishing
injected or modified DOM elements from genuine page elements. This is performed
by annotating web page DOM elements with a \emph{provenance label set} that
indicates the principal(s) responsible for adding or modifying that element,
both while the page is loading from the publisher as well as during normal
script and extension execution. These annotations serve as trustworthy,
fine-grained provenance indicators for web page content. \origintracer can be
easily integrated into any browser in order to inform users of extension-based
content modification. Since \origintracer identifies all types of content
injections, it is able to highlight all injected advertisements regardless of
their types (e.g., flash ads, banner ads, and text ads).

We implemented a prototype of \origintracer as a set of modifications to the
Chromium browser, and evaluated its effectiveness by conducting a user study.
The user study reveals that \origintracer produced a significantly greater
awareness of third-party content modification, and did not detract from the
users' browsing experience.

The rest of this chapter is organized as follows.
Section~\ref{adinjection:sec:background} outlines the necessary background on
browser extensions and ad injection. Section~\ref{adinjection:sec:design}
presents our approach to web content provenance, while
Section~\ref{adinjection:sec:impl} discusses the implementation of our prototype
system. An evaluation of the effectiveness, usability, and performance of our
prototype is presented in Section~\ref{adinjection:sec:analysis} and
Section~\ref{adinjection:sec:conclusion} summarizes the paper.

\section{Background}
\label{adinjection:sec:background}

In the following, we introduce background information on browser extensions,
present an overview of advertisement injection as a canonical example of
questionable content modification, and motivate our approach in this context.

\subsection{Browser Extensions}
\label{adinjection:sec:background:extensions}

Browser extensions are programs that extend the functionality of a web browser.
Today, extensions are typically implemented using a combination of HTML, CSS,
and JavaScript written against a browser-specific extension API. These APIs
expose the ability to modify the browser user interface in controlled ways,
manipulate HTTP headers, and modify web page content through the document object
model (DOM) API. An extension ecosystem is provided by almost all major browser
vendors; for instance, Google and Mozilla both host centralized repositories of
extensions that users can download at the Chrome Web Store and Mozilla Add-ons
sites, respectively.

\subsection{Advertisement Injection}
\label{adinjection:sec:background:ad-injection}

As web advertising grew in popularity, those in a position to modify web content
such as ISPs and browser extension authors realized that profit could be
realized by injecting or replacing ads in web pages. For instance, some ISPs
began to tamper with HTTP traffic in transit, injecting DOM elements into HTML
documents that added ISP's advertisements into pages visited by their
customers~\cite{isp_adinjection2,isp_adinjection1}. In a similar fashion,
browser extensions started modifying pages to inject DOM elements in order to
show ads to users without necessarily obtaining the user's prior consent. Ad
injection has evolved to become a common form of unrequested third-party content
injection on today's web~\cite{adinjection_profit}.

\begin{figure}[t]
    \centering
    \includegraphics[width=0.8\textwidth]{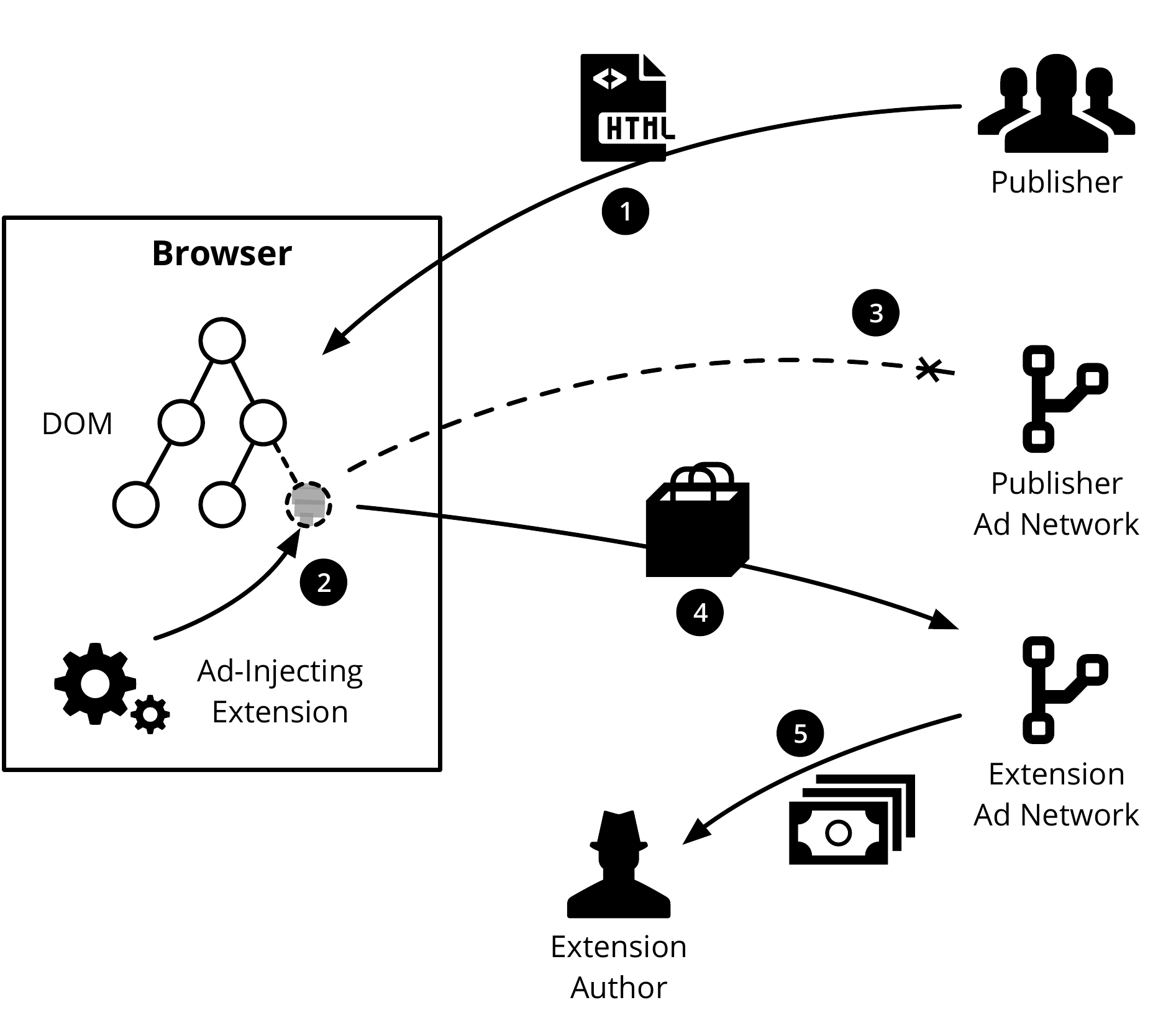}
    \caption{Overview of advertisement injection.
        \textbf{(1)}~The user accesses the publisher's site.
        \textbf{(2)}~An ad-injecting browser extension adds DOM elements to display ads to the user, and optionally removes existing ads.
        \textbf{(3)}~Ad revenue is diverted from the publisher.
        \textbf{(4)}~Ad impressions, clicks, and conversions are instead directed to the extension's ad network.
        \textbf{(5)}~Ad revenue flows to the extension author.
    }
    \label{adinjection:fig:adinjection}
\end{figure}

These practices have several effects on both publishers and users. On one hand,
ad injection diverts revenue from the publisher to the third party responsible
for the ad injection. If advertisements are the primary source of income for a
publisher, this can have a significant effect on their bottom line. If the
injected ads contain or reference undesired content (e.g., adult or political
topics), ad injection can also harm the reputation of the publisher from the
user's perspective. If the content injection is also malicious in nature, the
publisher's reputation can be further harmed in addition to exposing users to
security risks due to malware, phishing, and other threats. Prior work has shown
that users exposed to ad injection are more likely to be exposed to
``malvertising'' and traditional
malware~\cite{sp2015adinjection,www2015adinjection}.
Figure~\ref{adinjection:fig:adinjection} gives an overview of ad injection's
effect on the normal ad delivery process, while
Figure~\ref{adinjection:fig:indicator} shows an instance of ad injection on
\texttt{amazon.com} website.

\subsection{Motivation}
\label{adinjection:sec:background:motivation}

Recently, there have been efforts by browser vendors to remove ad-injecting
extensions from their repositories~\cite{google_adinjection}. Although
semi-automated central approaches have been successful in identifying
ad-injecting extensions, deceptive extensions can simply hide their ad injection
behaviors during the short period of analysis time. In addition, finding web
pages that trigger ad injection is a non-trivial task, and they can miss some
ad-injecting extensions. Moreover, there are extensions that are not provided
through the web stores, and users can get them from local marketplaces, which
may not examined the extensions properly. Hence, we believe that there is a need
for a protection tool to combat ad injection on the client side in addition to
centralized examination by browser vendors.

Furthermore, automatically determining whether third-party content modification
-- such as that due to ad injection -- should be allowed is not straightforward.
Benign extensions extensively modify web pages as part of their normal
functionality. To substantiate this, we examined five popular Chrome extensions
as of the time of writing; these are listed in
Table~\ref{adinjection:tab:popular_extensions}. Each of these extensions are
available for all major browsers, and all modify web pages (e.g., inject
elements) to implement their functionality. Therefore, automated approaches
based on this criterion run a high risk of false positives when attempting to
identify malicious or undesirable extensions.

\begin{table}[t]
    \centering
    \footnotesize
    \caption{Five popular Chrome extensions that modify web pages as part of their benign functionality}
    \begin{tabular}{lrc}
    \toprule
    \textbf{Extension} & \textbf{No. of Users} & \textbf{Injected Element} \\
    \midrule
    Adblock Plus & 10,000,000+ & \texttt{\textless iframe\textgreater} \\
    Google Translate & 6,000,000+ & \texttt{\textless div\textgreater} \\
    Tampermonkey & 5,800,000+ & \texttt{\textless img\textgreater} \\
    Evernote Web Clipper & 4,300,000+ & \texttt{\textless iframe\textgreater} \\
    Google Dictionary & 3,000,000+ & \texttt{\textless div\textgreater} \\
    \bottomrule
    \end{tabular}
    \label{adinjection:tab:popular_extensions}
\end{table}

Moreover, it is not enough to identify that advertisements, for instance, have
been injected by a third party. This is because some users \emph{might
legitimately desire} the content that is being added to web pages by the
extensions they install. To wit, it is primarily for this reason that a recent
purge of extensions from the Chrome Web Store did not encompass the entirety of
the extensions that were identified as suspicious in a previous study, as the
third-party content modification could not be clearly considered as
malicious~\cite{sp2015adinjection}. Instead, we claim that \emph{users
themselves} are best positioned to make the determination as to whether
third-party content modification is desired or not. An approach that proceeds
from this observation would provide sufficient, easily comprehensible
information to users in order to allow an informed choice as to whether content
is desirable or should be blocked. It should be noted that defending against
drive-by downloads and general malware is not the focus of this paper. Rather,
the goal is to highlight injected ads to increase likelihood that user will make
an informed choice to not click on them.

We envision that \origintracer could be used as a complementary approach to
existing techniques such as central approaches used by browser vendors. Also,
browser vendors can benefit from using our system in addition to end users to
detect the content modifications by extensions in a more precise and reliable
way. In the following sections, we present design and implementation of our
system.

\section{Design}
\label{adinjection:sec:design}

In this section, we describe an in-browser approach for identifying third-party
content modifications in web browsers. The approach adds \emph{fine-grained
provenance tracking} to the browser, at the level of individual DOM elements.
Provenance information is used in two ways:
\begin{inparaenum}[\itshape i)\upshape]
    \item to distinguish between content that originates from the web page
    publisher and content injected by an unassociated third party, and
    \item to indicate \emph{which} third party (e.g., extension) is responsible
    for content modifications using provenance indicators.
\end{inparaenum}
By integrating the approach directly into the browser, we guarantee the
trustworthiness of both the provenance information and the visual indicators.
That is, as the browser is already part of the trusted computing base (TCB) in
the web security model, we leverage this as the appropriate layer to compute
precise, fine-grained provenance information. Similarly, the browser holds
sufficient information to ensure that provenance indicators cannot be tampered
with or occluded by malicious extensions. While we consider malicious or
exploited browser plug-ins such as Flash Player outside our threat model, we
note that modern browsers take great pains to isolate plug-ins in least
privilege protection domains. We report separately on the implementation of the
approach in Section~\ref{adinjection:sec:impl}.

In the following, we present our approach to tracking and propagating content
provenance, and then discuss provenance indicators and remediation strategies.

\subsection{Content Provenance}
\label{adinjection:sec:design:provenance}

Web pages are composed of HTML that references resources such as stylesheets,
scripts, images, plug-ins such as Flash objects, or even other web pages loaded
inside frames. The document object model (DOM) is a natural structural
representation of a web page that can be manipulated through a standard API, and
serves as a suitable basis for provenance tracking. In particular, our system
tracks the provenance of each element \(e\) contained in a DOM. Provenance for a
DOM element is recorded as a set of labels \(\ell \in
\mathcal{P}\left(L\right)\), where the set of all labels \(L\) corresponds to a
generalization of standard web origins to include extensions. That is, instead
of the classic origin 3-tuple of \(\left\langle \texttt{scheme}, \texttt{host},
\texttt{port} \right\rangle\), we record

\begin{align*}
L &= \left\langle S, I, P, X \right\rangle \\
S &= \left\{ \texttt{scheme} \right\} \cup \left\{ \texttt{``extension''} \right\} \\
I &= \left\{ \texttt{host} \right\} \cup \left\{ \texttt{extension-identifier} \right\} \\
P &= \left\{ \texttt{port} \right\} \cup \left\{ \texttt{null} \right\} \\
X &= \left\{0, 1, 2, \ldots\right\}
\end{align*}

\begin{figure}[t]
    \centering
    \includegraphics[width=0.8\textwidth]{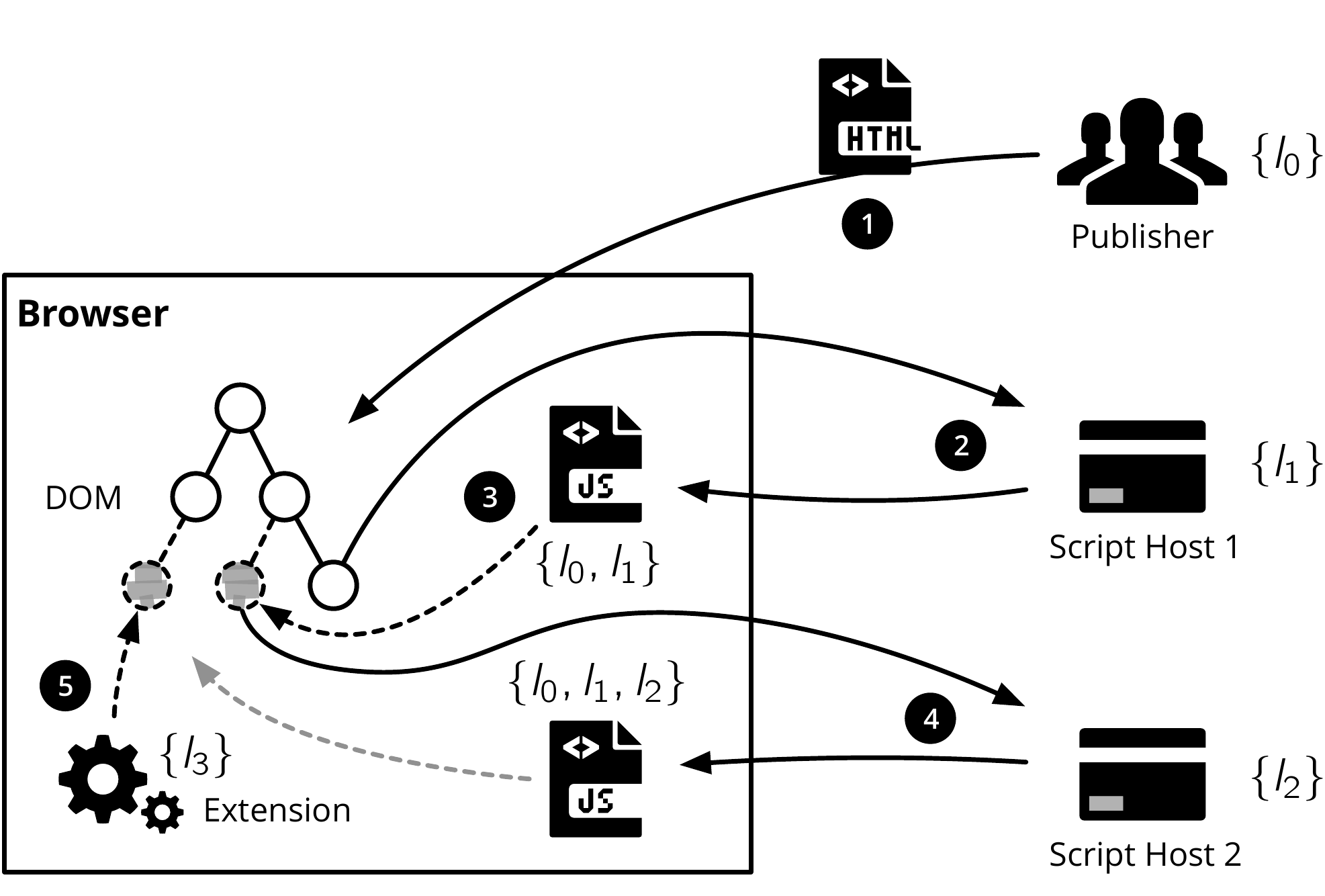}
    \caption{Element-granularity provenance tracking.
        \textbf{(1)}~Content loaded directly from the publisher is labeled
        with the publisher's origin, \(l_0\).
        \textbf{(2)}~An external script reference to origin \(l_1\) is performed.
        \textbf{(3)}~DOM modifications from \(l_1\)'s script are labeled with
        the label set \(\left\{l_0,l_1\right\}\).
        \textbf{(4)}~Further external script loads and subsequent DOM
        modifications induce updated label sets -- e.g.,
        \(\left\{l_0,l_1,l_2\right\}\).
        \textbf{(5)}~A DOM modification that originates from an extension
        produces provenance label sets \(\left\{l_0, l_1, l_2, l_3\right\}\) for
        the element}
    \label{adinjection:fig:provenance}
\end{figure}

In other words, a label is a 4-tuple that consists of a normal network scheme or
\texttt{extension}, a network host or a unique extension identifier, a port or
the special \texttt{null} value, and an index used to impose a global total
order on labels as described below. While browsers use different extension
identifiers, including randomly-generated identifiers, the exact representation
used is unimportant so long as there is a one-to-one mapping between extensions
and identifiers and their use is locally consistent within the browser. An
overview of provenance tracking is depicted in
Figure~\ref{adinjection:fig:provenance}.

\subsubsection{Static Publisher Provenance}

Content provenance tracking begins with a web page load. As the DOM is parsed by
the browser, each element is labeled with a singleton label set containing the
origin of the publisher, \(\left\{l_0\right\}\). Thus, static provenance
tracking is straightforward and equivalent to the standard use of origins as a
browser security context.

\subsubsection{Dynamic Publisher Provenance}

Content provenance becomes more interesting in the presence of dynamic code
execution. As JavaScript can add, modify, and remove DOM elements in an
arbitrary fashion using the DOM API exposed by the browser, it is necessary to
track these modifications in terms of provenance labels.

New provenance labels are created from the publisher's label set
\(\left\{l_0\right\}\) as follows. Whenever an external script is referenced
from the initial DOM resulting from the page load, a new label \(l_i, i \in
\left\{1,2,\ldots\right\}\) is generated from the origin of the script. All
subsequent DOM modifications that occur as a result of an external script loaded
from the initial DOM are recorded as \(\left\{l_0,l_i\right\}\). Successive
external script loads follow the expected inductive label generation process --
i.e., three successive external script loads from unique origins will result in
a label set \(\left\{l_0,l_i,l_j,l_k\right\}\). Finally, label sets contain
unique elements such that consecutive external script loads from a previously
accessed origin are not reflected in the label for subsequent DOM modifications.
For instance, if the web page publisher loads a script from the publisher's
origin, then any resulting DOM modifications will have a provenance label set of
\(\left\{l_0\right\}\) instead of \(\left\{l_0,l_0\right\}\). Content provenance
is propagated for three generic classes of DOM operations: element insertion,
modification, and deletion.

Element insertions produce an updated DOM that contains the new element labeled
with the current label set, and potentially generates a new label set if the
injected element is a script. Element modifications produce a DOM where the
modified element's label set is merged with the current label set. Finally,
element deletions simply remove the element from the DOM.

\subsubsection{Extension Provenance}

The third and final form of provenance tracking concerns content modifications
due to DOM manipulations by extensions. In this case, provenance propagation
follows the semantics for the above case of dynamic publisher provenance. Where
these two cases differ, however, is in the provenance label initialization.
While provenance label sets for content that originates, perhaps indirectly,
from the web page publisher contains the publisher's origin label \(l_0\),
content that originates from an extension is rooted in a label set initialized
with the \emph{extension's} label. In particular, content modifications that
originate from an extension \emph{are not labeled} by the publisher's origin. An
exception to this occurs when the extension, either directly or indirectly,
subsequently loads scripts from the publisher, or modifies an existing element
that originated from the publisher.

\subsection{Content Provenance Indicators}
\label{adinjection:sec:design:indicators}

With the fine-grained content provenance scheme described above, identifying the
principal responsible for DOM modifications is straightforward. For each
element, all that is required is to inspect its label set \(\ell\) to check
whether it contains the label of any extension.

A related, but separate, question is how best to relay this information to the
user. In this design, several options are possible on a continuum from simply
highlighting injected content without specific provenance information to
reporting the full ordered provenance chain from the root to the most recent
origin. The first option makes no use of the provenance chain, while the other
end of the spectrum is likely to overwhelm most users with too much information,
degrading the practical usefulness of provenance tracking. We suspect that a
reasonable balance between these two extremes is a summarization of the full
chain, for instance by reporting only the label of the corresponding extension.

Finally, if a user decides that the third-party content modification is
unwanted, another design parameter is how to act upon this decision. Possible
actions include blocking specific element modifications, removing the offending
extension, or reporting its behavior to a central authority. We report on the
specific design choices we made with respect to provenance indicators in the
presentation of our implementation in Section~\ref{adinjection:sec:impl}.

\section{Implementation}
\label{adinjection:sec:impl}

In this section, we present \origintracer, our prototype implementation for
identifying and highlighting extension-based web page content modifications. We
implemented \origintracer as a set of modifications to the Chromium
browser\footnote{https://github.com/sajjadium/OriginTracer}. In particular, we
modified both Blink and the extension engine to track the provenance of content
insertion, modification, and removal according to the semantics presented in
Section~\ref{adinjection:sec:design}. These modifications also implement
provenance indicators for suspicious content that does not originate from the
publisher. In total, our changes consist of approximately 900 SLOC for C++ and
several lines of JavaScript\footnote{SLOC were measured using David Wheeler's
SLOCCount~\cite{sloccount}.}. In the following, we provide more detail on the
integration of \origintracer into Chromium.

\subsection{Tracking Publisher Provenance}

A core component of \origintracer is responsible for introducing and propagating
provenance label sets for DOM elements. In the following, we discuss the
implementation of provenance tracking for publisher content.

\subsubsection{Tracking Static Elements}

As discussed in Section~\ref{adinjection:sec:design}, provenance label sets for
static DOM elements that comprise the HTML document sent by the publisher as
part of the initial page load are equivalent to the publisher's web origin -- in
our notation, \(l_0\). Therefore, minimal modifications to the HTML parser were
necessary to introduce these element annotations, which is performed in an
incremental fashion as the page is parsed.

\subsubsection{Tracking Dynamic Elements}

To track dynamic content modifications, this component of \origintracer must
also monitor JavaScript execution. When a \texttt{script} tag is encountered
during parsing of a page, Blink creates a new element and attaches it to the
DOM. Then, Blink obtains the JavaScript code (fetching it from network in the
case of remote script reference), submits the script to the V8 JavaScript engine
for execution, and pauses the parsing process until the script execution is
finished. During execution of the script, some new elements might be created
dynamically and inserted into the DOM. According to the provenance semantics,
these new elements inherit the label set of the script. In order to create new
elements in JavaScript, one can
\begin{inparaenum}[\itshape i)\upshape]
    \item use DOM APIs to create a new element and attach it to the web page's
    DOM, or
    \item write HTML tags directly into the page.
\end{inparaenum}
In the first method, to create a new element object, a canonical example is to
provide the tag name to the \texttt{createElement} function. Then, other
attributes of the newly created element are set -- e.g., after creating an
element object for an \texttt{a} tag, an address must be provided for its
\texttt{href} attribute. Finally, the new element should be attached to the DOM
tree as a child using \texttt{appendChild} or \texttt{insertBefore} functions.
In the second method, HTML is inserted directly into the web page using the
functions such as \texttt{write} and \texttt{writeln}, or by modifying the
\texttt{innerHTML} attribute. In cases where existing elements are modified
(e.g., changing an image's \texttt{src} attribute), the element inherits the
label set of the currently executing script as well. In order to have a complete
mediation of all DOM modifications to Web page, several classes in Blink
implementation were instrumented in order to assign provenance label sets for
newly created or modified elements using the label set applied to the currently
executing script.

\subsubsection{Handling Events and Timers}

An additional consideration for this \origintracer component is modifications to
event handlers and timer registrations, as developers make heavy use of event
and timer callbacks in modern JavaScript. For instance, such callbacks are used
to handle user interface events such as clicking on elements, hovering over
elements, or to schedule code after a time interval has elapsed. In practice,
this requires the registration of callback handlers via
\texttt{addEventListener} API for events, and \texttt{setTimeout} and
\texttt{setInterval} for timers. To mediate callbacks related to the addition
and firing of events and timers, we slightly modified the \texttt{EventTarget}
and \texttt{DOMTimer} classes in Blink, respectively. Specifically, we record
the mapping between the running scripts and their registered callback functions,
and then recover the responsible scripts for DOM modification during callback
execution.

\subsection{Tracking Extension Provenance}
\label{adinjection:sec:impl:extension}

Chromium's extension engine is responsible for loading extensions, checking
their permissions against those declared in the manifest file, injecting content
scripts, dispatching background scripts and content scripts to the V8 script
engine for execution, and providing a channel for communication between content
scripts and background page.

Chromium extensions can manipulate the web page's content by injecting
\emph{content scripts} into the web page or using the \texttt{webRequest} API.
Content scripts are JavaScript programs that can manipulate the web page using
the shared DOM, communicate with external servers via \texttt{XMLHttpRequest},
invoke a limited set of \texttt{chrome.*} APIs, and interact with their owning
extension's background page. By using \texttt{webRequest}, extensions are also
able to modify and block HTTP requests and responses in order to change the web
page's DOM.

In this work, we only track content modifications by content scripts and leave
identifying ad injection by \texttt{webRequest} for future engineering work.
Prior work, however, has mentioned that only 5\% of ad injection incidents
occurred via \texttt{webRequest}; instead, Chrome extensions mostly rely on
content scripts to inject advertisements~\cite{sp2015adinjection}. Moreover,
with modern websites becoming more complex, injecting stealthy advertisement
into the page using \texttt{webRequest} is not a trivial task.

\subsubsection{Tracking Content Script Injection and Execution}

To track elements created or modified during the execution of content scripts,
extension engine was modified to hook events corresponding to script injection
and execution. Content scripts can be inserted into the web page using different
methods. If a content script should be injected into every matched web page, it
must be registered in the extension manifest file using the
\texttt{content\_scripts} field. By providing different options for this field,
one can control when and where the content scripts be injected. Another method
is programmatic injection, which is useful when content scripts should be
injected in response to specific events (e.g., a user clicks the extension's
browser action). With programmatic injection, content scripts can be injected
using the \texttt{tabs.executeScript} API if the \texttt{tabs} permission is set
in the manifest file. Either way, content scripts have a provenance label set
initialized with the extension's label upon injection.

\subsubsection{Handling Callback Functions}

Chromium's extension engine provides a messaging API as a communication channel
between the background page and the content scripts. The background page and
content scripts can receive messages from each other by providing a callback
function for the \texttt{onMessage} or \texttt{onRequest} events, and can send
messages by invoking \texttt{sendMessage} or \texttt{sendRequest}. To track the
registration and execution of callback functions, the \texttt{send\_request} and
\texttt{event} modules were slightly modified in the extension engine.
Specifically, we added some code to map registered callbacks to their
corresponding content scripts in order to find the extension responsible for DOM
modification.

\subsection{Content Provenance Indicators}
\label{adinjection:sec:impl:indicators}

Given DOM provenance information, \origintracer must first
\begin{inparaenum}[\itshape i)\upshape]
    \item identify when suspicious content modifications -- e.g.,
    extension-based ad injection -- has occurred, and additionally
    \item communicate this information to the user in an easily comprehensible
    manner.
\end{inparaenum}

To implement the first requirement, our prototype monitors for content
modifications where a subtree of elements are annotated with label sets that
contains a particular extension's label. This check can be performed efficiently
by traversing the DOM and inspecting element label sets after a set of changes
have been performed on the DOM.

\begin{figure}[!t]
    \centering
    \includegraphics[width=0.8\textwidth]{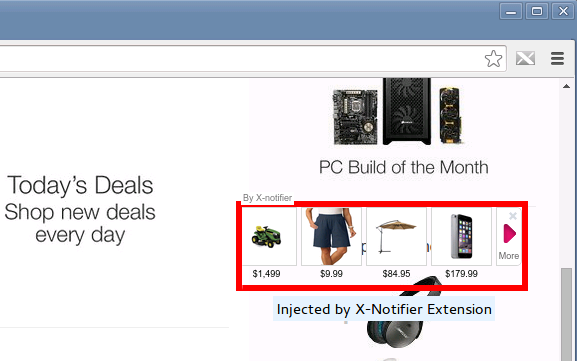}
    \caption{An example of indicator for an injected advertisement on \texttt{amazon.com} website}
    \label{adinjection:fig:indicator}
\end{figure}

There are several possible options to communicate content provenance as
mentioned in Section~\ref{adinjection:sec:design}. In our current prototype,
provenance is indicated using a configurable border color of the root element of
the suspicious DOM subtree. This border should be chosen to be visually distinct
from the existing color palette of the web page. Finally, a tooltip indicating
the root label is displayed when the user hovers their mouse over the DOM
subtree. An example is shown in Figure~\ref{adinjection:fig:indicator}. To
implement these features, \origintracer modifies \texttt{style} and
\texttt{title} attributes. In addition, since \origintracer highlights elements
in an online fashion, it must delay the addition of highlighting until the
element is attached to the page's DOM and is displayed. Therefore, modifications
were made to the \texttt{ContainerNode} class that is responsible for attaching
new elements to the DOM.

While we did not exhaustively explore the design space of content provenance
indicators in this work (e.g., selective blocking of extension-based DOM
modifications), we report on the usability of the prototype implementation in
our evaluation.

\section{Analysis}
\label{adinjection:sec:analysis}

In this section, we measure the effectiveness, usability, and performance of
content provenance indicators using the \origintracer prototype. In particular,
the questions we aim to answer with this evaluation are:

\begin{enumerate*}
    \item[\textbf{(Q1)}] \label{adinjection:eval:q1}
    How susceptible are users to injected content such as third-party
    advertisements? (\S\ref{adinjection:sec:analysis:susceptibility})
    \item[\textbf{(Q2)}] \label{adinjection:eval:q2}
    Do provenance indicators lead to a significant, measurable decrease in the
    likelihood of clicking on third-party content that originates from
    extensions? (\S\ref{adinjection:sec:analysis:effectiveness})
    \item[\textbf{(Q3)}] \label{adinjection:eval:q3}
    Are users likely to use the system during their normal web browsing?
    (\S\ref{adinjection:sec:analysis:usability})
    \item[\textbf{(Q4)}] \label{adinjection:eval:q4}
    Does integration of the provenance tracking system significantly degrade
    the users' browsing experience and performance of the browser on a
    representative sample of websites? (\S\ref{adinjection:sec:analysis:perf})
\end{enumerate*}

\paragraph{Ethics.}

As part of the evaluation, we performed two experiments involving users
unaffiliated with the project as described below. Due to the potential risk to
user confidentiality and privacy, we formulated an experimental protocol that
was approved by our university's institutional review board (IRB). This protocol
included safeguards designed to prevent exposing sensitive user data such as
account names, passwords, personal addresses, and financial information, as well
as to protect the anonymity of the study participants with respect to data
storage and reporting. While users were not initially told the purpose of some
of the experiments, all users were debriefed at the end of each trial as to the
true purpose of the study.

\subsection{Effectiveness}

Similar to prior work~\cite{chi2006phishing}, we performed a user study to
measure the effectiveness of content provenance in enabling users to more easily
identify unwanted third-party content. However, we performed the user study with
a significantly larger group of participants. The study population was composed
of 80 students that represent a range of technical sophistication. We conducted
an initial briefing prior to the experiments where we made it clear that we were
interested in honest answers.

\subsubsection{User Susceptibility to Ad Injection}
\label{adinjection:sec:analysis:susceptibility}

The goal of the first phase of the experiment was to measure whether users were
able to detect third-party content that was not intended for inclusion by the
publishers of web pages presented to them. Users were divided into two equal
sized groups of 40. In each group, users were first presented with three
unmodified Chromium browsers, each of which had a separate ad-injecting
extension installed: \texttt{Auto Zoom}, \texttt{Alpha Finder}, and
\texttt{X-Notifier} for the first group, and \texttt{Candy Zapper},
\texttt{uTorrent}, and \texttt{Gethoneybadger} for the second group. These
extensions were chosen because they exhibit a range of ad injection behaviors,
from subtle injections that blend into the publisher's web page to very obvious
pop-ups that are visually distinct from the publisher's content.

\begin{figure}[t]
    \centering
    \begin{subfigure}[t]{0.48\textwidth}
        \includegraphics[width=1\textwidth]{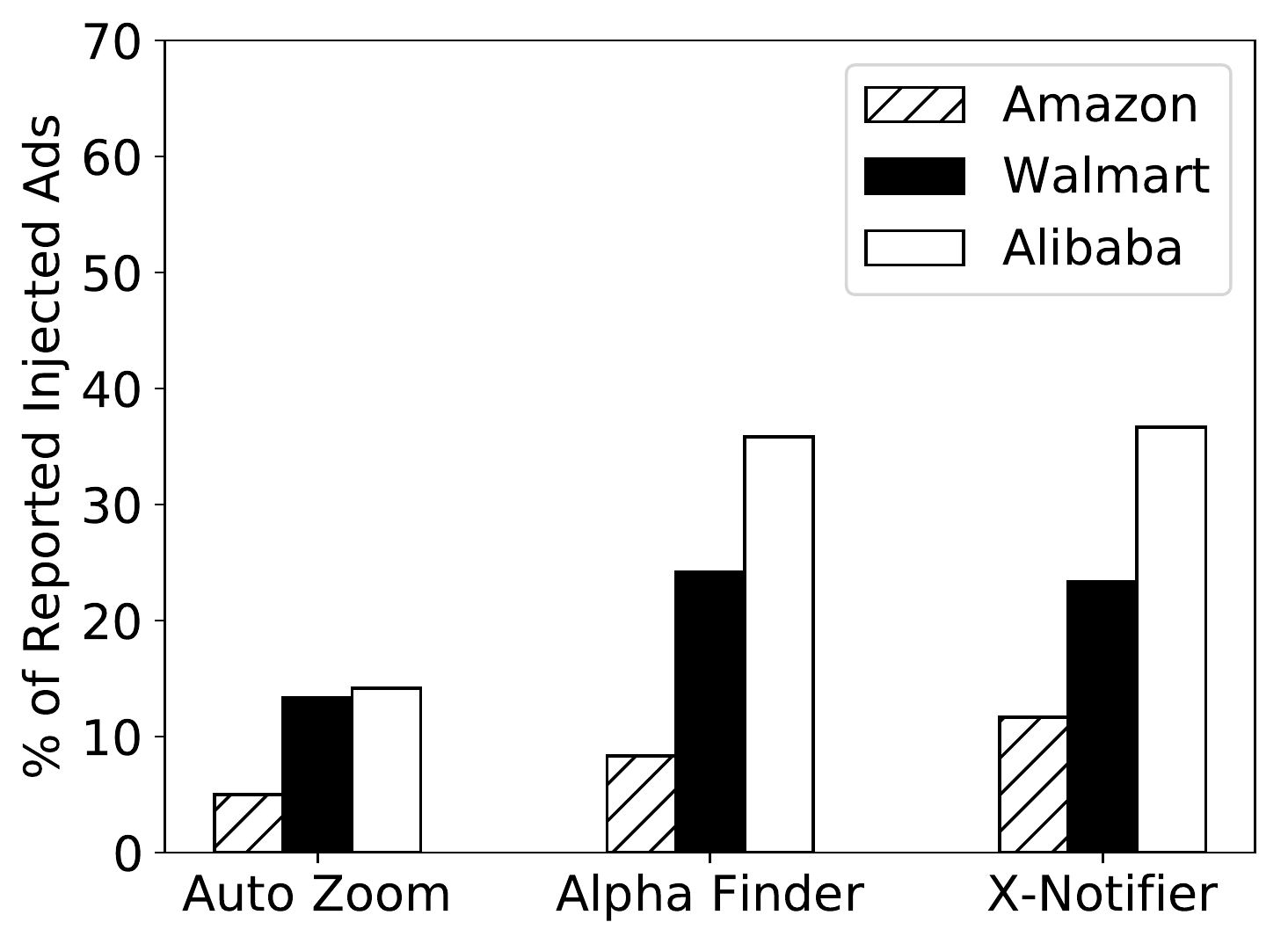}
        \caption{\scriptsize{Group 1.}}
        \label{adinjection:fig:user_study:1}
    \end{subfigure}
    \hfill
    \begin{subfigure}[t]{0.48\textwidth}
        \includegraphics[width=1\textwidth]{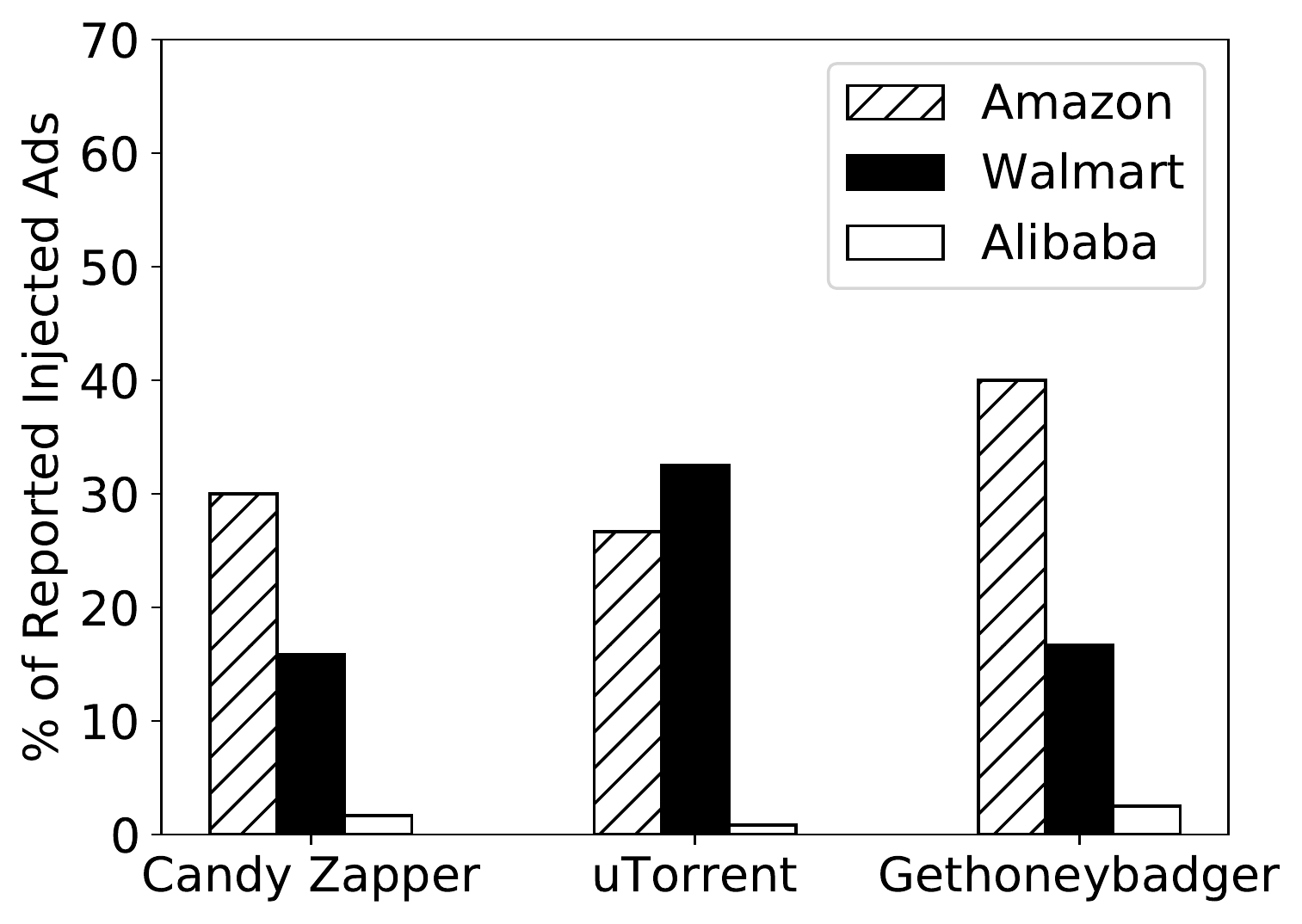}
        \caption{\scriptsize{Group 2.}}
        \label{adinjection:fig:user_study:2}
    \end{subfigure}
    \caption{Percentage of injected ads that are reported correctly by all the participants}
    \label{adinjection:fig:extensions_adinjection}
\end{figure}

Using each browser, the participants were asked to visit three popular retail
websites: \texttt{Amazon}, \texttt{Walmart}, and \texttt{Alibaba}. Each
ad-injecting extension monitors for visits to these websites, and each injects
three different types of advertisements into these sites. For each website, we
asked the participants to examine the page and tell us if they noticed any
content in the page that did not belong to the website -- in other words,
whether any content did not seem to originate from the publisher. For each
group, we aggregated the responses and presented the percentage of correctly
reported ad injection incidents for each extension in
Figure~\ref{adinjection:fig:extensions_adinjection}.

The results demonstrate that a significant number of Internet users often do not
recognize when ad injection occurs in the wild, even when told to look for
foreign content. For example, 34 participants did not recognize \emph{any}
injected ads out of the three that were added to Amazon website by
\texttt{Auto Zoom} extension. Comparatively more users were able to identify ads
injected by \texttt{Alpha Finder} and \texttt{X-Notifier}. We suspect the reason
for this is because these extensions make use of pop-up advertisements that are
easier to recognize as out-of-place. However, a significant number of users
nevertheless failed to note these pop-up ads, and even after prompting stated
that they thought these ads were part of the publisher's content. More
generally, across all websites and extensions, many participants failed to
identify any injected ads whatsoever.

\begin{figure}[!t]
    \centering
    \begin{subfigure}[t]{0.48\textwidth}
        \includegraphics[width=1\textwidth]{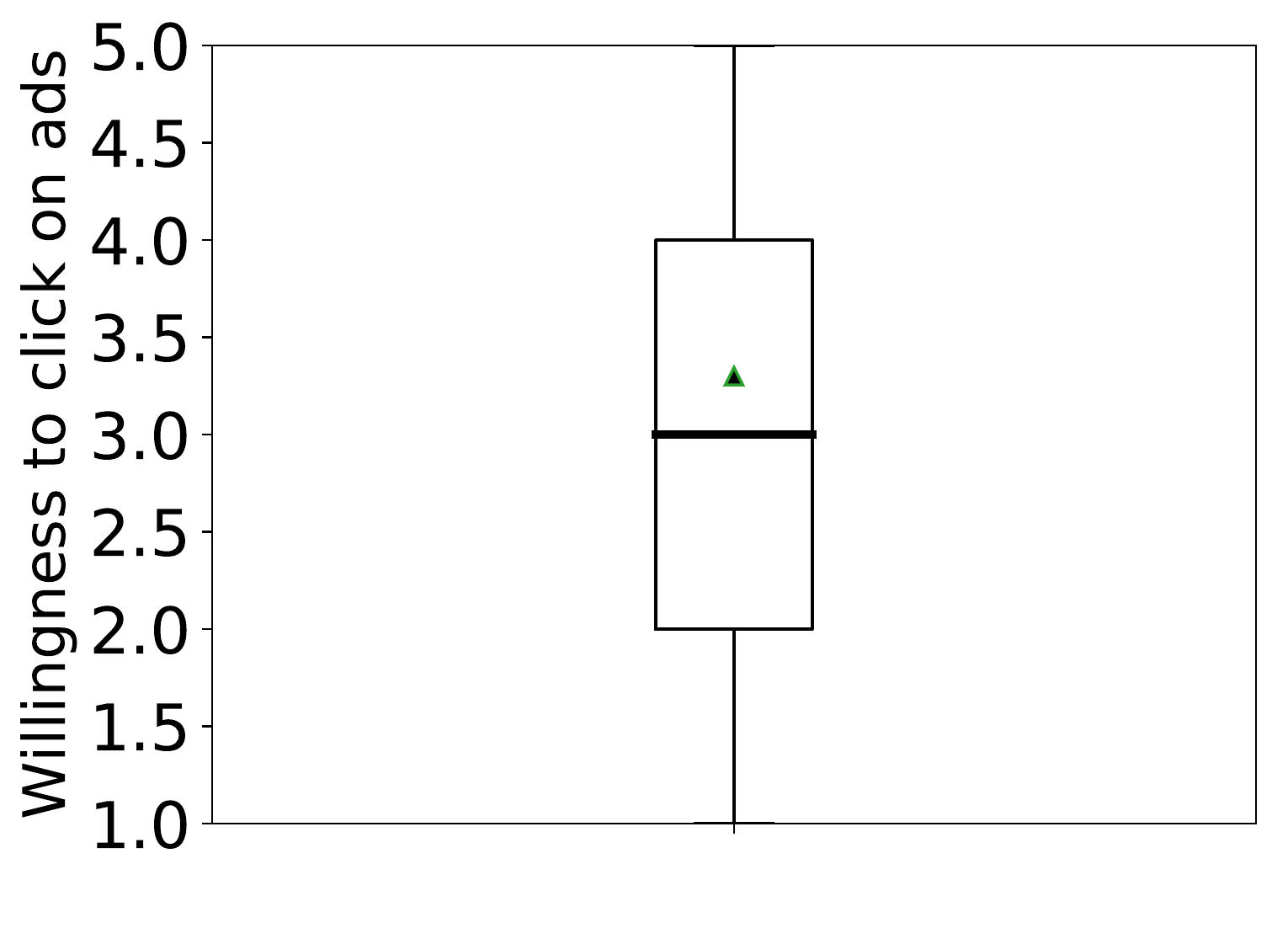}
        \caption{\scriptsize{Susceptibility to ad injection}}
        \label{adinjection:fig:user_study:susceptibility}
    \end{subfigure}
    \begin{subfigure}[t]{0.48\textwidth}
        \includegraphics[width=1\textwidth]{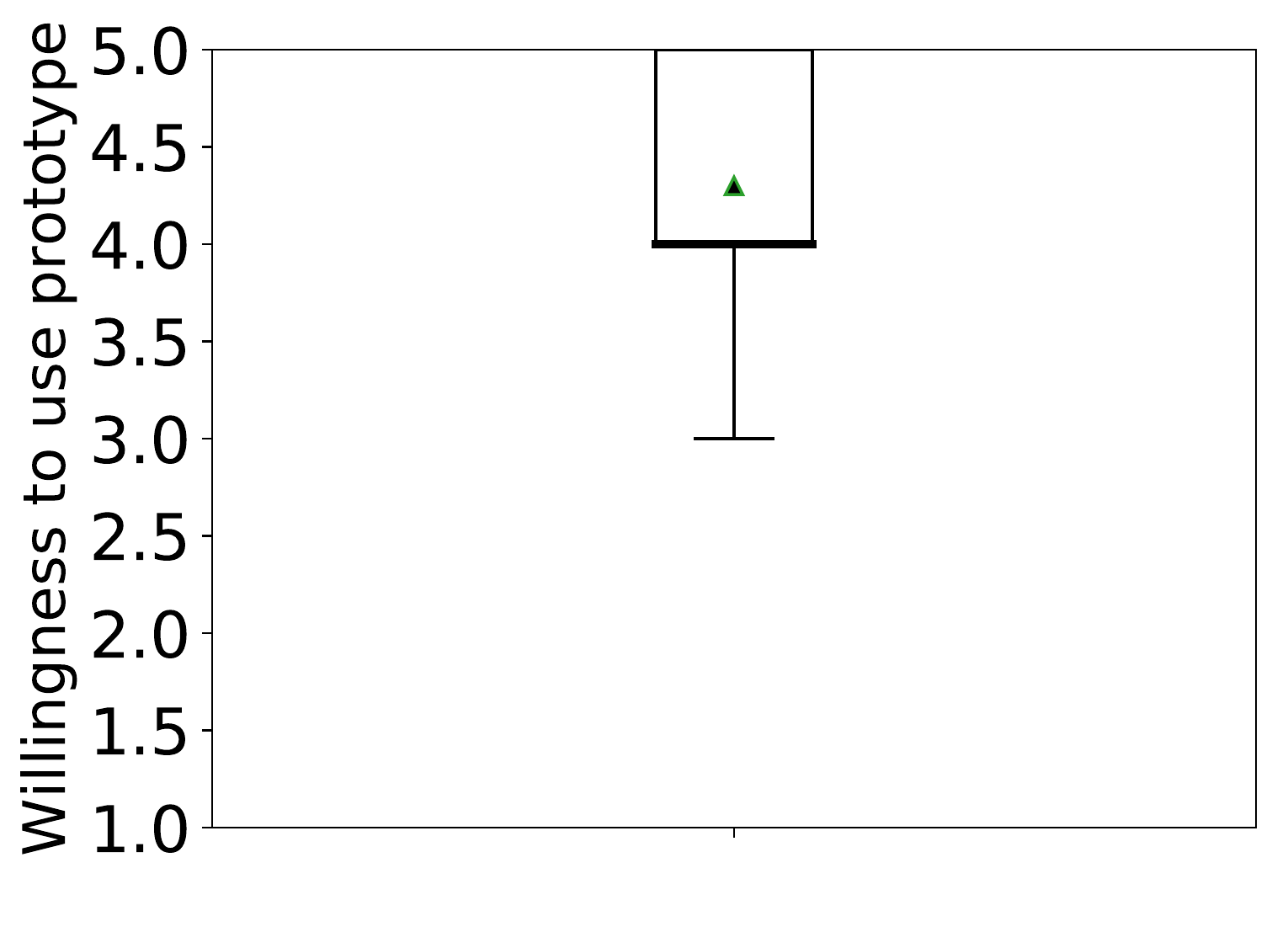}
        \caption{\scriptsize{Usability of content provenance}}
        \label{adinjection:fig:user_study:usability}
    \end{subfigure}
    \begin{subfigure}[t]{0.48\textwidth}
        \includegraphics[width=1\textwidth]{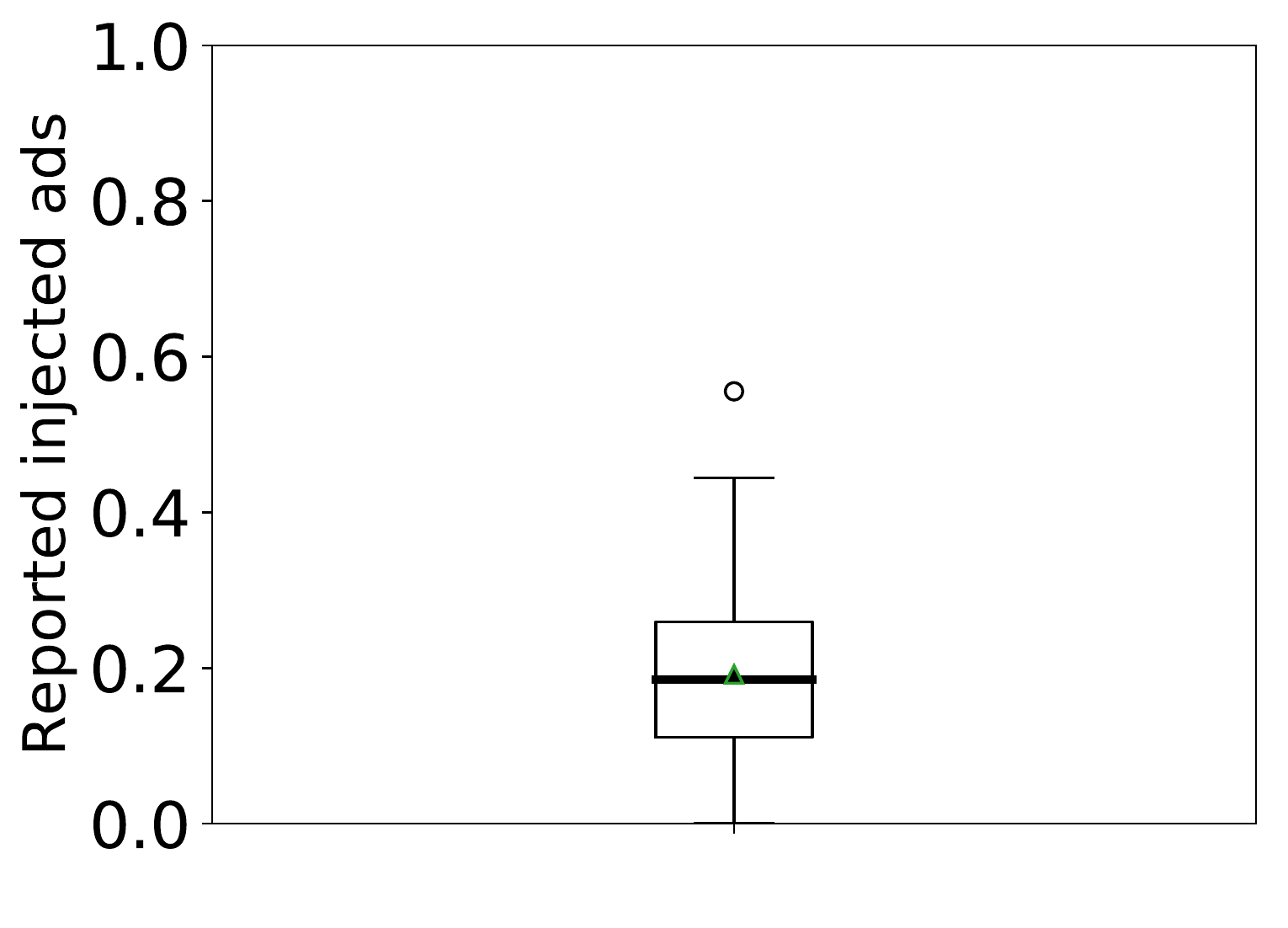}
        \caption{\scriptsize{Unassisted identification of injected ads}}
        \label{adinjection:fig:user_study:identification_unassisted}
    \end{subfigure}
    \begin{subfigure}[t]{0.48\textwidth}
        \includegraphics[width=1\textwidth]{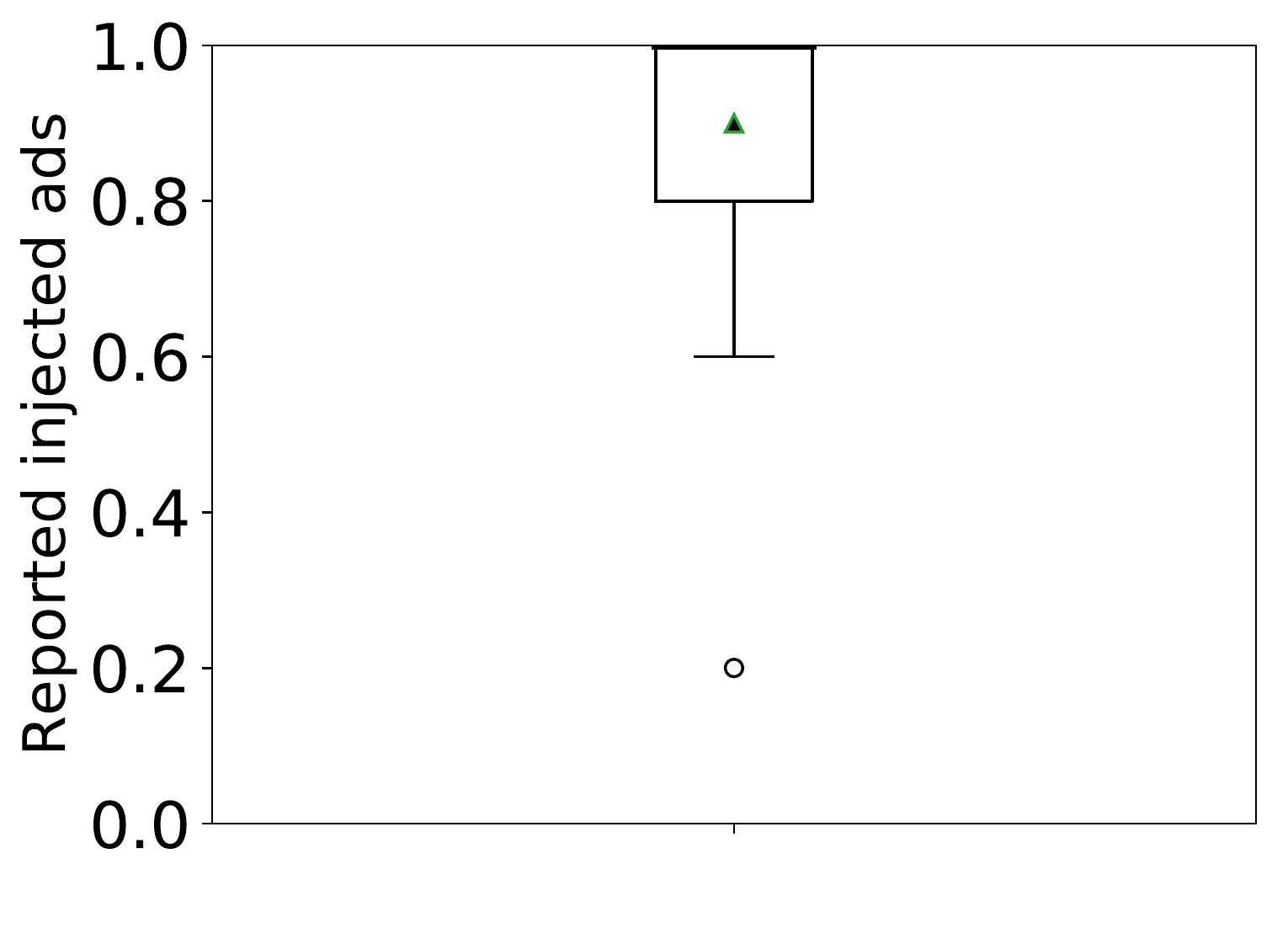}
        \caption{\scriptsize{Assisted identification of injected ads}}
        \label{adinjection:fig:user_study:identification_assisted}
    \end{subfigure}
    \caption{User study results. For each boxplot, the box represents the
    boundaries of the first and third quartiles. The band within each box
    is the median, while the triangle is the mean. The whiskers represent
    1.5 IQR boundaries, and outliers are represented as a circle}
    \label{adinjection:fig:user_study}
\end{figure}

We then asked each participant whether they would click on ads in general to
measure the degree of trust that users put into the contents on the publisher's
page. Specifically, we asked participants to rate the likelihood of clicking on
ads on a scale from one to five, where one means that they would never click on
an ad while five means that they would definitely click on an ad. We aggregated
the responses and present the results in
Figure~\ref{adinjection:fig:user_study:susceptibility}.

These results show that a significant number of users, roughly half,
\emph{would} click on advertisements that might not originate from the
publisher, but that were instead injected by an extension. This demonstrates the
effectiveness of ad injection as a mechanism for diverting revenue from
publishers to extension authors. It also shows the potential effectiveness of
malicious extensions in using content modifications to expose users to
traditional malware.

\subsubsection{Effectiveness of Content Provenance Indicators}
\label{adinjection:sec:analysis:effectiveness}

After the first phase of the experiment, we briefly explained the purpose of
\origintracer and content provenance to the participants. Then, for each
participant in each group, we picked one of the three ad-injecting extensions in
which, the participant did not detect most of the injected ads and installed it
on a Chromium instance equipped with \origintracer. Then, each participant was
asked to visit one of the three retail websites by his choice and identify
third-party content modifications -- i.e., injected ads -- with the help of
provenance indicators. The results (normalized to \([0,1]\)) for unassisted and
assisted identification of injected ads are shown in
Figure~\ref{adinjection:fig:user_study:identification_unassisted} and
Figure~\ref{adinjection:fig:user_study:identification_assisted}, respectively.
Unassisted identification is the aggregated number of reported ad injections
without any assistance in the presence of three ad-injecting extensions across
three retail websites, and assisted identification is the number of reported
injected ads with the help of content provenance indicators.

These results clearly imply that users are more likely to recognize the presence
of third-party content modifications using provenance indicators. To confirm
statistical significance, we performed a hypothesis test where the null
hypothesis is that provenance indicators do not assist in identifying
third-party content modifications, while the alternative hypothesis is that
provenance indicators do assist in identifying such content. Using a paired
t-test, we obtain a p-value of \(4.9199 \times 10^{-7}\), sufficient to reject
the null hypothesis at a 1\% significance level. The outliers in assisted
identification are due to the fact that our ad highlighting technique was not
identifiable by a small number of participants. We believe that using different
visual highlighting techniques would make it easier for users to identify the
injected ads.

Finally, we asked each participant how likely they would be to use the content
provenance system in their daily web browsing. We asked participants to rate
this likelihood on a scale from one to five, where one means they would never
use the system and five means that they would always use it. The results are
shown in Figure~\ref{adinjection:fig:user_study:usability}, and indicate that
most users would be willing to use a content provenance system. The reason
behind the outliers is because a few of the participants stated that they do not
need our system since they would not click on any advertisements. However, we
note that it can be difficult to distinguish between advertisements and other
legitimate content (e.g., products in retail sites) and, consequently, users
might be lured into clicking on ad content injected by extensions.

\subsubsection{Summary}

From this user study, we draw several conclusions. First, we confirm that in
many cases users are unable to distinguish injected third-party content from
publisher content. We also show that because users place trust in publishers,
they will often click on injected ads, and thus they tend to be susceptible to
ad injection. Our data shows that content provenance assists in helping users
distinguish between trusted publisher content and injected third-party content
that should not be trusted. Finally, we show that many users would be willing to
use the system based on their experience in this study.

\subsection{Usability}
\label{adinjection:sec:analysis:usability}

We conducted another experiment on a separate population of users to measure the
usability of the \origintracer prototype. The user population was composed of 13
students with different technical background. We presented the participants with
\origintracer integrated into Chromium~43, and asked them to browse the web for
several hours, visiting any websites of their choice. For privacy reasons,
however, we asked users to avoid browsing websites that require a login or that
involve sensitive subject matter (e.g., adult or financial websites). In
addition, for each user, we randomly selected 50 websites from the Alexa Top 500
that satisfy our user privacy constraints and asked the user to visit them. In
particular, each participant was asked to browse at least three levels down from
the home page and visit external links contained in each site. Finally, to gain
some assurance that \origintracer would not break benign extensions, we
configured the browser with the five high-profile extensions list in
Table~\ref{adinjection:tab:popular_extensions}.

During the browsing session, the browser was modified to record the number of
URLs visited. We also asked participants to record the number of pages in which
they encountered one of two types of errors. Type~I errors are those where the
browser crashed, system error messages were displayed, pages would not load, or
the website was completely unusable for some other reason. Type~II errors
include non-catastrophic errors that impact usability but did not preclude it --
e.g., the page took an abnormally long time to load, or the appearance of the
page was not as expected. We also asked users to report any broken functionality
for the benign extensions described above as well.

Out of close to 2,000 URLs, two catastrophic errors and 27 non-catastrophic
errors were encountered. However, we note that the majority of URLs rendered and
executed correctly. In addition, none of the participants reported any broken
extensions. We therefore conclude that the proposed approach is compatible with
modern browsers and benign extensions, and further work would very likely allow
the prototype to execute completely free of errors.

\subsection{Performance}
\label{adinjection:sec:analysis:perf}

To measure the performance overhead of \origintracer, we configured both an
unmodified Chromium browser and the prototype to automatically visit the Alexa
Top 1K. The Alexa Top 1K covers many popular websites and is weighted towards
being representative of the sites that people use most often. By using this test
set, we ensured that each browser visited a broad spectrum of websites that
include both static and dynamic content, and especially websites that make heavy
use of third-party components and advertisements. Moreover, we configured both
browser instances with the five benign extensions discussed in
Section~\ref{adinjection:sec:background} that change the DOM to
measure performance in the presence of extensions. A more detailed evaluation
would analyze more pages on these websites to garner a more realistic
representation, but that is beyond the scope of the current work.

We built a crawler based on Selenium~\cite{selenium} to automatically visit the
entire list of websites and recorded the total elapsed time from the beginning
of the browsing process until the entire list of websites was visited.
Specifically, our crawler moves to the next website in the list when the current
website is fully loaded, signified by the firing of the \texttt{onload} event.
In order to account for fluctuations in browsing time due to network delays and
the dynamic nature of advertisements, we repeated the experiment 10 times and
measured the average elapsed time. The average elapsed time for browsing the
home pages of the Alexa Top 1K websites measured in this way is 3,457 seconds
for the unmodified browser and 3,821 seconds for \origintracer. Therefore,
\origintracer incurred a 10.5\% overhead on browsing time on average. We also
measured the delay imposed by \origintracer on startup time by launching the
browser 10 times and measuring the average launch time. \origintracer did not
cause any measurable overhead on startup time.

While this overhead is not insignificant, we note that our user study in
Section~\ref{adinjection:sec:analysis:usability} indicates that many users would
be willing to trade off actual perceived performance overhead against the
security benefits provided by the system. Moreover, this prototype is just a
proof-of-concept implementation of our system and there is still room for
optimizing the implementation to decrease the page load time.

\section{Chapter Summary}
\label{adinjection:sec:conclusion}

In this chapter, we evaluated a prototype implementation of web content
provenance tracking, a modified version of Chromium we call \origintracer,
through a user study that demonstrated a statistically significant improvement
in the ability of users to identify unwanted third-party content. Our
performance evaluation shows a modest overhead on a large representative sample
of popular websites, while our user experiments indicate that users are willing
to trade off a slight decrease in performance for more insight into the sources
of web content that they browse. We also performed a comprehensive study on the
content modifications performed by ad-injecting extensions in the wild.

\chapter{Analysis of Style Injection by Relative Path Overwrite}
\label{chap:rpo}

\section{Introduction}

Cross-Site Scripting (XSS)~\cite{owasp_xss} attacks are one of the most common
threats on the Web. While XSS has traditionally been understood as the
attacker's capability to inject script into a site and have it executed by the
victim's web browser, more recent work has shown that script injection is not a
necessary precondition for effective attacks. By injecting Cascading Style Sheet
(CSS) directives, for instance, attackers can carry out so-called
\textit{scriptless} attacks~\cite{ccs2012scriptless} and exfiltrate secrets from
a site.

The aforementioned injection attacks typically arise due to the lack of
separation between code and data~\cite{ccs2013dedacota}, and more specifically,
insufficient sanitization of untrusted inputs in web applications. While script
injection attacks are more powerful than those based on style injection, they
are also more well-known as a threat, and web developers are comparatively more
likely to take steps to make them more difficult. From an attacker's point of
view, style injection attacks may be an option in scenarios where script
injection is not possible.

There are many existing techniques of how style directives could be injected
into a site~\cite{ccs2012scriptless,ccs2010cross_origin_css}. A relatively
recent class of attacks is Relative Path Overwrite (RPO), first proposed in a
blog post by Gareth Heyes~\cite{rpo} in 2014. These attacks exploit the semantic
disconnect between web browsers and web servers in interpreting relative paths
(\textit{path confusion}). More concretely, in certain settings an attacker can
manipulate a page's URL in such a way that the web server still returns the same
content as for the benign URL. However, using the manipulated URL as the base,
the web browser incorrectly expands relative paths of included resources, which
can lead to resources being loaded despite not being intended to be included by
the developer. Depending on the implementation of the site, different variations
of RPO attacks may be feasible. For example, an attacker could manipulate the
URL to make the page include user-generated content hosted on the same
domain~\cite{rpo_techniques}. When an injection vulnerability is present in a
page, an attacker could manipulate the URL such that the web page references
itself as the stylesheet, which turns a simple text injection vulnerability into
a style sink~\cite{rpo}. Among these attack instantiations, the latter variant
has preconditions that are comparatively frequently met by sites. Our work
focuses on this variant of RPO.

In this chapter, we present the first in-depth study of style injection
vulnerability using RPO. We extract pages using relative-path stylesheets from
the Common Crawl dataset~\cite{common_crawl}, automatically test if style
directives can be injected using RPO, and determine whether they are interpreted
by the browser. Out of 31 million pages from 222 thousand Alexa Top 1\,M
sites~\cite{alexa_top_1m} in the Common Crawl that use relative-path
stylesheets, we find that 377\,k pages (12\,k sites) are vulnerable; 11\,k pages
on 1\,k sites can be exploited in Chrome, and nearly 55\,k pages on over 3\,k
sites can be exploited in Internet Explorer.

The rest of this chapter is organized as follows.
Section~\ref{rpo:sec:background} outlines the necessary background on cross-site
scripting, scriptless attacks, and relative path overwrite.
Section~\ref{rpo:sec:methodology} presents the design and implementation of our
measurement methodology, while Section~\ref{rpo:sec:analysis} presents our
findings. Finally, we summarize the chapter in Section~\ref{rpo:sec:summary}.

\section{Background}
\label{rpo:sec:background}

The general threat model of Relative Path Overwrite (RPO) resembles that of
Cross-Site Scripting (XSS). Typically, the attacker's goal is to steal sensitive
information from a third-party site or make unauthorized transactions on the
site, such as gaining access to confidential financial information or
transferring money out of a victim's account.

The attacker carries out the attack against the site indirectly, by way of a
victim that is an authorized user of the site. The attacker can trick the victim
into following a crafted link, such as when the victim visits a domain under the
attacker's control and the page automatically opens the manipulated link, or
through search engine poisoning, deceptive shortened links, or through means of
social engineering.

\subsection{Cross-Site Scripting}

Many sites have vulnerabilities that let attackers inject malicious script.
Dynamic sites frequently accept external inputs that can be controlled by an
attacker, such as data in URLs, cookies, or forms. While the site developer's
aim would have been to render the input as text, lack of proper sanitization can
result in the input being executed as script~\cite{owasp_xss_defense}. The
inclusion of unsanitized inputs could occur on the server side or client side,
and in a persistent \textit{stored} or volatile \textit{reflected}
way~\cite{owasp_xss}. To the victim's web browser, the code appears as
originating from the first-party site, thus it is given full access to the
session data in the victim's browser. Thereby, the attacker bypasses protections
of the Same-Origin Policy.

\subsection{Scriptless Attacks}

Cross-Site Scripting is perhaps the most well-known web-based attack, against
which many sites defend by filtering user input. Client-side security mechanisms
such as browser-based XSS filters~\cite{www2010xss_auditor} and Content Security
Policy~\cite{www2010csp,w3c_csp} also make it more challenging for attackers to
exploit injection vulnerabilities for XSS. This has led attackers (and
researchers) to investigate potential alternatives, such as \textit{scriptless}
attacks. These attacks allow sniffing users' browsing
histories~\cite{dsn2014scriptless_timing,w2sp2010history_sniffing}, exfiltrating
arbitrary content~\cite{css_font_face}, reading HTML
attributes~\cite{sexy_assassin_css,csp_nonce_bypass}, and bypassing Clickjacking
defenses~\cite{sexy_assassin_css}. In the following, we highlight two types of
scriptless attacks proposed in the literature. Both assume that an attacker
cannot inject or execute script into a site. Instead, the attacker abuses
features related to Cascading Style Sheets (CSS).

Heiderich et al.~\cite{ccs2012scriptless} consider scenarios where an attacker
can inject CSS into the context of the third-party page so that the style
directives are interpreted by the victim's browser when displaying the page.
That is, the injection sink is either located inside a style context, or the
attacker can inject markup to create a style context around the malicious CSS
directives. While the CSS standard is intended for styling and layout purposes
such as defining sizes, colors, or background images and as such does not
contain any traditional scripting capabilities, it does provide some
context-sensitive features that, in combination, can be abused to extract and
exfiltrate data. If the secret to be extracted is not displayed, such as a token
in a hidden form field or link URL, the attacker can use the CSS attribute
accessor and content property to extract the secret and make it visible as text,
so that style directives can be applied to it. Custom attacker-supplied fonts
can change the size of the secret text depending on its value. Animation
features can be used to cycle through a number of fonts in order to test
different combinations. Media queries or the appearance of scrollbars can be
used to implement conditional style, and data exfiltration by loading a
different URL for each condition from the attacker's server. Taken together,
Heiderich et al. demonstrate that these techniques allow an attacker to steal
credit card numbers or CSRF tokens~\cite{owasp_csrf_defense} without script
execution.

Rather than using layout-based information leaks to exfiltrate data from a page,
Huang et al.~\cite{ccs2010cross_origin_css} show how syntactically lax parsing
of CSS can be abused to make browsers interpret an HTML page as a
``stylesheet.'' The attack assumes that the page contains two injection sinks,
one before and one after the location of the secret in the source code. The
attacker injects two CSS fragments such as
\texttt{\{\}*\{background:url('//attacker.com/?} and \texttt{');\}}, which make
the secret a part of the URL that will be loaded from the attacker's server when
the directive is interpreted. It is assumed that the attacker cannot inject
markup, thus the injected directive is not interpreted as style when the site is
conventionally opened in a browser. However, the CSS standard mandates that
browsers be very forgiving when parsing CSS, skipping over parts they do not
understand~\cite{css_syntax}. In practice, this means that an attacker can set
up a site that loads the vulnerable third-party site \textit{as a stylesheet}.
When the victim visits the attacker's site while logged in, the victim's browser
loads the third-party site and interprets the style directive, causing the
secret to be sent to the attacker. To counter this attack, modern browsers do
not load documents with non-CSS content types and syntax errors as stylesheets
when they originate from a different domain than the including page. Yet,
attacks based on tolerant CSS parsing are still feasible when both the including
and the included page are loaded from the same domain. Relative Path Overwrite
attacks can abuse such a scenario~\cite{rpo_gadgets}.

\subsection{Relative Path Overwrite}
\label{rpo:sec:background:rpo}

Relative Path Overwrite vulnerabilities can occur in sites that use relative
paths to include resources such as scripts or stylesheets. Before a web browser
can issue a request for such a resource to the server, it must expand the
relative path into an absolute URL. For example, assume that a web browser has
loaded an HTML document from \texttt{http://example.com/rpo/test.php} which
references a remote stylesheet with the relative path \texttt{dist/styles.css}.
Web browsers treat URLs as file system-like paths, that is, \texttt{test.php}
would be assumed to be a file within the parent directory \texttt{rpo/}, which
would be used as the starting point for relative paths, resulting in the
absolute URL \texttt{http://example.com/rpo/dist/styles.css}.

However, the browser's interpretation of the URL may be very different from how
the web server resolves the URL to determine which resource should be returned
to the browser. The URL may not correspond to an actual server-side file system
structure at all, or the web server may internally rewrite parts of the URL. For
instance, when a web server receives a request for
\texttt{http://example.com/rpo/test.php/} with an added trailing slash, it may
still return the same HTML document corresponding to the \texttt{test.php}
resource. Yet, to the browser this URL would appear to designate a directory
(without a file name component), thus the browser would request the stylesheet
from \texttt{http://example.com/rpo/test.php/dist/styles.css}. Depending on the
server configuration, this may either result in an error since no such file
exists, or the server may interpret \texttt{dist/styles.css} as a parameter to
the script \texttt{test.php} and return the HTML document. In the latter case,
the HTML document includes itself as a stylesheet. Provided that the document
contains a (text) injection vulnerability, attackers can carry out the
scriptless attacks; since the stylesheet inclusion is same-origin, the document
load is permitted.

\subsection{Preconditions for RPO Style Attacks}

For the purpose of this work, we focus on a generic type of RPO attack because
its preconditions are less specific and are likely met by a larger number of
sites. More formally, we define a page as \textit{vulnerable} if:

\begin{itemize}

\item The page includes at least one stylesheet using a relative path.

\item The server is set up to serve the same page even if the URL is manipulated
by appending characters that browsers interpret as path separators.

\item The page reflects style directives injected into the URL or cookie. Note
that the reflection can occur in an arbitrary location within the page, and
markup or script injection are not necessary.

\item The page does not contain a \texttt{<base>} HTML tag before relative paths
that would let the browser know how to correctly expand them.

\end{itemize}

This attack corresponds to style injection by means of a page that references
itself as a stylesheet (PRSSI). Since the ``stylesheet'' self-reference is, in
fact, an HTML document, web servers would typically return it with a
\texttt{text/html} content type. Browsers in standards-compliant mode do not
attempt to parse documents with a content type other than CSS even if referenced
as a stylesheet, causing the attack to fail. However, web browsers also support
\textit{quirks mode} for backwards compatibility with non-standards compliant
sites~\cite{browser_modes_doctype}; in this mode, browsers ignore the content
type and parse the document according to the inclusion context only.

We define a vulnerable page as \textit{exploitable} if the injected style is
interpreted by the browser--that is, if an attacker can force browsers to render
the page in quirks mode. This can occur in two alternative ways:

\begin{itemize}

\item The vulnerable HTML page specifies a \textit{document type} that causes
the browser to use quirks mode instead of standards mode. The document type
indicates the HTML version and dialect used by the page;
Section~\ref{rpo:sec:analysis:doctypes} provides details on how the major web
browsers interpret the document types we encountered during our study.

\item Even if the page specifies a document type that would usually result in
standards mode being used, quirks mode parsing can often be enforced in Internet
Explorer~\cite{prssi}. Framed documents inherit the parsing mode from the parent
document, thus an attacker can create an attack page with an older document type
and load the vulnerable page into a frame. This trick only works in Internet
Explorer, however, and it may fail if the vulnerable page uses any anti-framing
technique, or if it specifies an explicit value for the \texttt{X-UA-Compatible}
HTTP header (or equivalent).

\end{itemize}

Our measurement methodology in Section~\ref{rpo:sec:methodology} tests how often
these preconditions hold in the wild in order to quantify the vulnerability and
exploitability of pages with respect to RPO attacks.

\section{Methodology}
\label{rpo:sec:methodology}

Our methodology consists of three main phases. We seed our system with pages
from the Common Crawl archive to extract \textit{candidate} pages that include
at least one stylesheet using a relative path. To determine whether these
candidate pages are \textit{vulnerable}, we attempt to inject style directives
by requesting variations of each page's URL to cause \textit{path confusion} and
test whether the generated response reflects the injected style directives.
Finally, we test how often vulnerable pages can be \textit{exploited} by
checking whether the reflected style directives are parsed and used for
rendering in a web browser.

\paragraph{Ethics.}

One ethical concern is that the injected CSS might be stored on the server
instead of being reflected in the response, and it could break sites as a
result. We took several cautionary steps in order to minimize any damaging side
effects on sites we probed. First, we did not try to login to the site, and we
only tested RPO on the publicly available version of the page. In addition, we
only requested pages by tainting different parts of the URL, and did not submit
any forms. Moreover, we did not click on any button or link in the page in order
to avoid triggering JavaScript events. These steps significantly decrease the
chances that injected CSS will be stored on the server. In order to minimize the
damaging side effects in case our injected CSS was stored, the injected CSS is
not a valid style directive, and even if it is stored on the server, it will not
have any observable effect on the page. In addition, experiment resulted in the
discovery of vulnerable content management systems (CMSes) used world-wide, and
we contacted them so they can fix the issue. We believe the real-world
experiments that we conducted were necessary in order to measure the risk posed
by these vulnerabilities and inform site owners of potential risks to their
users.

\subsection{Candidate Identification}
\label{rpo:sec:methodology:candidate}

For finding the initial seed set of candidate pages with relative-path
stylesheets, we leverage the Common Crawl from August 2016, which contains more
than 1.6 billion pages. By using an existing dataset, we can quickly identify
candidate pages without creating any web crawl traffic. We use a Java HTML
parser to filter any pages containing only inline CSS or stylesheets referenced
by absolute URLs, leaving us with over 203 million pages on nearly 6 million
sites. For scalability purposes, we further reduce the set of candidate pages in
two steps:

\begin{enumerate}

\item We retain only pages from sites listed in the Alexa Top 1 million ranking,
which reduces the number of candidate pages to 141 million pages on 223 thousand
sites. In doing so, we bias our result toward popular sites--that is, sites
where attacks could have a larger impact because of the higher number of
visitors.

\item We observed that many sites use templates customized through query strings
or path parameters. We expect these templates to cause similar vulnerability and
exploitability behavior for their instantiations, thus we can speed up our
detection by grouping URLs using the same template, and testing only one random
representative of each group.

In order to group pages, we replace all the values of query parameters with
constants, and we also replace any number identifier in the path with a
constant. We group pages that have the same abstract URL as well as the same
document type in the Common Crawl dataset. Table~\ref{rpo:tab:grouped_webpages}
illustrates this process.

\end{enumerate}

\begin{table}[t]
  \footnotesize
    \caption{Sample URL grouping.}
    \centering
    \begin{tabular}{ll}
    \toprule
    \textbf{Group By} & \textbf{URL} \\
    \midrule
    \multirow{2}{*}{Query Parameter} & http://example.com/?lang=\textbf{en} \\
    & http://example.com/?lang=\textbf{fr} \\
    \midrule
    \multirow{2}{*}{Path Parameter} & http://example.com/\textbf{028} \\
    & http://example.com/\textbf{142} \\
    \bottomrule
    \end{tabular}
    \label{rpo:tab:grouped_webpages}
\end{table}

Since our methodology contains a step during which we actively test whether a
vulnerability can be exploited, we remove from the candidate set all pages
hosted on sites in \texttt{.gov}, \texttt{.mil}, \texttt{.army}, \texttt{.navy},
and \texttt{.airforce}. The final candidate set consists of 137 million pages
(31 million page groups) on 222 thousand sites.

\subsection{Vulnerability Detection}
\label{rpo:sec:methodology:vulnerable}

To determine whether a candidate page is vulnerable, we implemented a
lightweight crawler based on the Python Requests API. At a high level, the
crawler simulates how a browser expands relative paths and tests whether style
directives can be injected into the resources loaded as stylesheets using path
confusion.

For each page group from the candidate set, the crawler randomly selects one
representative URL and mutates it according to a number of techniques explained
below. Each of these techniques aims to cause path confusion and taints page
inputs with a style directive containing a long unique, random string. The
crawler requests the mutated URL from the server and parses the response
document, ignoring resources loaded in frames. If the response contains a
\texttt{<base>} tag, the crawler considers the page not vulnerable since the
\texttt{<base>} tag, if used correctly, can avoid path confusion. Otherwise, the
crawler extracts all relative stylesheet paths from the response and expands
them using the mutated URL of the main page as the base, emulating how browsers
treat relative paths (see Section~\ref{rpo:sec:background:rpo}). The crawler
then requests each unique stylesheet URL until one has been found to reflect the
injected style in the response.

The style directive we inject to test for reflection vulnerabilities is shown in
the legend of Figure~\ref{rpo:fig:taint_techniques}. The payload begins with an
encoded newline character, as we observed that the presence of a newline
character increases the probability of a successful injection. We initially use
\texttt{\%0A} as the newline character, but also test \texttt{\%0C} and
\texttt{\%0D} in case of unsuccessful injection. The remainder of the payload
emulates the syntax of a simple CSS directive and mainly consists of a randomly
generated string used to locate the payload in the body of the server response.
If the crawler finds a string match of the injected unique string, it considers
the page vulnerable.

In the following, we describe the various URL mutation techniques we use to
inject style directives. All techniques also use RPO so that instead of the
original stylesheet files, browsers load different resources that are more
likely to contain an injection vulnerability. Conceptually, the RPO approaches
we use assume some form of server-side URL rewriting as described in
Section~\ref{rpo:sec:background:rpo}. That is, the server internally resolves a
crafted URL to the same script as the ``clean'' URL. Under that assumption, the
path confusion caused by RPO would result in the page referencing itself as the
stylesheet when loaded in a web browser. However, this assumption is only
conceptual and not necessary for the attack to succeed. For servers that do not
internally rewrite URLs, our mutated URLs likely cause error responses since the
URLs do not correspond to actual files located on these servers. Error responses
are typically HTML documents and may contain injection sinks, such as when they
display the URL of the file that could not be found. As such, server-generated
error responses can be used for the attack in the same way as regular pages.

Our URL mutation techniques differ in how they attempt to cause path confusion
and inject style directives by covering different URL conventions used by a
range of web application platforms.

\begin{figure}
    \centering

    \begin{subfigure}[t]{1\textwidth}
        \begin{lstlisting}
        http://domain/dir/page.asp
        http://domain/dir/page.asp@@/@@**PAYLOAD**@@//@@
        http://domain/dir/page.asp@@/@@**PAYLOAD**@@/@@style.css
        \end{lstlisting}
        \caption{Path Parameter (Simple)}
        \label{rpo:fig:taint:parameter_simple}
    \end{subfigure}

    \begin{subfigure}[t]{1\textwidth}
        \begin{lstlisting}
        http://domain/page.php/param
        http://domain/page.php/**PAYLOAD**param@@//@@
        http://domain/page.php/**PAYLOAD**param@@/@@style.css
        \end{lstlisting}
        \caption{Path Parameter (PHP or ASP)}
        \label{rpo:fig:taint:parameter_php}
    \end{subfigure}

    \begin{subfigure}[t]{1\textwidth}
        \begin{lstlisting}
        http://domain/dir/page.jsp;param
        http://domain/dir/page.jsp;**PAYLOAD**param@@//@@
        http://domain/dir/page.jsp;**PAYLOAD**param@@/@@style.css
        \end{lstlisting}
        \caption{Path Parameter (JSP)}
        \label{rpo:fig:taint:parameter_jsp}
    \end{subfigure}

    \begin{subfigure}[t]{1\textwidth}
        \begin{lstlisting}
        http://domain/dir/page.aspx
        http://domain/**PAYLOAD**@@/..%2F@@dir/**PAYLOAD**@@/..%2F@@page.aspx@@//@@
        http://domain/**PAYLOAD**@@/..%2F@@dir/**PAYLOAD**@@/..%2F@@page.aspx@@/@@style.css
        \end{lstlisting}
        \caption{Encoded Path}
        \label{rpo:fig:taint:path}
    \end{subfigure}

    \begin{subfigure}[t]{1\textwidth}
        \begin{lstlisting}
        http://domain/dir/page.html?key=value
        http://domain/dir/page.html@@%3F@@key=**PAYLOAD**value@@//@@
        http://domain/dir/page.html@@%3F@@key=**PAYLOAD**value@@/@@style.css
        \end{lstlisting}
        \caption{Encoded Query}
        \label{rpo:fig:taint:query}
    \end{subfigure}

    \begin{subfigure}[t]{1\textwidth}
        \begin{lstlisting}
        http://domain/dir/page.php?key=value
        http://domain/dir/page.php@@//@@?key=value
        http://domain/dir/page.php@@/@@style.css

        Original Cookie: name=val
         Crafted Cookie: name=**PAYLOAD**val
        \end{lstlisting}
        \caption{Cookie}
        \label{rpo:fig:taint:cookie}
    \end{subfigure}

    \caption{Various techniques of \redbold{path confusion} and \bluebold{style
    injection}. In each example, the first URL corresponds to the regular page, and
    the second one to the page URL crafted by the attacker. Each HTML page is
    assumed to reference a stylesheet at \protect\texttt{../style.css}, resulting in the
    browser expanding the stylesheet path as shown in the third URL.
    \bluebold{PAYLOAD} corresponds to
    \bluebold{\texttt{\%0A\{\}body\{background:NONCE\}}} (simplified), where
    \texttt{NONCE} is a randomly generated string.}

    \label{rpo:fig:taint_techniques}
\end{figure}

\subsubsection{Path Parameter}

A number of web frameworks such as PHP, ASP, or JSP can be configured to use URL
schemes that encode script input parameters as a directory-like string following
the name of the script in the URL. Figure~\ref{rpo:fig:taint:parameter_simple}
shows a generic example where there is no parameter in the URL. Since the
crawler does not know the name of valid parameters, it simply appends the
payload as a subdirectory to the end of the URL. In this case, content injection
can occur if the page reflects the page URL or referrer into the response. Note
that in the example, we appended two slashes so that the browser does not remove
the payload from the URL when expanding the stylesheet reference to the parent
directory (\texttt{../style.css}). In the actual crawl, we always appended
twenty slashes to avoid having to account for different numbers of parent
directories. We did not observe relative paths using large numbers of
\texttt{../} to reference stylesheets, thus we are confident that twenty slashes
suffice for our purposes.

Different web frameworks handle path parameters slightly differently, which is
why we distinguish a few additional cases. If parameters are present in the URL,
we can distinguish these cases based on a number of regular expressions that we
generated. For example, parameters can be separated by slashes
(Figure~\ref{rpo:fig:taint:parameter_php}, PHP or ASP) or semicolons (
Figure~\ref{rpo:fig:taint:parameter_jsp}, JSP). When the crawler detects one of
these known schemes, it injects the payload into each parameter. Consequently,
in addition to URL and referrer reflection, injection can also be successful
when any of the parameters is reflected in the page.

\subsubsection{Encoded Path}

This technique targets web servers such as IIS that decode encoded slashes in
the URL for directory traversal, whereas web browsers do not. Specifically, we
use \texttt{\%2F}, an encoded version of `\texttt{/}', to inject our payload
into the URL in such a way that the canonicalized URL is equal to the original
page URL (see Figure~\ref{rpo:fig:taint:path}). Injection using this technique
succeeds if the page reflects the page URL or referrer into its output.

\subsubsection{Encoded Query}

Similar to the technique above, we replace the URL query delimiter `\texttt{?}'
with its encoded version \texttt{\%3F} so that web browsers do not interpret it
as such. In addition, we inject the payload into every value of the query
string, as can be seen in Figure~\ref{rpo:fig:taint:query}. CSS injection
happens if the page reflects either the URL, referrer, or any of the query
values in the HTML response.

\subsubsection{Cookie}

Since stylesheets referenced by a relative path are located in the same origin
as the referencing page, its cookies are sent when requesting the stylesheet.
CSS injection may be possible if an attacker can create new cookies or tamper
with existing ones (a strong assumption compared to the other techniques), and
if the page reflects cookie values in the response. As shown in
Figure~\ref{rpo:fig:taint:cookie}, the URL is only modified by adding slashes to
cause path confusion. The payload is injected into each cookie value and sent by
the crawler as an HTTP header.

\subsection{Exploitability Detection}
\label{rpo:sec:methodology:exploitable}

Once a page has been found to be vulnerable to style injection using RPO, the
final step is to verify whether the reflected CSS in the response is evaluated
by a real browser. To do so, we built a crawler based on Google Chrome, and used
the Remote Debugging Protocol~\cite{debugging_protocol} to drive the browser and
record HTTP requests and responses. In addition, we developed a Chrome extension
to populate the cookie header in CSS stylesheet requests with our payload.

In order to detect exploitable pages, we crawled all the pages from the previous
section that had at least one reflection. Specifically, for each page we checked
which of the techniques in Figure~\ref{rpo:fig:taint_techniques} led to
reflection, and crafted the main URL with a CSS payload. The CSS payload used to
verify exploitability is different from the simple payload used to test
reflection. Specifically, the style directive is prefixed with a long sequence
of \texttt{\}} and \texttt{]} characters to close any preceding open curly
braces or brackets that may be located in the source code of the page, since
they might prevent the injected style directive from being parsed correctly. The
style directive uses a randomly-generated URL to load a background image for the
HTML body. We determine whether the injected style is evaluated by checking the
browser's network traffic for an outgoing HTTP request for the image.

\subsubsection{Overriding Document Types}

Reflected CSS is not always interpreted by the browser. One possible explanation
is the use of a modern document type in the page, which does not cause the
browser to render the page in quirks mode. Under certain circumstances, Internet
Explorer allows a parent page to force the parsing mode of a framed page into
quirks mode~\cite{prssi}. To test how often this approach succeeds in practice,
we also crawled vulnerable pages with Internet Explorer~11 by framing them while
setting \texttt{X-UA-Compatible} to \texttt{IE=EmulateIE7} via a \texttt{meta}
tag in the attacker's page.

\subsection{Limitations}

RPO is a class of attacks and our methodology covers only a subset of them. We
target RPO for the purpose of style injection using an HTML page referencing
itself (or, accidentally, an error page) as the stylesheet. In terms of style
injection, our crawler only looks for reflection, not stored injection of style
directives. Furthermore, manual analysis of a site might reveal more
opportunities for style injection that our crawler fails to detect
automatically.

For efficiency reasons, we seed our analysis with an existing Common Crawl
dataset. We do not analyze the vulnerability of pages not contained in the
Common Crawl seed, which means that we do not cover all sites, and we do not
fully cover all pages within a site. Consequently, the results presented in this
paper should be seen as a lower bound. If desired, our methodology can be
applied to individual sites in order to analyze more pages.

\section{Analysis}
\label{rpo:sec:analysis}

\begin{table}[t]
    \centering
    \footnotesize

    \caption{Narrowing down the Common Crawl to the candidate set used in our
    analysis (from left to right)}

    \begin{tabular}{lrrr}
    \toprule
    &
    \multicolumn{1}{r}{\textbf{Relative CSS}} &
    \multicolumn{1}{c}{\textbf{Alexa Top 1M}} &
    \multicolumn{1}{c}{\textbf{Candidate Set}} \\
    \midrule
    All Pages & 203,609,675 & 141,384,967 & 136,793,450 \\
    Tested Pages & 53,725,270 & 31,448,446 & 30,991,702 \\
    Sites & 5,960,505 & 223,212 & 222,443 \\
    Doc. Types & 9,833 & 2,965 & 2,898 \\
    \bottomrule
    \end{tabular}
    \label{rpo:tab:dataset_stats}
\end{table}

For the purposes of our analysis, we gradually narrow down the seed data from
the Common Crawl to pages using relative style paths in the Alexa Top 1\,M,
reflecting injected style directives under RPO, and being exploitable due to
quirks mode rendering.

Table~\ref{rpo:tab:dataset_stats} shows a summary of our dataset. \textit{Tested
Pages} refers to the set of randomly selected pages from the page groups as
discussed in Section~\ref{rpo:sec:methodology:candidate}. For brevity, we are
referring to \textit{Tested Pages} wherever we mention pages in the remainder of
the paper.

\subsection{Relative Stylesheet Paths}
\label{rpo:sec:analysis:relative}

\begin{figure}[t]
    \centering
    \includegraphics[width=0.8\textwidth]{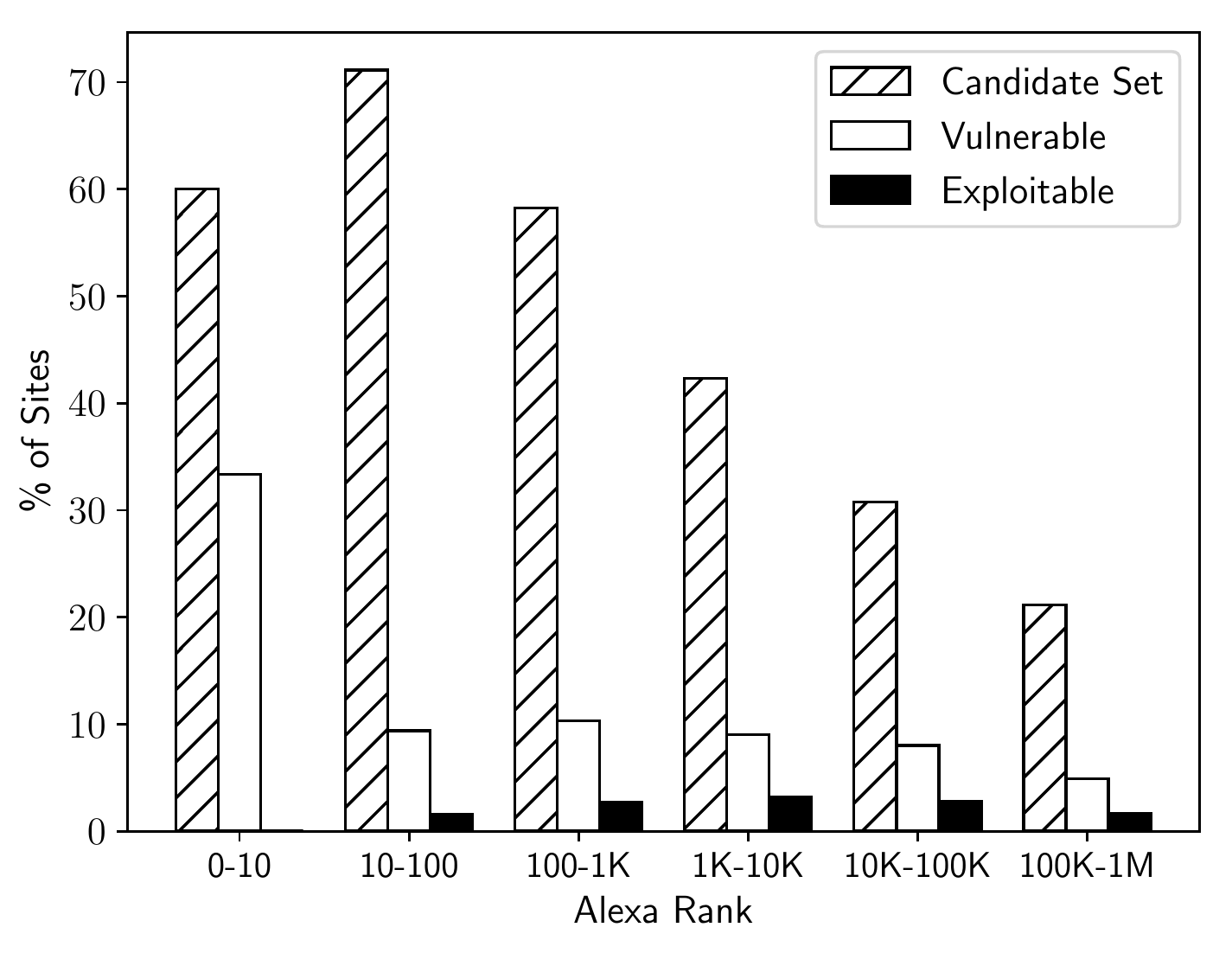}
    \caption{Percentage of the Alexa site ranking in our candidate set
             (exponentially increasing bucket size).}
    \label{rpo:fig:analysis:alexa_rank}
\end{figure}

\begin{figure}[t]
    \centering
    \includegraphics[width=0.8\textwidth]{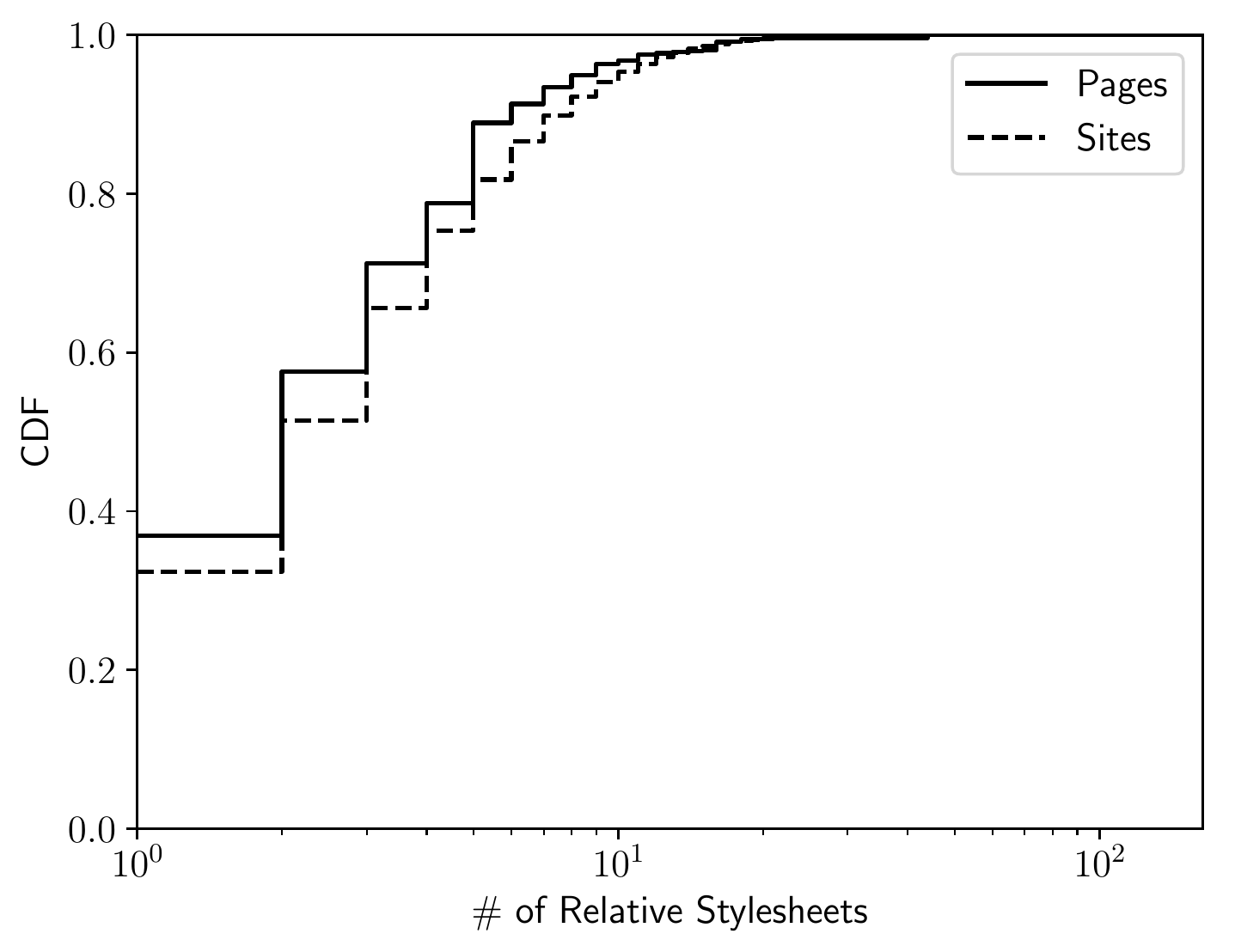}
    \caption{CDF of total and maximum number of relative stylesheets per web
             page and site, respectively.}
    \label{rpo:fig:analysis:relative_stylesheets}
\end{figure}

To assess the extent to which our Common Crawl-seeded candidate set covers sites
of different popularity, consider the hatched bars in
Figure~\ref{rpo:fig:analysis:alexa_rank}. Six out of the ten largest sites
according to Alexa are represented in our candidate set. That is, they are
contained in the Common Crawl, and have relative style paths. The figure shows
that our candidate set contains a higher fraction of the largest sites and a
lower fraction of the smaller sites. Consequently, our results better represent
the most popular sites, which receive most visitors, and most potential victims
of RPO attacks.

While all the pages in the candidate set contain at least one relative
stylesheet path, Figure~\ref{rpo:fig:analysis:relative_stylesheets} shows that
63.1\,\% of them contain multiple relative paths, which increases the chances of
finding a successful RPO and style injection point.

\subsection{Vulnerable Pages}
\label{rpo:sec:analysis:vulnerable}

\begin{table}[t]
    \footnotesize
    \centering
    \caption{Vulnerable pages and sites in the candidate set}
    \begin{tabular}{lrr}
    \toprule
    \textbf{Technique} &
    \multicolumn{1}{c}{\textbf{Pages}} &
    \multicolumn{1}{c}{\textbf{Sites}} \\

    \midrule

    Path Parameter & 309,079 (1.0\%) & 9,136 (4.1\%) \\
    Encoded Path & 53,502 (0.2\%) & 1,802 (0.8\%) \\
    Encoded Query & 89,757 (0.3\%) & 1,303 (0.6\%) \\
    Cookie & 15,656 (\textless 0.1\%) & 1,030 (0.5\%) \\

    \midrule

    Total & 377,043 (1.2\%) & 11,986 (5.4\%) \\

    \bottomrule
    \end{tabular}
    \label{rpo:tab:vulnerable_result}

\end{table}

We consider a candidate page vulnerable if one of the style injection techniques
of Section~\ref{rpo:sec:methodology:vulnerable} succeeds. In other words, the
server's response should reflect the injected payload. Furthermore, we
conservatively require that the response not contain a \texttt{base} tag since a
correctly configured base tag can prevent path confusion.

Table~\ref{rpo:tab:vulnerable_result} shows that 1.2\,\% of pages are vulnerable
to at least one of the injection techniques, and 5.4\,\% of sites contain at
least one vulnerable page. The path parameter technique is most effective
against pages, followed by the encoded query and the encoded path techniques.
Sites that are ranked higher according to Alexa are more likely to be
vulnerable, as shown in Figure~\ref{rpo:fig:analysis:alexa_rank}, where
vulnerable and exploitable sites are relative to the candidate set in each
bucket. While one third of the candidate set in the Top~10 (two out of six
sites) is vulnerable, the percentage oscillates between 8 and 10\,\% among the
Top~100\,k. The candidate set is dominated by the smaller sites in the ranks
between 100\,k and 1\,M, which have a vulnerability rate of 4.9\,\% and push
down the average over the entire ranking.

A \texttt{base} tag in the server response can prevent path confusion because it
indicates how the browser should expand relative paths. We observed a number of
inconsistencies with respect to its use. At first, 603 pages on 60 sites
contained a \texttt{base} tag in their response; however, the server response
after injecting our payload did not contain the tag anymore, rendering these
pages potentially exploitable. Furthermore, Internet Explorer's implementation
of the \texttt{base} tag appears to be broken. When such a tag is present,
Internet Explorer fetches two URLs for stylesheets---one expanded according to
the base URL specified in the tag, and one expanded in the regular, potentially
``confused'' way of using the page URL as the base. In our experiments, Internet
Explorer always applied the ``confused'' stylesheet, even when the one based on
the \texttt{base} tag URL loaded faster. Consequently, \texttt{base} tags do not
appear to be an effective defense against RPO in Internet Explorer (They seem to
work as expected in other browsers, including Edge).

\subsection{Exploitable Pages}
\label{rpo:sec:analysis:exploitable}

\begin{table}[t]
    \footnotesize
    \centering
    \caption{Exploitable pages and sites in the candidate set (IE using framing)}
    \begin{tabular}{lrrrr}
    \toprule
    \multirow{2}{*}{\textbf{Technique}} &
    \multicolumn{2}{c}{\textbf{Chrome}} &
    \multicolumn{2}{c}{\textbf{Internet Explorer}} \\

    \cmidrule[0.5pt](lr){2-3}
    \cmidrule[0.5pt](lr){4-5}

    &
    \multicolumn{1}{c}{\textbf{Pages}} &
    \multicolumn{1}{c}{\textbf{Sites}} &
    \multicolumn{1}{c}{\textbf{Pages}} &
    \multicolumn{1}{c}{\textbf{Sites}}
    \\

    \midrule

    Path Parameter & 6,048 (\textless 0.1\%) & 1,025 (0.5\%) & 52,344 (0.2\%) & 3,433 (1.5\%) \\
    Encoded Path & 3 (\textless 0.1\%) & 2 (\textless 0.1\%) & 24 (\textless 0.1\%) & 5 (\textless 0.1\%) \\
    Encoded Query & 23 (\textless 0.1\%) & 20 (\textless 0.1\%) & 137 (\textless 0.1\%) & 43 (\textless 0.1\%) \\
    Cookie & 4,722 (\textless 0.1\%) & 81 (\textless 0.1\%) & 2,447 (\textless 0.1\%) & 238 (0.1\%) \\

    \midrule

    Total & 10,781 (<0.1\%) & 1,106 (0.5\%) & 54,853 (0.2\%) & 3,645 (1.6\%) \\

    \bottomrule
    \end{tabular}
    \label{rpo:tab:exploitable_result}
\end{table}

To test whether a vulnerable page was exploitable, we opened it in Chrome,
injected a style payload with an image reference (randomly generated URL), and
checked if the image was indeed loaded. This test succeeded for 2.9\,\% of
vulnerable pages; 0.5\,\% of sites in the candidate set had at least one
exploitable page (Table~\ref{rpo:tab:exploitable_result}).

In the following, we explore various factors that may impact whether a
vulnerable page can be exploited, and we show how some of these partial defenses
can be bypassed.

\subsubsection{Document Types}
\label{rpo:sec:analysis:doctypes}

\begin{table}[t]
    \centering
    \footnotesize
    \caption{Quirks mode document types by browser}
    \begin{tabular}{lccc}
    \toprule

    \textbf{Browser} &
    \textbf{Version} &
    \textbf{OS} &
    \multicolumn{1}{c}{\textbf{Doc. Types}} \\

    \midrule
    Chrome & 55 & Ubuntu 16.04 & 1,378 (31.9\,\%) \\
    Opera & 42 & Ubuntu 16.04 & 1,378 (31.9\,\%) \\
    Safari & 10 & macOS Sierra & 1,378 (31.9\,\%) \\
	\addlinespace
    Firefox & 50 & Ubuntu 16.04 & 1,326 (30.7\,\%) \\
	\addlinespace
    Edge & 38 & Windows 10 & 1,319 (30.5\,\%) \\
    IE & 11 & Windows 7 & 1,319 (30.5\,\%) \\
    \bottomrule
    \end{tabular}
    \label{rpo:tab:doctypes_browsers}
\end{table}

HTML document types play a significant role in RPO-based style injection attacks
because browsers typically parse resources with a non-CSS content type in a CSS
context only when the page specifies an ancient or non-standard HTML document
type (or none at all). The pages in our candidate set contain a total of 4,318
distinct document types. However, the majority of these unique document types
are not standardized and differ from the standardized ones only by small
variations, such as forgotten spaces or misspellings.

To determine how browsers interpret these document types (i.e., whether they
cause them to render a page in standards or quirks mode), we designed a
controlled experiment. For each unique document type, we set up a local page
with a relative stylesheet path and carried out an RPO attack to inject CSS
using a payload similar to what we described in
Section~\ref{rpo:sec:methodology:exploitable}. We automatically opened the local
page in Chrome, Firefox, Edge, Internet Explorer, Safari, and Opera, and we kept
track of which document type caused the injected CSS to be parsed and the
injected background image to be downloaded.

Table~\ref{rpo:tab:doctypes_browsers} contains the results of this experiment.
Even though the exact numbers vary among browsers, roughly a third of the unique
document types we encountered result in quirks mode rendering. Not surprisingly,
both Microsoft products Edge and Internet Explorer exhibit identical results,
whereas the common Webkit ancestry of Chrome, Opera, and Safari also show
identical results. Overall, 1,271 (29.4\,\%) of the unique document types force
all the browsers into quirks mode, whereas 1,378 (31.9\,\%) of them cause at
least one browser to use quirks mode rendering.
Table~\ref{rpo:tab:top_quirksmode_doctypes} shows the most frequently used
document types that force all the browsers into quirks mode, which includes the
absence of a document type declaration in the page.

\begin{table}[t]
    \centering
    \footnotesize
    \caption{Most frequent document types causing all browsers to render
    in quirks mode, as well as the sites that use at least one such
    document type}
    \begin{tabular}{lrr}
    \toprule
    
    \textbf{Doc. Type (shortened)} &
    \multicolumn{1}{c}{\textbf{Pages}} &
    \multicolumn{1}{c}{\textbf{Sites}} \\

    \midrule

    (none) & 1,818,595 (5.9\,\%) & 56,985 (25.6\,\%) \\
    "-//W3C//DTD HTML 4.01 Transitional//EN" & 721,884 (2.3\,\%) & 18,648 (8.4\,\%) \\
    "-//W3C//DTD HTML 4.0 Transitional//EN" & 385,656 (1.2\,\%) & 11,566 (5.2\,\%) \\
    "-//W3C//DTD HTML 3.2 Final//EN" & 22,019 (<0.1\,\%) & 1,175 (0.5\,\%) \\
    "-//W3C//DTD HTML 3.2//EN" & 10,839 (<0.1\,\%) & 927 (0.4\,\%) \\

    \midrule

    All & 3,046,449 (9.6\,\%) & 71,597 (32.2\,\%) \\

    \bottomrule
    \end{tabular}
    \label{rpo:tab:top_quirksmode_doctypes}
\end{table}

\begin{figure}[t]
    \centering
    \includegraphics[width=0.8\textwidth]{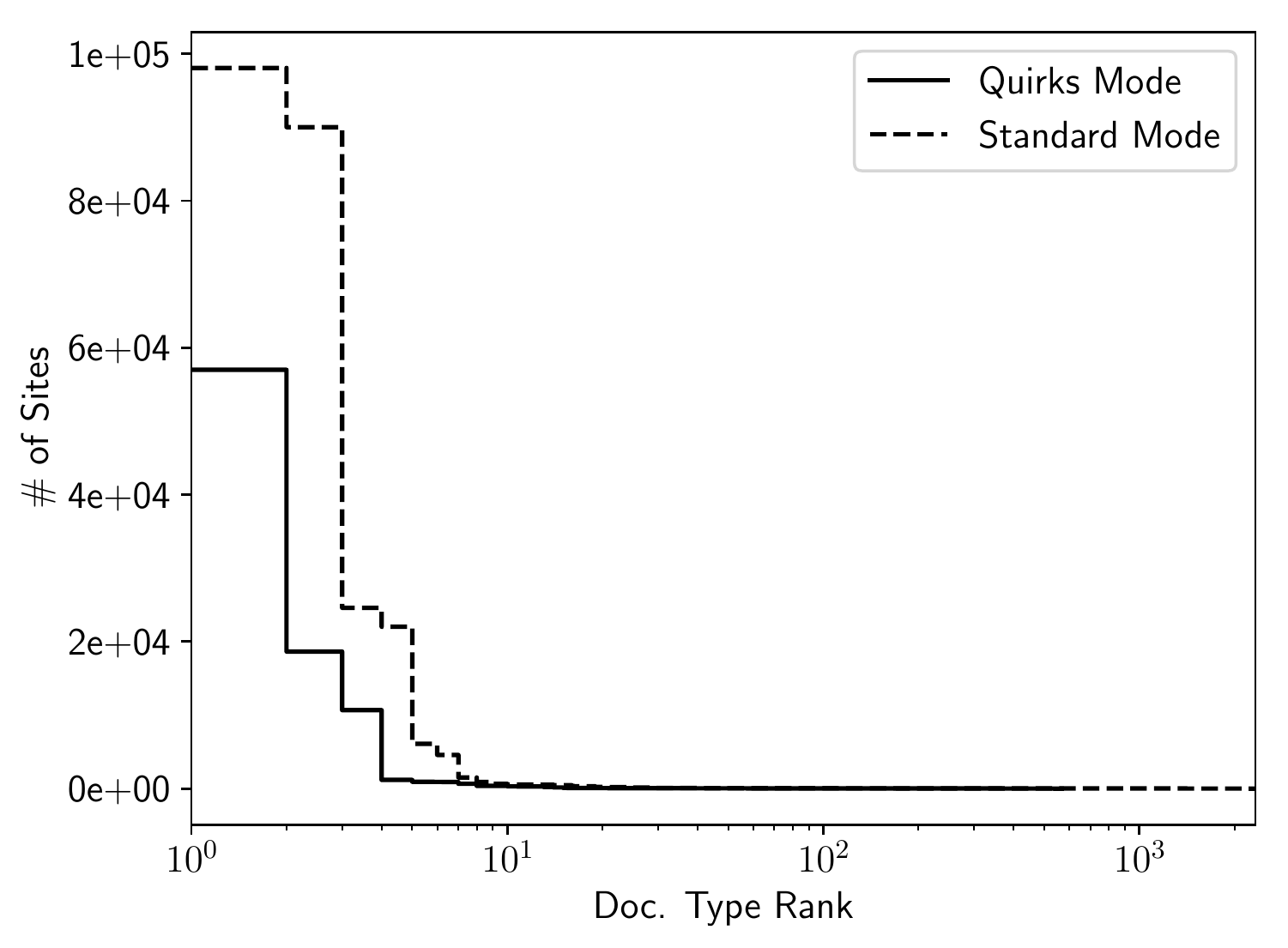}
    \caption{Number of sites containing at least one page with a certain
             document type (ordered by doctype rank).}
    \label{rpo:fig:analysis:doctypes_rank_sites}
\end{figure}

To test how often Internet Explorer allows a page's document type to be
overridden when loading it in an \texttt{iframe}, we created another controlled
experiment using a local attack page framing the victim page, as outlined in
Section~\ref{rpo:sec:methodology:exploitable}. Using Internet Explorer~11, we
loaded our local attack page for each unique document type inside the frame, and
tested if the injected CSS was parsed. While Internet Explorer parsed the
injected CSS for 1,319 (30.5\,\%) of the document types in the default setting,
the frame override trick caused CSS parsing for 4,248 (98.4\,\%) of the unique
document types.

While over one thousand document types result in quirks mode, and around three
thousand document types cause standards mode parsing, the number of document
types that have been standardized is several orders of magnitude smaller. In
fact, only a few (standardized) document types are used frequently in pages,
whereas the majority of unique document types are used very rarely.
Figure~\ref{rpo:fig:analysis:doctypes_rank_sites} shows that only about ten
standards and quirks mode document types are widely used in sites. Furthermore,
only about 9.6\,\% of pages in the candidate set use a quirks mode document
type; on the remaining pages, potential RPO style injection vulnerabilities
cannot be exploited because the CSS would not be parsed (unless Internet
Explorer is used). However, when grouping pages in the candidate set by site,
32.2\,\% of sites contain at least one page rendered in quirks mode
(Table~\ref{rpo:tab:doctypes_summary}), which is one of the preconditions for
successful RPO.

\begin{table}[t]
    \centering
    \footnotesize
    \caption{Summary of document type usage in sites}
    \begin{tabular}{lrr}
    \toprule
    \textbf{Doc. Type} &
    \multicolumn{1}{c}{\textbf{At Least One Page}} &
    \multicolumn{1}{c}{\textbf{All Pages}} \\
    \midrule

    None & 56,985 (25.6\%) & 19,968 (9.0\%) \\
    Quirks & 27,794 (12.5\%) & 7,720 (3.5\%) \\
    None or Quirks & 71,597 (32.2\%) & 30,040 (13.5\%) \\
    \addlinespace
    Standards & 192,403 (86.5\%) & 150,846 (67.8\%) \\
    
    \bottomrule
    \end{tabular}
    \label{rpo:tab:doctypes_summary}
\end{table}

\subsubsection{Internet Explorer Framing}

We showed above that by loading a page in a frame, Internet Explorer can be
forced to disregard a standards mode document type that would prevent
interpretation of injected style. To find out how often this technique can be
applied for successful RPO attacks, we replicated our Chrome experiment in
Internet Explorer, this time loading each vulnerable page inside a frame. Around
14.5\,\% of vulnerable pages were exploitable in Internet Explorer, five times
more than in Chrome (1.6\,\% of the sites in the candidate set as shown in
Table~\ref{rpo:tab:exploitable_result}).

Figure~\ref{rpo:fig:analysis:alexa_rank} shows the combined exploitability
results for Chrome and Internet Explorer according to the rank of the site.
While our methodology did not find any exploitable vulnerability on the six
highest-ranked sites in the candidate set, between 1.6\,\% and 3.2\,\% of
candidate sites in each remaining bucket were found to be exploitable. The
highest exploitability rate occurred in the ranks 1\,k through 10\,k.

Broken down by injection technique, the framing trick in Internet Explorer
results in more exploitable pages for each technique except for cookie injection
(Table~\ref{rpo:tab:exploitable_result}). One possible explanation for this
difference is that the Internet Explorer crawl was conducted one month after the
Chrome crawl, and sites may have changed in the meantime. Furthermore, we
observed two additional impediments to successful exploitation in Internet
Explorer that do not apply to Chrome. The framing technique is susceptible to
frame-busting methods employed by the framed pages, and Internet Explorer
implements an anti-MIME-sniffing header that Chrome appears to ignore. We
analyze these issues below.

\subsubsection{Anti-Framing Techniques}

Some sites use a range of techniques to prevent other pages from loading them in
a frame~\cite{w2sp2010frame_busting}. One of these techniques is the
\texttt{X-Frame-Options} header. It accepts three different values:
\texttt{DENY}, \texttt{SAMEORIGIN}, and \texttt{ALLOW-FROM} followed by a
whitelist of URLs.

In the vulnerable dataset, 4,999 pages across 391 sites use this header
correctly and as a result prevent the attack. However, 1,900 pages across 34
sites provide incorrect values for this header, and we successfully attack 552
pages on 2 sites with Internet Explorer.

A related technique is the \texttt{frame-ancestors} directive provided by
Content Security Policy. It defines a (potentially empty) whitelist of URLs
allowed to load the current page in a frame, similar to \texttt{ALLOW-FROM}.
However, it is not supported by Internet Explorer, thus it cannot be used to
prevent the attack.

Furthermore, developers may use JavaScript code to prevent framing of a page.
Yet, techniques exist to bypass this
protection~\cite{owasp_clickjacking_defence}. In addition, the attacker can use
the HTML 5 \texttt{sandbox} attribute in the \texttt{iframe} tag and omit the
\texttt{allow-top-navigation} directive to render JavaScript frame-busting code
ineffective. However, we did not implement any of these techniques to allow
framing, which means that more vulnerable pages could likely be exploited in
practice.

\subsubsection{MIME Sniffing}

A consequence of self-reference in the type of RPO studied in this paper is that
the HTTP content type of the fake ``stylesheet'' is \texttt{text/html} rather
than the expected \texttt{text/css}. Because many sites contain misconfigured
content types, many browsers attempt to infer the type based on the request
context or file extension (\textit{MIME sniffing}), especially in quirks mode.
In order to ask the browser to disable content sniffing and refuse interpreting
data with an unexpected or wrong type, sites can set the header
\texttt{X-Content-Type-Options: nosniff}~\cite{sp2009contentsniff,firefox_mime_sniff,content_type_options}.

To determine whether the injected CSS is still being parsed and executed in
presence of this header while the browser renders in quirks mode, we ran an
experiment similar to Section~\ref{rpo:sec:analysis:doctypes}. For each browser
in Table~\ref{rpo:tab:doctypes_browsers}, we extracted the document types in
which the browser renders in quirks mode, and for each of them, we set up a
local page with a relative stylesheet path. We then opened the page in the
browser, launched an RPO attack, and monitored if the injected CSS was executed.

Only Firefox, Internet Explorer, and Edge respected this header and did not
interpret injected CSS in any of the quirks mode document types. The remaining
browsers did not block the stylesheet even though the content type was not
\texttt{text/css}. With an additional experiment, we confirmed that Internet
Explorer blocked our injected CSS payload when \texttt{nosniff} was set, even in
the case of the framing technique.

Out of all the vulnerable pages, 96,618 pages across 232 sites had a
\texttt{nosniff} response header; 23 pages across 10 sites were confirmed
exploitable in Chrome but not in Internet Explorer, since the latter browser
respects the header while the former does not.

\subsection{Content Management Systems}
\label{rpo:sec:analysis:cmses}

While analyzing the exploitable pages in our dataset, we noticed that many
appeared to belong to well-known CMSes. Since these web applications are
typically installed on thousands of sites, fixing RPO weaknesses in these
applications could have a large impact.

To identify CMSes, we visited all exploitable pages using
Wappalyzer~\cite{wappalyzer}. Additionally, we detected two CMSes that were not
supported by Wappalyzer. Overall, we identified 23 CMSes on 41,288 pages across
1,589 sites. Afterwards, we manually investigated whether the RPO weakness
stemmed from the CMS by installing the latest version of each CMS (or using the
online demo), and testing whether exploitable paths found in our dataset were
also exploitable in the CMS. After careful analysis, we confirmed four CMSes to
be exploitable in their most recent version that are being used by 40,255 pages
across 1,197 sites.

Out of the four exploitable CMSes, one declares no document type and one uses a
quirks mode document type. These two CMSes can be exploited in Chrome, whereas
the remaining two can be exploited with the framing trick in Internet Explorer.
Beyond the view of our Common Crawl candidate set, Wappalyzer detected nearly
32\,k installations of these CMSes across the Internet, which suggests that many
more sites could be attacked with RPO. We reported the RPO weaknesses to the
vendors of these CMSes using recommended notification
techniques~\cite{usenixsec2016vulnnotify1,usenixsec2016vulnnotify2,weis2017vulnnotify}.
Thus far, we heard back from one of the vendors, who acknowledged the
vulnerability and are going to take the necessary steps to fix the issue.
However, we have not received any response from the other vendors.

\subsection{Mitigation Techniques}
\label{rpo:sec:mitigation}

Relative path overwrites rely on the web server and the web browser interpreting
URLs differently. HTML pages can use only absolute (or root-relative) URLs,
which removes the need for the web browser to expand relative paths.
Alternatively, when the HTML page contains a \texttt{<base>} tag, browsers are
expected to use the URL provided therein to expand relative paths instead of
interpreting the current document's URL. Both methods can remove ambiguities and
render RPO impossible if applied correctly. Specifically, base URLs must be set
according to the server's content routing logic. If developers choose to
calculate base URLs dynamically on the server side rather than setting them
manually to constant values, there is a risk that routing-agnostic algorithms
could be confused by manipulated URLs and re-introduce attack opportunities by
instructing browsers to use an attacker-controlled base URL. Furthermore,
Internet Explorer does not appear to implement this tag correctly.

Web developers can reduce the attack surface of their sites by eliminating any
injection sinks for strings that could be interpreted as a style directive.
However, doing so is challenging because in the attack presented in this paper,
style injection does not require a specific sink type and does not need the
ability of injecting markup. Injection can be accomplished with relatively
commonly used characters, that is, alphanumeric characters and
\texttt{()\{\}/"}. Experience has shown that despite years of efforts, even
context-sensitive and more special character-intensive XSS injection is still
possible in many sites, which leads us to believe that style injection will be
similarly difficult to eradicate. Even when all special characters in user input
are replaced by their corresponding HTML entities and direct style injection is
not possible, more targeted RPO attack variants referencing existing files may
still be feasible. For instance, it has been shown that user uploads of
seemingly benign profile pictures can be used as ``scripts'' (or
stylesheets)~\cite{rpo_techniques}.

Instead of preventing RPO and style injection vulnerabilities, the most
promising approach could be to avoid exploitation. In fact, declaring a modern
document type that causes the HTML document to be rendered in standards mode
makes the attack fail in all browsers except for Internet Explorer. Web
developers can harden their pages against the frame-override technique in
Internet Explorer by using commonly recommended HTTP headers:
\texttt{X-Content-Type-Options} to disable ``content type sniffing'' and always
use the MIME type sent by the server (which must be configured correctly),
\texttt{X-Frame-Options} to disallow loading the page in a frame, and
\texttt{X-UA-Compatible} to turn off Internet Explorer's compatibility view.

\section{Chapter Summary}
\label{rpo:sec:summary}

In this chapter, we showed that over 5\,\% of sites in the intersection of the
Common Crawl and the Alexa Top 1M are vulnerable to at least one injection
technique. While the number of exploitable sites depends on the browser and is
much smaller in relative terms, it is still consequential in absolute terms with
thousands of affected sites. RPO is a class of attacks, and our automated
crawler tested for only a subset of conceivable attacks. Therefore, the results
of our study should be seen as a lower bound; the true number of exploitable
sites is likely higher.

\chapter{Conclusion}
\label{chap:conclusion}

In this thesis, I developed three systems to measure and reduce the security
risks of content inclusions for website publishers as well as their users. More
importantly, our novel techniques are complementary to the existing defenses and
users can browse websites with a higher confidence.

In chapter~\ref{chap:inclusion}, we presented \excision as a complementary
system to other defensive approaches such as CSP and Google Safe Browsing.
\excision incrementally constructs an \emph{inclusion tree} for a given web page
and automatically prevents loading malicious resources by classifying their
\emph{inclusion sequences} using a set of pre-built models. \excision detected a
significant number of malicious third-party content in the wild and was also
able to detect previously unknown malicious inclusions while not impacting
users' browsing experience negatively.

In chapter~\ref{chap:adinjection}, we introduced fine-grained \emph{web content
provenance} tracking and demonstrated its use for identifying unwanted
third-party content (e.g., \emph{injected advertisements}) through
\origintracer, our prototype implementation. Due to the highly interconnected
structure of the web and the oftentimes obscure nature of its trust
relationships, we believe that surfacing this information in the form of
provenance is a generally useful capability, and can be applied in other novel
ways in order to lead to safer and more informed web browsing. Our evaluation
suggests that \origintracer can be used as a complementary system to ad blocking
systems such as AdblockPlus~\cite{adblockplus} and Ghostery~\cite{ghostery}.

In chapter~\ref{chap:rpo}, we presented a systematic study of \emph{style
injection} by \emph{relative path overwrite (RPO)} in the wild. We discussed a
range of factors that prevent a vulnerability from being exploited, and found
that simple countermeasures exist to mitigate RPO. We also linked many
exploitable pages to installations of Content Management Systems (CMSes), and
notified the vendors. Compared to XSS, it is much more challenging to avoid
injection of style directives. Yet, developers have at their disposal a range of
simple mitigation techniques that can prevent their sites from being exploited
in web browsers.

\section{Publications}

This thesis is written based on the following three published papers:

\begin{itemize}

\item \textbf{Chapter~\ref{chap:inclusion}:} Sajjad Arshad, Amin Kharraz,
William Robertson, \ul{Include Me Out: In-Browser Detection of Malicious
Third-Party Content Inclusions}, \textit{Financial Cryptography and Data
Security (FC), 2016}\footnote{https://github.com/sajjadium/Excision}

\item \textbf{Chapter~\ref{chap:adinjection}:} Sajjad Arshad, Amin Kharraz,
William Robertson, \ul{Identifying Extension-based Ad Injection via Fine-grained
Web Content Provenance}, \textit{Research in Attacks, Intrusions and Defenses
(RAID), 2016}\footnote{https://github.com/sajjadium/OriginTracer}

\item \textbf{Chapter~\ref{chap:rpo}:} Sajjad Arshad, Seyed Ali Mirheidari,
Tobias Lauinger, Bruno Crispo, Engin Kirda, William Robertson, \ul{Large-Scale
Analysis of Style Injection by Relative Path Overwrite}, \textit{The Web
Conference (WWW), 2018}

\end{itemize}

Our \emph{inclusion tree} crawler has also been evolving, called
\texttt{DeepCrawling}\footnote{https://github.com/sajjadium/DeepCrawling}, and
was utilized in topics such as tracking and privacy, and web security:

\begin{itemize}

\item Muhammad Ahmad Bashir, Sajjad Arshad, William Robertson, Christo Wilson,
\ul{Tracing Information Flows Between Ad Exchanges Using Retargeted Ads},
\textit{USENIX Security Symposium,
2016}\footnote{http://personalization.ccs.neu.edu/Projects/Retargeting/}

\item Muhammad Ahmad Bashir, Sajjad Arshad, Christo Wilson, \ul{Recommended For
You: A First Look at Content Recommendation Networks}, \textit{ACM Internet
Measurement Conference (IMC),
2016}\footnote{http://personalization.ccs.neu.edu/Projects/Recommended/}

\item Tobias Lauinger, Abdelberi Chaabane, Sajjad Arshad, William Robertson,
Christo Wilson, Engin Kirda, \ul{Thou Shalt Not Depend on Me: Analysing the Use
of Outdated JavaScript Libraries on the Web}, \textit{Network and Distributed
System Security Symposium (NDSS),
2017}\footnote{https://seclab.ccs.neu.edu/static/projects/javascript-libraries/}

\item Muhammad Ahmad Bashir, Sajjad Arshad, Engin Kirda, William Robertson,
Christo Wilson, \ul{How Tracking Companies Circumvented Ad Blockers Using
WebSockets}, \textit{ACM Internet Measurement Conference (IMC), 2018}

\end{itemize}

The author has also been involved with topics other than web security including
malware detection and binary analysis:

\begin{itemize}

\item Amin Kharraz, Sajjad Arshad, Collin Muliner, William Robertson, Engin
Kirda, \ul{UNVEIL: A Large-Scale, Automated Approach to Detecting Ransomware},
\textit{USENIX Security Symposium, 2016}

\item Reza Mirzazade farkhani, Saman Jafari, Sajjad Arshad, William Robertson,
Engin Kirda, Hamed Okhravi, \ul{On the Effectiveness of Type-based Control Flow
Integrity}, \textit{Annual Computer Security Applications Conference (ACSAC),
2018}\footnote{https://github.com/sajjadium/typed-cfi}

\end{itemize}

\pagebreak

\bibliographystyle{plain}
\bibliography{dissertation}

\begin{thebibliography}{100}

\bibitem{google_adinjection}
The ad injection economy.
\newblock
  \url{http://googleonlinesecurity.blogspot.com/2015/05/new-research-ad-injection-economy.html}.

\bibitem{adblockplus}
{Adblock Plus}.
\newblock \url{https://adblockplus.org/}.

\bibitem{adsafe}
{ADsafe}.
\newblock \url{http://www.adsafe.org/}.

\bibitem{csp_content_scripts}
{CSP in Content Scripts}.
\newblock
  \url{https://developer.chrome.com/extensions/contentSecurityPolicy#interactions}.

\bibitem{ghostery}
Ghostery.
\newblock \url{https://www.ghostery.com/en/}.

\bibitem{phantomjs}
{PhantomJS}.
\newblock \url{http://phantomjs.org/}.

\bibitem{sloccount}
{SLOCCount}.
\newblock \url{http://www.dwheeler.com/sloccount/}.

\bibitem{virustotal}
{VirtusTotal}.
\newblock \url{https://www.virustotal.com/}.

\bibitem{cors_spec}
{Cross-Origin Resource Sharing (CORS)}.
\newblock \url{http://www.w3.org/TR/cors/}, 2014.

\bibitem{csp_spec}
{Content Security Policy 1.1}.
\newblock
  \url{https://dvcs.w3.org/hg/content-security-policy/raw-file/tip/csp_specification.dev.html},
  2015.

\bibitem{debugging_protocol}
Chrome remote debugging protocol.
\newblock \url{https://chromedevtools.github.io/devtools-protocol/}, 2018.

\bibitem{asiaccs2012flashover}
Steven~Van Acker, Nick Nikiforakis, Lieven Desmet, Wouter Joosen, and Frank
  Piessens.
\newblock {FlashOver}: Automated discovery of cross-site scripting
  vulnerabilities in rich internet applications.
\newblock In {\em ACM Symposium on Information, Computer and Communications
  Security ({ASIACCS})}, 2012.

\bibitem{alexa_top_1m}
Alexa.
\newblock Top sites.
\newblock \url{http://www.alexa.com/topsites}, 2016.

\bibitem{sp2009contentsniff}
Adam Barth, Juan Caballero, and Dawn Song.
\newblock Secure content sniffing for web browsers, or how to stop papers from
  reviewing themselves.
\newblock In {\em {IEEE} Symposium on Security and Privacy ({S\&P})}, 2009.

\bibitem{tr2008chromium_security_architecture}
Adam Barth, Collin Jackson, Charles Reis, and {The Google Chrome Team}.
\newblock The security architecture of the chromium browser.
\newblock Technical report, 2008.

\bibitem{www2010xss_auditor}
Daniel Bates, Adam Barth, and Collin Jackson.
\newblock Regular expressions considered harmful in client-side xss filters.
\newblock In {\em International World Wide Web Conference ({WWW})}, 2010.

\bibitem{ndss2015infoflows}
Lujo Bauer, Shaoying Cai, Limin Jia, Timothy Passaro, Michael Stroucken, and
  Yuan Tian.
\newblock Run-time monitoring and formal analysis of information flows in
  {Chromium}.
\newblock In {\em Network and Distributed System Security Symposium ({NDSS})},
  2015.

\bibitem{ndss2011exposure}
Leyla Bilge, Engin Kirda, Christopher Kruegel, and Marco Balduzzi.
\newblock {EXPOSURE}: Finding malicious domains using passive {DNS} analysis.
\newblock In {\em Network and Distributed System Security Symposium ({NDSS})},
  2011.

\bibitem{dimva2008xss_guard}
Prithvi Bisht and V.~N. Venkatakrishnan.
\newblock {XSS-GUARD}: Precise dynamic prevention of cross-site scripting
  attacks.
\newblock In {\em Detection of Intrusions and Malware, and Vulnerability
  Assessment ({DIMVA})}, 2008.

\bibitem{burpsuite}
{Burp Suite}.
\newblock \url{https://portswigger.net/burp/}, 2017.

\bibitem{weis2017vulnnotify}
Orcun Cetin, Carlos Ganan, Maciej Korczynski, and Michel van Eeten.
\newblock Make notifications great again: Learning how to notify in the age of
  large-scale vulnerability scanning.
\newblock In {\em Workshop on the Economics of Information Security ({WEIS})},
  2017.

\bibitem{usenixsec2007sif}
Stephen Chong, K.~Vikram, and Andrew~C. Myers.
\newblock {SIF}: Enforcing confidentiality and integrity in web applications.
\newblock In {\em {USENIX} Security Symposium}, 2007.

\bibitem{isp_adinjection}
Devin Coldewey.
\newblock Marriott puts an end to shady ad injection service.
\newblock
  \url{http://techcrunch.com/2012/04/09/marriott-puts-an-end-to-shady-ad-injection-service/},
  2012.

\bibitem{isp_adinjection2}
Devin Coldewey.
\newblock Marriott puts an end to shady ad injection service.
\newblock
  \url{http://techcrunch.com/2012/04/09/marriott-puts-an-end-to-shady-ad-injection-service/},
  2012.

\bibitem{www2010jsand}
Marco Cova, Christopher Kruegel, and Giovanni Vigna.
\newblock Detection and analysis of drive-by-download attacks and malicious
  {JavaScript} code.
\newblock In {\em International World Wide Web Conference ({WWW})}, 2010.

\bibitem{common_crawl}
Common Crawl.
\newblock \url{https://commoncrawl.org/}, August 2016.

\bibitem{rpo_iis}
Soroush Dalili.
\newblock Non-root-relative path overwrite ({RPO}) in {IIS} and {.Net}
  applications.
\newblock
  \url{https://soroush.secproject.com/blog/2015/02/non-root-relative-path-overwrite-rpo-in-iis-and-net-applications/},
  2015.

\bibitem{chi2006phishing}
Rachna Dhamija, J.~D. Tygar, and Marti Hearst.
\newblock Why phishing works.
\newblock In {\em Proceedings of the {SIGCHI} Conference on Human Factors in
  Computing Systems ({CHI})}, 2006.

\bibitem{acsac2009sabre}
Mohan Dhawan and Vinod Ganapathy.
\newblock Analyzing information flow in {JavaScript}-based browser extensions.
\newblock In {\em Annual Computer Security Applications Conference ({ACSAC})},
  2009.

\bibitem{acsac2011adsentry}
Xinshu Dong, Minh Tran, Zhenkai Liang, and Xuxian Jiang.
\newblock {AdSentry}: Comprehensive and flexible confinement of
  {JavaScript}-based advertisements.
\newblock In {\em Annual Computer Security Applications Conference ({ACSAC})},
  2011.

\bibitem{ccs2013dedacota}
Adam Doupe, Weidong Cui, Mariusz~H. Jakubowski, Marcus Peinado, Christopher
  Kruegel, and Giovanni Vigna.
\newblock {deDacota}: Toward preventing server-side {XSS} via automatic code
  and data separation.
\newblock In {\em {ACM} Conference on Computer and Communications Security
  ({CCS})}, 2013.

\bibitem{sosp2005asbestos}
Petros Efstathopoulos, Maxwell Krohn, Steve VanDeBogart, Cliff Frey, David
  Ziegler, Eddie Kohler, David Mazieres, Frans Kaashoek, and Robert Morris.
\newblock Labels and event processes in the asbestos operating system.
\newblock In {\em {ACM} Symposium on Operating Systems Principles ({SOSP})},
  2005.

\bibitem{usenixatc2007spyware_analysis}
Manuel Egele, Christopher Kruegel, Engin Kirda, Heng Yin, and Dawn Song.
\newblock Dynamic spyware analysis.
\newblock In {\em {USENIX} Annual Technical Conference ({ATC})}, 2007.

\bibitem{usenixwebapps2011app_permissions}
Adrienne~Porter Felt, Kate Greenwood, and David Wagner.
\newblock The effectiveness of application permissions.
\newblock In {\em {USENIX} Conference on Web Application Development
  ({WebApps})}, 2011.

\bibitem{ndss2010capabilityleaks}
Matthew Finifter, Joel Weinberger, and Adam Barth.
\newblock Preventing capability leaks in secure {JavaScript} subsets.
\newblock In {\em Network and Distributed System Security Symposium ({NDSS})},
  2010.

\bibitem{middleware2012spade}
Ashish Gehani and Dawood Tariq.
\newblock {SPADE}: Support for provenance auditing in distributed environments.
\newblock In {\em International Middleware Conference}, 2012.

\bibitem{osdi2012hails}
Daniel~B. Giffin, Amit Levy, Deian Stefan, David Terei, David Mazieres, John~C.
  Mitchell, and Alejandro Russo.
\newblock Hails: Protecting data privacy in untrusted web applications.
\newblock In {\em {USENIX} Symposium on Operating Systems Design and
  Implementation (OSDI)}, 2012.

\bibitem{blackhatusa2017webcache}
Omer Gil.
\newblock Web cache deception attack.
\newblock In {\em {Black Hat USA}}, 2017.

\bibitem{webcache_deception}
Omer Gil.
\newblock Web cache deception attack.
\newblock
  \url{http://omergil.blogspot.com/2017/02/web-cache-deception-attack.html},
  2017.

\bibitem{gsb}
{Google, Inc.}
\newblock {Google Safe Browsing API}.
\newblock \url{https://developers.google.com/safe-browsing/}, 2015.

\bibitem{sp2008op}
Chris Grier, Shuo Tang, and Samuel~T. King.
\newblock Secure web browsing with the {OP} web browser.
\newblock In {\em {IEEE} Symposium on Security and Privacy ({Oakland})}, 2008.

\bibitem{usenixsec2009gatekeeper}
Salvatore Guarnieri and Benjamin Livshits.
\newblock {GATEKEEPER}: Mostly static enforcement of security and reliability
  policies for {JavaScript} code.
\newblock In {\em {USENIX} Security Symposium}, 2009.

\bibitem{sp2011ibex}
Arjun Guha, Matthew Fredrikson, Benjamin Livshits, and Nikhil Swamy.
\newblock Verified security for browser extensions.
\newblock In {\em {IEEE} Symposium on Security and Privacy ({Oakland})}, 2011.

\bibitem{propr2007provenance_web}
Andreas Harth, Axel Polleres, and Stefan Decker.
\newblock Towards a social provenance model for the web.
\newblock In {\em Workshop on Principles of Provenance ({PrOPr})}, 2007.

\bibitem{ldow2009provenance_web}
Olaf Hartig.
\newblock Provenance information in the web of data.
\newblock In {\em Workshop on Linked Data on the Web ({LDOW})}, 2009.

\bibitem{ccs2009sprov}
Ragib Hasan, Radu Sion, and Marianne Winslett.
\newblock {SPROV} 2.0: A highly configurable platform-independent library for
  secure provenance.
\newblock In {\em {ACM} Conference on Computer and Communications Security
  ({CCS})}, 2009.

\bibitem{ccs2012scriptless}
Mario Heiderich, Marcus Niemietz, Felix Schuster, Thorsten Holz, and J\"org
  Schwenk.
\newblock Scriptless attacks - stealing the pie without touching the sill.
\newblock In {\em {ACM} Conference on Computer and Communications Security
  ({CCS})}, 2012.

\bibitem{esorics2017dompurify}
Mario Heiderich, Christopher Sp\"ath, and J\"org Schwenk.
\newblock Dompurify: Client-side protection against xss and markup injection.
\newblock In {\em European Conference on Research in Computer Security
  ({ESORICS})}, 2017.

\bibitem{sexy_assassin_css}
Gareth Heyes.
\newblock The sexy assassin: Tactical exploitation using {CSS}.
\newblock
  \url{https://docs.google.com/viewer?url=www.businessinfo.co.uk/labs/talk/The_Sexy_Assassin.ppt},
  2009.

\bibitem{rpo}
Gareth Heyes.
\newblock {RPO}.
\newblock \url{http://www.thespanner.co.uk/2014/03/21/rpo/}, 2014.

\bibitem{sacmat2010e2eaccctrl}
Boniface Hicks, Sandra Rueda, Dave King, Thomas Moyer, Joshua Schiffman, Yogesh
  Sreenivasan, Patrick McDaniel, and Trent Jaeger.
\newblock An architecture for enforcing end-to-end access control over web
  applications.
\newblock In {\em {ACM} Symposium on Access Control Models and Technologies
  ({SACMAT})}, 2010.

\bibitem{ccs2010cssattack}
Lin-Shung Huang, Zack Weinberg, Chris Evans, and Collin Jackson.
\newblock Protecting browsers from cross-origin {CSS} attacks.
\newblock In {\em Proceedings of the {ACM} Conference on Computer and
  Communications Security ({CCS})}, 2010.

\bibitem{ccs2010cross_origin_css}
Lin-Shung Huang, Zack Weinberg, Chris Evans, and Collin Jackson.
\newblock Protecting browsers from cross-origin {CSS} attacks.
\newblock In {\em {ACM} Conference on Computer and Communications Security
  ({CCS})}, 2010.

\bibitem{usenixsec2015webeval}
Nav Jagpal, Eric Dingle, Jean-Philippe Gravel, Panayiotis Mavrommatis, Niels
  Provos, Moheeb~Abu Rajab, and Kurt Thomas.
\newblock Trends and lessons from three years fighting malicious extensions.
\newblock In {\em {USENIX} Security Symposium}, 2015.

\bibitem{w2sp2010history_sniffing}
Artur Janc and Lukasz Olejnik.
\newblock Feasibility and real-world implications of web browser history
  detection.
\newblock In {\em Web 2.0 Security and Privacy ({W2SP})}, 2010.

\bibitem{icdcs2010escudo}
Karthick Jayaraman, Wenliang Du, Balamurugan Rajagopalan, and Steve~J. Chapin.
\newblock {ESCUDO}: A fine-grained protection model for web browsers.
\newblock In {\em 30th {IEEE} International Conference on Distributed Computing
  Systems ({ICDCS})}, 2010.

\bibitem{sec2011deseo}
John~P. John, Fang Yu, Yinglian Xie, Arvind Krishnamurthy, and Martin Abadi.
\newblock {deSEO}: Combating search-result poisoning.
\newblock In {\em {USENIX} Security Symposium}, 2011.

\bibitem{usenixsec2014hulk}
Alexandros Kapravelos, Chris Grier, Neha Chachra, Chris Kruegel, Giovanni
  Vigna, and Vern Paxson.
\newblock Hulk: Eliciting malicious behavior in browser extensions.
\newblock In {\em {USENIX} Security Symposium}, 2014.

\bibitem{commacm2014tangled_web}
Christoph Kern.
\newblock Securing the tangled web.
\newblock {\em Communications of the {ACM}}, 57, no. 9:38--47, 2014.

\bibitem{firefox_mime_sniff}
Christoph Kerschbaumer.
\newblock Mitigating {MIME} confusion attacks in firefox.
\newblock
  \url{https://blog.mozilla.org/security/2016/08/26/mitigating-mime-confusion-attacks-in-firefox/},
  2016.

\bibitem{prssi}
James Kettle.
\newblock Detecting and exploiting path-relative stylesheet import ({PRSSI})
  vulnerabilities.
\newblock \url{http://blog.portswigger.net/2015/02/prssi.html}, 2015.

\bibitem{css_font_face}
Masato Kinugawa.
\newblock {CSS} based attack: Abusing unicode-range of {@font-face}.
\newblock
  \url{http://mksben.l0.cm/2015/10/css-based-attack-abusing-unicode-range.html},
  2015.

\bibitem{sosp2007flume}
Maxwell Krohn, Alexander Yip, Micah Brodsky, Natan Cliffer, M.~Frans Kaashoek,
  Eddie Kohler, and Robert Morris.
\newblock Information flow control for standard os abstractions.
\newblock In {\em Symposium on Operating Systems Principles ({SOSP})}, 2007.

\bibitem{isp_adinjection1}
Greg Kumparak.
\newblock Real evil: {ISP} inserted advertising.
\newblock
  \url{http://techcrunch.com/2007/06/23/real-evil-isp-inserted-advertising/},
  2007.

\bibitem{csp_nonce_bypass}
Sebastian Lekies.
\newblock How to bypass {CSP} nonces with {DOM} {XSS}.
\newblock
  \url{http://sirdarckcat.blogspot.com/2016/12/how-to-bypass-csp-nonces-with-dom-xss.html},
  2016.

\bibitem{ccs2017script_gadgets}
Sebastian Lekies, Krzysztof Kotowicz, Samuel Grob, Eduardo A.~Vela Nava, and
  Martin Johns.
\newblock Code-reuse attacks for the web: Breaking cross-site scripting
  mitigations via script gadgets.
\newblock In {\em {ACM} Conference on Computer and Communications Security
  ({CCS})}, 2017.

\bibitem{blackhatusa2017script_gadgets}
Sebastian Lekies, Krzysztof Kotowicz, and Eduardo~Vela Nava.
\newblock Breaking xss mitigations via script gadgets.
\newblock In {\em {Black Hat USA}}, 2017.

\bibitem{ccs2013domxss}
Sebastian Lekies, Ben Stock, and Martin Johns.
\newblock 25 million flows later - large-scale detection of {DOM}-based {XSS}.
\newblock In {\em {ACM} Conference on Computer and Communications Security
  ({CCS})}, 2013.

\bibitem{ecml1998naivebayes}
David~D. Lewis.
\newblock Naive (bayes) at forty: The independence assumption in information
  retrieval.
\newblock In {\em European Conference on Machine Learning ({ECML})}, 1998.

\bibitem{usenixsec2016vulnnotify1}
Frank Li, Zakir Durumeric, Jakub Czyz, Mohammad Karami, Michael Bailey, Damon
  McCoy, Stefan Savage, and Vern Paxson.
\newblock You've got vulnerability: Exploring effective vulnerability
  notifications.
\newblock In {\em {USENIX} Security Symposium}, 2016.

\bibitem{sp2013linchpins}
Zhou Li, Sumayah Alrwais, Yinglian Xie, Fang Yu, and XiaoFeng Wang.
\newblock Finding the linchpins of the dark web: a study on topologically
  dedicated hosts on malicious web infrastructures.
\newblock In {\em {IEEE} Symposium on Security and Privacy ({Oakland})}, 2013.

\bibitem{ccs2012madtracer}
Zhou Li, Kehuan Zhang, Yinglian Xie, Fang Yu, and XiaoFeng Wang.
\newblock Knowing your enemy: Understanding and detecting malicious web
  advertising.
\newblock In {\em {ACM} Conference on Computer and Communications Security
  ({CCS})}, 2012.

\bibitem{raid2007spyshield}
Zhuowei Li, XiaoFeng Wang, and Jong~Youl Choi.
\newblock {SpyShield}: Preserving privacy from spy add-ons.
\newblock In {\em International Conference on Recent Advances in Intrusion
  Detection ({RAID})}, 2007.

\bibitem{dsn2014scriptless_timing}
Bin Liang, Wei You, Liangkun Liu, Wenchang Shi, and Mario Heiderich.
\newblock Scriptless timing attacks on web browser privacy.
\newblock In {\em {IEEE/IFIP} International Conference on Dependable Systems
  and Networks ({DSN})}, 2014.

\bibitem{ndss2012chrome_extensions}
Lei Liu, Xinwen Zhang, Guanhua Yan, and Songqing Chen.
\newblock Chrome extensions: Threat analysis and countermeasures.
\newblock In {\em Network and Distributed System Security Symposium ({NDSS})},
  2012.

\bibitem{blackhatasia2014domxss}
Nera W.~C. Liu and Albert Yu.
\newblock Ultimate {DOM} based {XSS} detection scanner on cloud.
\newblock In {\em {Black Hat Asia}}, 2014.

\bibitem{usenixsec2010adjail}
Mike~Ter Louw, Karthik~Thotta Ganesh, and V.N. Venkatakrishnan.
\newblock {AdJail}: Practical enforcement of confidentiality and integrity
  policies on web advertisements.
\newblock In {\em {USENIX} Security Symposium}, 2010.

\bibitem{springer2008browserspy}
Mike~Ter Louw, Jin~Soon Lim, and V.~N. Venkatakrishnan.
\newblock Enhancing web browser security against malware extensions.
\newblock {\em Journal in Computer Virology}, 4(3):179--195, 2008.

\bibitem{sp2009blueprint}
Mike~Ter Louw and V.N. Venkatakrishnan.
\newblock {BLUEPRINT}: Robust prevention of cross-site scripting attacks for
  existing browsers.
\newblock In {\em {IEEE} Symposium on Security and Privacy ({S\&P})}, 2009.

\bibitem{csf2009langisojs}
Sergio Maffeis and Ankur Taly.
\newblock Language-based isolation of untrusted {JavaScript}.
\newblock In {\em {IEEE} Computer Security Foundations Symposium ({CSF})},
  2009.

\bibitem{noscript}
Giorgio Maone.
\newblock {NoScript}.
\newblock \url{https://noscript.net/}, 2009.

\bibitem{adinjection_profit}
Ginny Marvin.
\newblock Google study exposes "tangled web" of companies profiting from ad
  injection.
\newblock \url{http://marketingland.com/ad-injector-study-google-127738}, 2015.

\bibitem{content_type_options}
{MDN}.
\newblock {X-Content-Type-Options}.
\newblock
  \url{https://developer.mozilla.org/en-US/docs/Web/HTTP/Headers/X-Content-Type-Options},
  2018.

\bibitem{sp2010conscript}
Leo~A. Meyerovich and Benjamin Livshits.
\newblock {ConScript}: Specifying and enforcing fine-grained security policies
  for {JavaScript} in the browser.
\newblock In {\em {IEEE} Symposium on Security and Privacy ({Oakland})}, 2010.

\bibitem{cvlist}
{Microsoft}.
\newblock Understanding the compatibility view list.
\newblock
  \url{https://msdn.microsoft.com/en-us/library/gg699485(v=vs.85).aspx}, 2015.

\bibitem{websci2010provenance_web}
Luc Moreau.
\newblock The foundations for provenance on the web.
\newblock {\em Foundations and Trends in Web Science}, 2(2--3):99--241,
  February 2010.

\bibitem{popl1999jflow}
Andrew~C. Myers.
\newblock {JFlow}: Practical mostly-static information flow control.
\newblock In {\em Symposium on Principles of Programming Languages ({POPL})},
  1999.

\bibitem{ndss2009dsi}
Yacin Nadji, Prateek Saxena, and Dawn Song.
\newblock Document structure integrity: A robust basis for cross-site scripting
  defense.
\newblock In {\em Network and Distributed System Security Symposium ({NDSS})},
  2009.

\bibitem{ndss2009doc_integrity}
Yacin Nadji, Prateek Saxena, and Dawn Song.
\newblock Document structure integrity: A robust basis for cross-site scripting
  defense.
\newblock In {\em Network and Distributed System Security Symposium ({NDSS})},
  2009.

\bibitem{usenixsec2015webwitness}
Terry Nelms, Roberto Perdisci, Manos Antonakakis, and Mustaque Ahamad.
\newblock {WebWitness}: Investigating, categorizing, and mitigating malware
  download paths.
\newblock In {\em {USENIX} Security Symposium}, 2015.

\bibitem{ccs2012jsinclusion}
Nick Nikiforakis, Luca Invernizzi, Alexandros Kapravelos, Steven~Van Acker,
  Wouter Joosen, Christopher Kruegel, Frank Piessens, , and Giovanni Vigna.
\newblock You are what you include: Large-scale evaluation of remote
  {JavaScript} inclusions.
\newblock In {\em {ACM} Conference on Computer and Communications Security
  ({CCS})}, 2012.

\bibitem{www2014shortening}
Nick Nikiforakis, Federico Maggi, Gianluca Stringhini, M~Rafique, Wouter
  Joosen, Christopher Kruegel, Frank Piessens, Giovanni Vigna, and Stefano
  Zanero.
\newblock Stranger danger: Exploring the ecosystem of ad-based {URL} shortening
  services.
\newblock In {\em International World Wide Web Conference ({WWW})}, 2014.

\bibitem{ccs2008soma}
Terri Oda, Glenn Wurster, P.~C. van Oorschot, and Anil Somayaji.
\newblock {SOMA}: Mutual approval for included content in web pages.
\newblock In {\em {ACM} Conference on Computer and Communications Security
  ({CCS})}, 2008.

\bibitem{owasp_xss}
{OWASP}.
\newblock Cross-site scripting ({XSS}).
\newblock \url{https://www.owasp.org/index.php/Cross-site_Scripting_(XSS)},
  2016.

\bibitem{owasp_clickjacking_defence}
{OWASP}.
\newblock Clickjacking defense cheat sheet.
\newblock
  \url{https://www.owasp.org/index.php/Clickjacking_Defense_Cheat_Sheet}, 2017.

\bibitem{owasp_csrf_defense}
{OWASP}.
\newblock Cross-site request forgery (csrf) prevention cheat sheet.
\newblock
  \url{https://www.owasp.org/index.php/Cross-Site_Request_Forgery_(CSRF)_Prevention_Cheat_Sheet},
  2017.

\bibitem{owasp_xss_defense}
{OWASP}.
\newblock {XSS} (cross site scripting) prevention cheat sheet.
\newblock
  \url{https://www.owasp.org/index.php/XSS_(Cross_Site_Scripting)_Prevention_Cheat_Sheet},
  2017.

\bibitem{asiaccs2009lightjs}
Phu~H. Phung, David Sands, and Andrey Chudnov.
\newblock Lightweight self-protecting {JavaScript}.
\newblock In {\em {ACM} Symposium on Information, Computer, and Communications
  Security ({ASIACCS})}, 2009.

\bibitem{acsac2012hifi}
Devin~J. Pohly, Stephen McLaughlin, and Kevin Butler.
\newblock {Hi-Fi}: Collecting high-fidelity whole-system provenance.
\newblock In {\em Annual Computer Security Applications Conference ({ACSAC})},
  2012.

\bibitem{ieee1989hmm}
Lawrence~R. Rabiner.
\newblock A tutorial on {Hidden Markov Models} and selected applications in
  speech recognition.
\newblock {\em Proceedings of the {IEEE}}, 77(2):257--285, 1989.

\bibitem{dsn2015segugio}
Babak Rahbarinia, Roberto Perdisci, and Manos Antonakakis.
\newblock Segugio: Efficient behavior-based tracking of new malware-control
  domains in large isp networks.
\newblock In {\em {IEEE/IFIP} International Conference on Dependable Systems
  and Networks ({DSN})}, 2015.

\bibitem{usenixosdi2006browsershield}
Charles Reis, John Dunagan, Helen~J. Wang, Opher Dubrovsky, and Saher Esmeir.
\newblock {BrowserShield}: Vulnerability-driven filtering of dynamic {HTML}.
\newblock In {\em {USENIX} Symposium on Operating Systems Design and
  Implementation ({OSDI})}, 2006.

\bibitem{nsdi2008tripwires}
Charles Reis, Steven~D. Gribble, Tadayoshi Kohno, and Nicholas~C. Weaver.
\newblock Detecting in-flight page changes with web {Tripwires}.
\newblock In {\em {USENIX} Symposium on Networked Systems Design and
  Implementation ({NSDI})}, 2008.

\bibitem{xss_filter}
David Ross.
\newblock {IE} 8 {XSS} filter architecture / implementation.
\newblock
  \url{https://blogs.technet.microsoft.com/srd/2008/08/19/ie-8-xss-filter-architecture-implementation/},
  2008.

\bibitem{w2sp2010frame_busting}
Gustav Rydstedt, Elie Bursztein, Dan Boneh, and Collin Jackson.
\newblock Busting frame busting: a study of clickjacking vulnerabilities on
  popular sites.
\newblock In {\em {IEEE} Oakland Web 2.0 Security and Privacy ({W2SP})}, 2010.

\bibitem{ccs2011csas}
Mike Samuel, Prateek Saxena, and Dawn Song.
\newblock Context-sensitive auto-sanitization in web templating languages using
  type qualifiers.
\newblock In {\em {ACM} Conference on Computer and Communications Security
  ({CCS})}, 2011.

\bibitem{selenium}
{Selenium Contributors}.
\newblock Selenium: Web browser automation.
\newblock \url{http://www.seleniumhq.org/}.

\bibitem{browser_modes_doctype}
Henri Sivonen.
\newblock Activating browser modes with doctype.
\newblock \url{https://hsivonen.fi/doctype/}, 2013.

\bibitem{ndss2013postman}
Sooel Son and Vitaly Shmatikov.
\newblock The postman always rings twice: Attacking and defending {postMessage}
  in {HTML5} websites.
\newblock In {\em Network and Distributed System Security Symposium ({NDSS})},
  2013.

\bibitem{www2010csp}
Sid Stamm, Brandon Sterne, and Gervase Markham.
\newblock Reining in the web with content security policy.
\newblock In {\em International World Wide Web Conference ({WWW})}, 2010.

\bibitem{usenixsec2014client_side_xss}
Ben Stock, Sebastian Lekies, Tobias Mueller, Patrick Spiegel, and Martin Johns.
\newblock Precise client-side protection against {DOM}-based cross-site
  scripting.
\newblock In {\em {USENIX} Security Symposium}, 2014.

\bibitem{usenixsec2016vulnnotify2}
Ben Stock, Giancarlo Pellegrino, Christian Rossow, Martin Johns, and Michael
  Backes.
\newblock Hey, you have a problem: On the feasibility of large-scale web
  vulnerability notification.
\newblock In {\em {USENIX} Security Symposium}, 2016.

\bibitem{imc2011adexchange}
Brett Stone-Gross, Ryan Stevens, Richard Kemmerer, Christopher Kruegel,
  Giovanni Vigna, and Apostolis Zarras.
\newblock Understanding fraudulent activities in online ad exchanges.
\newblock In {\em Internet Measurement Conference ({IMC})}, 2011.

\bibitem{ccs2013spiderweb}
Gianluca Stringhini, Christopher Kruegel, and Giovanni Vigna.
\newblock Shady paths: Leveraging surfing crowds to detect malicious web pages.
\newblock In {\em {ACM} Conference on Computer and Communications Security
  ({CCS})}, 2013.

\bibitem{usenixosdi2010illinois}
Shuo Tang, Haohui Mai, and Samuel~T. King.
\newblock Trust and protection in the {Illinois} browser operating system.
\newblock In {\em {USENIX} Symposium on Operating Systems Design and
  Implementation ({OSDI})}, 2010.

\bibitem{rpo_techniques}
Takeshi Terada.
\newblock A few {RPO} exploitation techniques.
\newblock \url{https://www.mbsd.jp/Whitepaper/rpo.pdf}, 2015.

\bibitem{sp2015adinjection}
Kurt Thomas, Elie Bursztein, Chris Grier, Grant Ho, Nav Jagpal, Alexandros
  Kapravelos, Damon McCoy, Antonio Nappa, Vern Paxson, Paul Pearce, Niels
  Provos, and Moheeb~Abu Rajab.
\newblock Ad injection at scale: Assessing deceptive advertisement
  modifications.
\newblock In {\em {IEEE} Symposium on Security and Privacy ({Oakland})}, 2015.

\bibitem{acns2012leaktracker}
Minh Tran, Xinshu Dong, Zhenkai Liang, and Xuxian Jiang.
\newblock Tracking the trackers: Fast and scalable dynamic analysis of web
  content for privacy violations.
\newblock In {\em Proceedings of the 10th international conference on Applied
  Cryptography and Network Security ({ACNS})}, pages 418--435, 2012.

\bibitem{css_syntax}
{W3C}.
\newblock Css syntax and basic data types.
\newblock \url{http://www.w3.org/TR/CSS2/syndata.html}, 2011.

\bibitem{w3c_csp}
{W3C}.
\newblock Content security policy level 2.
\newblock \url{https://www.w3.org/TR/CSP2/}, 2015.

\bibitem{usenixsec2009gazelle}
Helen~J. Wang, Chris Grier, Alexander Moshchuk, Samuel~T. King, Piali
  Choudhury, and Herman Venter.
\newblock The multi-principal {OS} construction of the {Gazelle} web browser.
\newblock In {\em {USENIX} Security Symposium}, 2009.

\bibitem{wappalyzer}
Wappalyzer.
\newblock Identify technologies on websites.
\newblock \url{https://www.wappalyzer.com/}, 2017.

\bibitem{ccs2016cspisdead}
Lukas Weichselbaum, Michele Spagnuolo, Sebastian Lekies, and Artur Janc.
\newblock Csp is dead, long live csp! on the insecurity of whitelists and the
  future of content security policy.
\newblock In {\em {ACM} Conference on Computer and Communications Security
  ({CCS})}, 2016.

\bibitem{esorics2011xss}
Joel Weinberger, Prateek Saxena, Devdatta Akhawe, Matthew Finifter, Richard
  Shin, and Dawn Song.
\newblock An empirical analysis of {XSS} sanitization in web application
  frameworks.
\newblock In {\em European Conference on Research in Computer Security
  ({ESORICS})}, 2011.

\bibitem{raid2014csp}
Michael Weissbacher, Tobias Lauinger, and William Robertson.
\newblock Why is {CSP} failing? trends and challenges in {CSP} adoption.
\newblock In {\em International Conference on Recent Advances in Intrusion
  Detection ({RAID})}, 2014.

\bibitem{domtree}
{World Wide Web Consortium (W3C)}.
\newblock What is the document object model?
\newblock \url{http://www.w3.org/TR/DOM-Level-2-Core/introduction.html}.

\bibitem{www2015adinjection}
Xinyu Xing, Wei Meng, Udi Weinsberg, Anmol Sheth, Byoungyoung Lee, Roberto
  Perdisci, and Wenke Lee.
\newblock Unraveling the relationship between ad-injecting browser extensions
  and malvertising.
\newblock In {\em International World Wide Web Conference ({WWW})}, 2015.

\bibitem{csscripting}
{XSS Jigsaw}.
\newblock {CSS}: Cascading style scripting.
\newblock \url{http://blog.innerht.ml/cascading-style-scripting/}, 2015.

\bibitem{rpo_gadgets}
{XSS Jigsaw}.
\newblock {RPO} gadgets.
\newblock \url{http://blog.innerht.ml/rpo-gadgets/}, 2016.

\bibitem{imc2014malvertising}
Apostolis Zarras, Alexandros Kapravelos, Gianluca Stringhini, Thorsten Holz,
  Christopher Kruegel, and Giovanni Vigna.
\newblock The dark alleys of madison avenue: Understanding malicious
  advertisements.
\newblock In {\em Proceedings of the Internet Measurement Conference ({IMC})},
  2014.

\bibitem{nsdi2008dstar}
Nickolai Zeldovich, Silas Boyd-Wickizer, and David Mazieres.
\newblock Security distributed systems with information flow control.
\newblock In {\em {USENIX} Symposium on Networked Systems Design and
  Implementation ({NSDI})}, 2008.

\end{thebibliography}

\end{document}